\documentclass[english,11pt,a4paper]{article}

\pdfoutput=1

\usepackage{jcappub}
\usepackage{amsmath}
\usepackage{amsfonts}
\usepackage{amssymb}
\usepackage{graphicx, rotating}
\usepackage{epstopdf}
\usepackage{epsfig}
\usepackage{latexsym}
\usepackage{multirow}
\usepackage{slashed}
\usepackage[top=2.5 cm, bottom=2.5 cm, left=2.5 cm, right=2.5 cm]{geometry}
\usepackage{amstext}
\usepackage{array}  
\usepackage{nicefrac}
\usepackage{bbold}

\usepackage[utf8]{inputenc}
\usepackage{bm}
\usepackage{leftidx}
\usepackage[numbers,sort&compress]{natbib}

\usepackage{makecell}

\usepackage{colortbl}

\usepackage{float}

\usepackage{feynmp-auto}

\newcommand{\be}{\begin{equation}}
\newcommand{\ee}{\end{equation}}
\newcommand{\ba}{\begin{array}}
\newcommand{\ea}{\end{array}}

\newcommand{\erf}{\text{erf}}

\subheader{ULB-TH/25-05}
\arxivnumber{2507.02063}
\title{Primordial high energy neutrinos: theoretical/observational constraints and sharp spectral features}

\author{Nicolas Grimbaum Yamamoto}
\author{and Thomas Hambye}
\affiliation{Service de Physique Th\'eorique, Universit\'e Libre de Bruxelles, Boulevard du Triomphe, CP225, 1050 Brussels, Belgium}
\emailAdd{nicolas.grimbaum@ulb.be}
\emailAdd{thomas.hambye@ulb.be}

\abstract{
Among the few ways that allow or could allow us to probe the early Universe from the observation of a flux of primordial particles, there is one possibility which  has been  little studied: the observation today of high energy neutrinos which could have been emitted shortly after the Big Bang, from the decay or annihilation of early universe relics.
We perform a general study of such a possibility. To this end, we first emphasise that these neutrinos could display various kinds of sharp spectral features, resulting from the primary energy spectrum at emission, and from how this spectrum is smoothed by redshift and radiative correction effects. 
Next, we determine the ranges of mass (from a fraction of eV all the way to the Planck scale) and lifetime of the source particles along which we do not/we do expect that the sharp spectral feature will be altered by interactions of the neutrinos on their way to the detector, mainly with the cosmic neutrino background or between themselves.
We also study the theoretical (i.e.~mainly BBN and CMB) and observational constraints which hold on such a possibility. This allows us to delineate the regions of parameter space (mass, lifetime and abundance) that are already excluded, hopeless for future observation or, instead, which could lead to the observation of such neutrinos in the near future. 
}

\begin{document}

\maketitle
\flushbottom

\section{Introduction}

There are very few types of known particles that, after having been produced shortly after the big bang, could reach us little altered: basically only photons and neutrinos (putting apart gravitational waves). CMB photons have been observed in great details. 21 centimetre photons, which concern a more recent epoch,
are expected to be clearly observed in the near future.
Cosmic neutrinos instead are yet to be observed, as they have very low energies today. Beyond these three Standard Model (SM) predicted fluxes of primordial particles, an important question is whether other fluxes of primordial particles could be observed. It is not excluded that beyond the SM physics could lead to observable fluxes of primordial particles, potentially with much higher energies today. 
The case of primordial high energy neutrinos (which, for brevity, we will call ``PHENUs" all along this work) is especially interesting because neutrinos interact relatively little on their way to us, much less than for instance high energy photons.
A simple way they could be produced is from the decay (or even annihilation, see below) of beyond the SM particles, that would have occurred when the temperature of the primordial plasma is below the neutrino decoupling temperature ($T_\text{dec}^\nu\simeq 1$~MeV), so that they do not thermalise with the plasma.
It is curious that such a possibility has been relatively little studied in the literature. 
Ref.~\cite{Frampton:1980rs} considers the possibility of a source particle decaying into 2 neutrinos and Ref.~\cite{Gondolo:1991rn} considers it further, in particular from the perspective of early experimental constraints.
Refs.~\cite{Kanzaki:2007pd,Ema:2013nda, Ema:2014ufa,McKeen:2018xyz,Jaeckel:2020oet} also consider the decay of a relic into neutrinos, in general for specific cases and energy ranges, for instance in the context of the high energy PeV neutrinos observed by the IceCube experiment \cite{IceCube:2013low}.\footnote{Note that primordial high energy neutrinos could also be produced from very different processes, such as from early evaporation of primordial black holes, leading to a quite different physics, see e.g.~\cite{Halzen:1995hu, Lunardini:2019zob}. Additionally, the evaporation of primordial black holes could be at the origin of the initial abundance of the heavy relic which then subsequently decays into neutrinos, as described for the two-body case in Ref.~\cite{Wu:2024uxa}.}
To our knowledge a general systematic study of this scenario has not yet been undertaken, related to the following questions: 
\begin{itemize} 
\item what are the various sharp spectral features they could lead to? 
\item what are the regions of parameters (mass $m_P$, lifetime $\tau_P$ and abundance $\Omega_P$ of the source particle for the decay case) for which interactions with the medium (elastic and inelastic scatterings of a PHENU with a cosmic neutrino background (C$\nu$B) neutrino or of 2 PHENUs) are relevant or irrelevant? (i.e.\ within which the sharp spectral features are expected to be altered or instead unaltered).
\item what are the general experimental and theoretical constraints which apply on such a scenario all the way from C$\nu$B energies to the Planck scale?
\item and from there what are the ranges of values of source particle parameters which are already experimentally excluded, which lead to a signal that could not be realistically observed or which could instead be observed soon?
\end{itemize}
These are the questions we will study in this work.
A general systematic study of the way the sharp spectral features are affected by interactions with the medium all the way to the detector in regions where these interactions are relevant, including all relevant processes, has also not been performed so far. Such an involved study is left for a subsequent work.

\section{Types of processes and associated sharp spectral features} \label{sec-spectra}

\subsection{\texorpdfstring{Neutrino lines: $P \rightarrow \nu \bar{\nu}$ or $P \rightarrow \nu \nu$}{Neutrino lines: P → νν̅ or P → νν}}

\begin{figure}[t!]
    \centering
\includegraphics[width=0.45\textwidth]{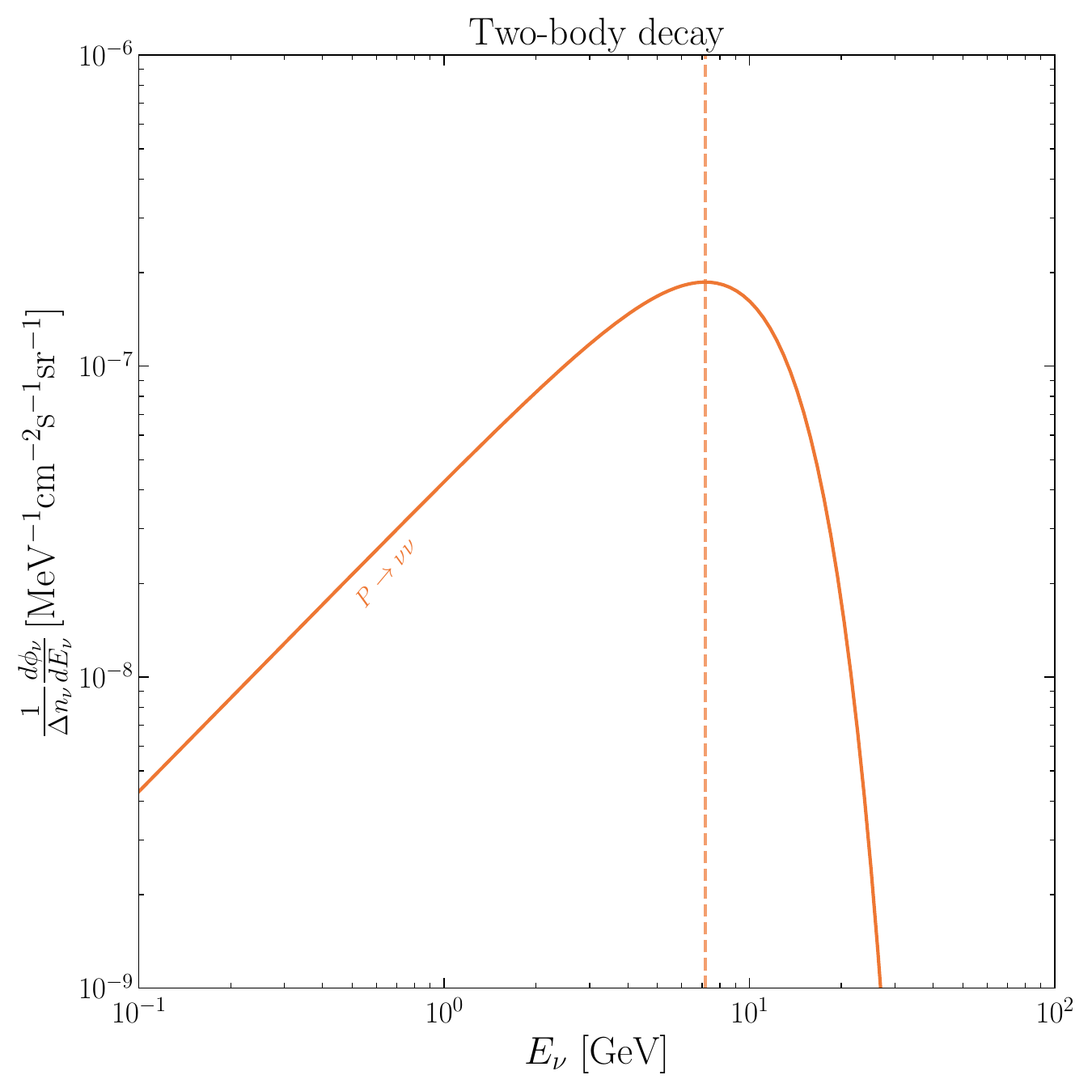}
    \includegraphics[width = 0.45\textwidth]{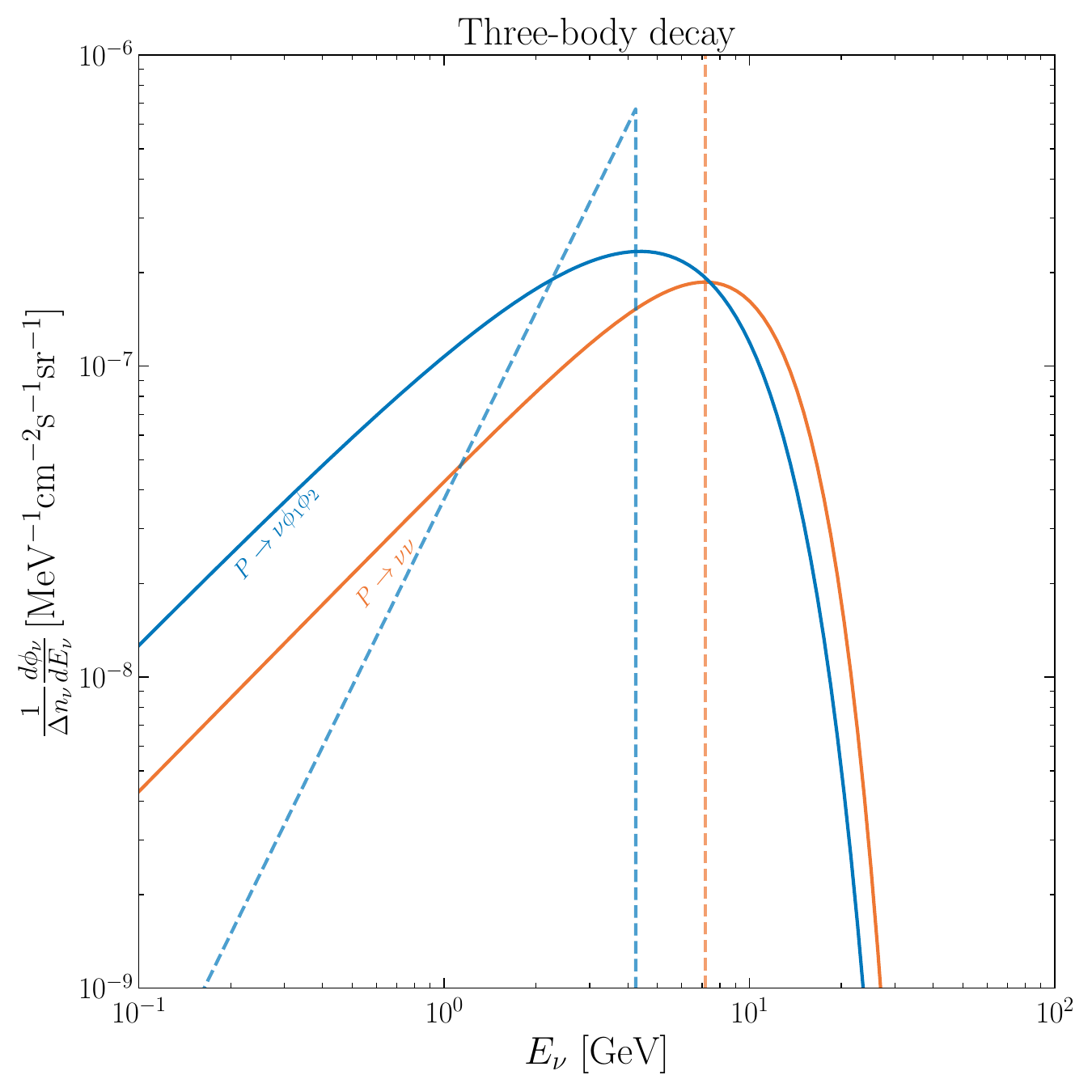}\\
    \includegraphics[width=0.45\linewidth]{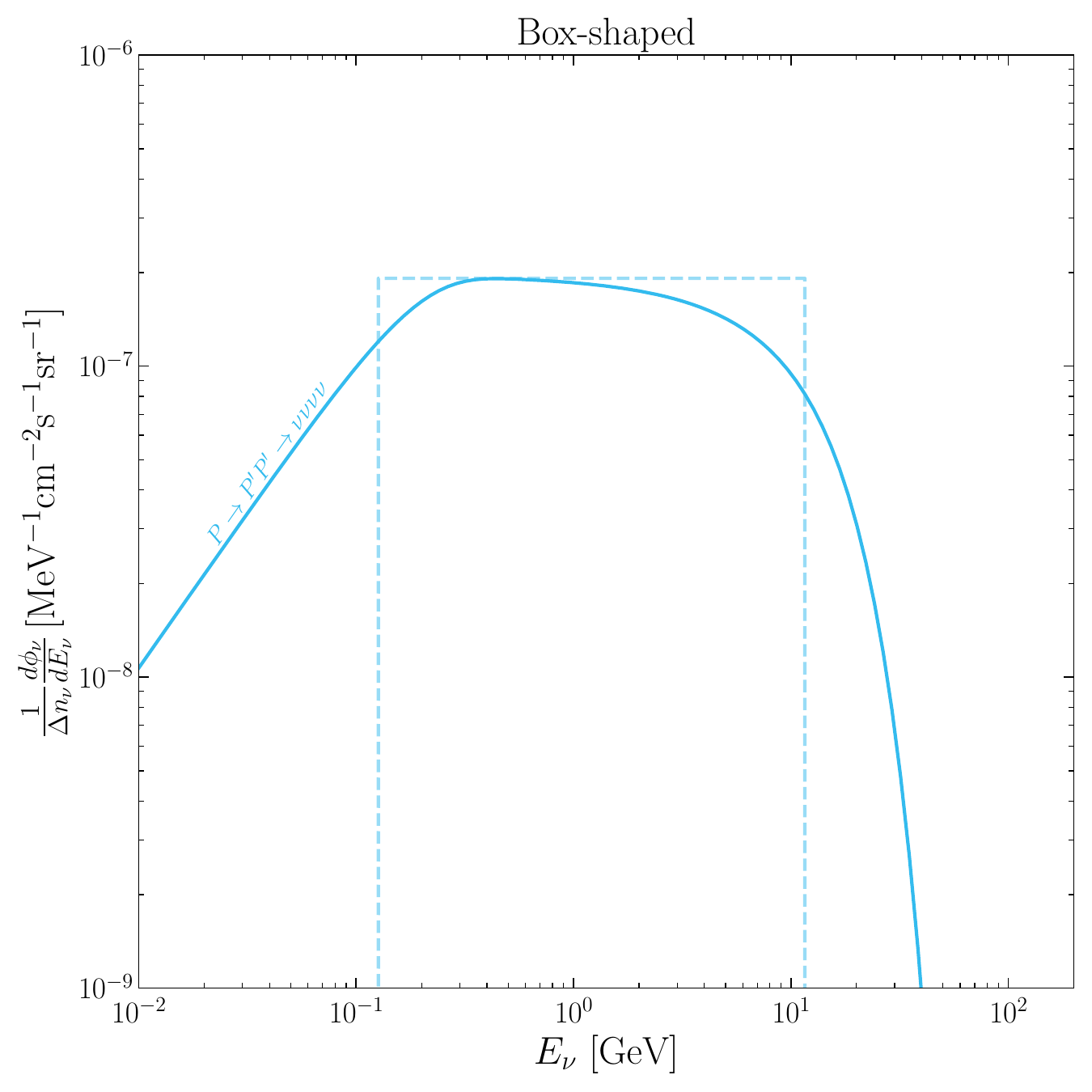}
    \includegraphics[width=0.45\linewidth]{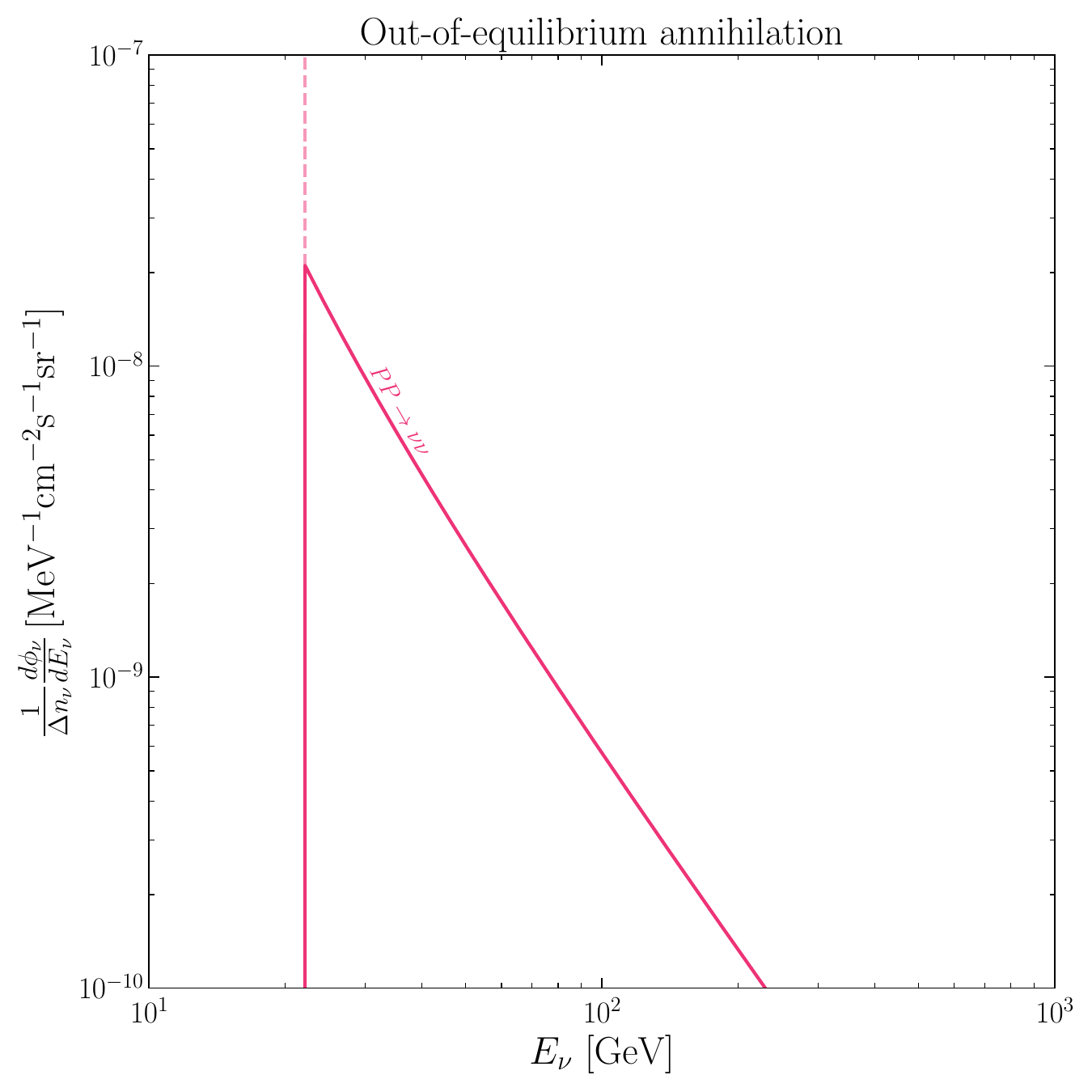}
    \caption{Neutrino flux as a function of the observed energy today for the 4 types of sharp spectral features considered. The upper panels give it for a two-body decay and a 3-body decay of a particle of mass $m_P = 10^{14}$ MeV and lifetime $\tau_P = 1$ s. The lower panels give it for an example of box-shaped spectrum and for an 
    out-of-equilibrium annihilation. Also given (in dashed lines) are the original spectra at source, rescaled in energy by an arbitrary factor (which we take so that the energy spectrum maxima coincide, explicitly showing the effect of redshift on the spectrum).}
    \label{fig:neutrino_flux}
\end{figure}

A non-relativistic particle of mass $m_P$ decaying into 2 neutrinos with lifetime $\tau_P$ produces a number of neutrino per unit volume and unit time equal to 
\begin{equation}
\frac{dn_\nu}{dt}+3Hn_\nu= 2 n_P \frac{1}{\tau_P} e^{-t/\tau_P}\,,
\label{dnudtline}
\end{equation}
with energy $m_P/2$, where $n_P$ is the number density of the decaying $P$ particle there would be in the absence of decay. Equivalently, this means that the comoving neutrino number density $a^3n_\nu$ (with $a$ the scale factor which we take to be unity today) scales as 
\begin{equation}
    \frac{d(a^3n_\nu)}{dt}= 2 a^3 n_P \frac{1}{\tau_P} e^{-t/\tau_P}\,.
    \label{dnudtlinecomoving}
\end{equation}
The number density $n_P$ can be related to $\Omega_P^0$, defined as the ratio of the energy density that $P$ would have today (if it did not decay) to the present critical energy density $\rho_{\text{crit}}^0$: $n_P= \Omega_P^0 \rho_{\text{crit}}^0/(m_Pa^3)=f_P\Omega_{DM}^0 \rho_{\text{crit}}^0/(m_Pa^3)$. 
In the following, to parametrise the abundance we will use the quantity $f_P$ which is the ratio of the source particle abundance that the particle would have today with the DM abundance today, $f_P\equiv \Omega_P^0/\Omega_{DM}^0$ with $\Omega_{DM}h^2=0.120$ and $h=0.674$ \cite{Planck:2018vyg}.
The number density of neutrinos this will lead to in the detector today is a factor of $a^3$ smaller than at production due to expansion.
Because of the resulting redshift, these neutrinos produced at time $t_\text{inj}$, corresponding to a redshift $z_\text{inj}$, will reach the detector with energy $m_P/(2 (1+z_\text{inj}))$. 
If the decay arises during the radiation (matter) dominated period the relations between time and scale factor are
\begin{eqnarray}
t&=&t_r a^2=\frac{t_r}{(1+z)^2}=\frac{1}{2H}, \quad t_r=\frac{1}{2 H_0 \sqrt{\Omega^0_{R}}}=2.4\cdot 10^{19}s\quad \hbox{(radiation)},\label{times} \\
t&=&t_m a^{3/2}=\frac{t_m}{(1+z)^{3/2}}=\frac{2}{3 H}, \quad t_m= \frac{2}{3 H_0 \sqrt{\Omega^0_{M}}}=5.5\cdot 10^{17}s\quad \hbox{(matter)},\label{times_matter}
\end{eqnarray}
with $(1+z)=a^{-1}$. 
In the following, the notation ``$E_\nu$'' will always refer to the energy of the neutrino observed in the detector today.
Thus, the neutrino flux crossing the detector with an energy between a value $E_\nu$ and $E_\nu+d E_\nu$ is $a^3$ times the number of neutrinos which have been produced per unit volume, with redshift between $1+z_\text{inj}=m_P/(2 E_\nu)$ and $1+z_\text{inj}+dz_\text{inj}=m_P/(2 (E_\nu+dE_\nu))$ (so that $dz_\text{inj}=-m_P dE_\nu / (2E^2_\nu$)), that is to say during a time interval $dt_\text{inj}=dE_\nu/(H_\text{inj} E_\nu)$ (where $H_\text{inj}$ refers to the value at the considered neutrino injection time). Multiplying this time interval by the flux at source, eq.~(\ref{dnudtline}), and by the $a^3$ dilution factor,
gives the differential flux in the detector (see also \cite{Gondolo:1991rn,McKeen:2018xyz}) 

\begin{equation}
\frac{d N_\nu}{dE_\nu}= 4\pi \frac{d \phi_\nu}{dE_\nu}=\frac{2 \,\Omega_P^0 \,\rho_{\text{crit}}^0}{m_P H_\text{inj} \tau_P E_\nu} e^{- t_{\text{inj}}/\tau_P} \,\Theta\left(\frac{m_P}{2} - E_\nu\right)\,,
\label{dnudERad}
\end{equation}
where $\phi_\nu$ is the neutrino flux per unit solid angle. 
This equation is valid independently of whether the emission occurs before or after the radiation-matter equality time. 
Nevertheless the energy spectrum differs in each case, and thus can be distinguished experimentally, as $H_\text{inj}$ in this equation is to be taken at injection time $t_{\text{inj}}$, see eqs.~(\ref{times}) and \eqref{times_matter}: $H_\text{inj}=m_P^2/(8 t_r E_\nu^2)$ ($H_\text{inj}=\frac{2}{3 t_m}(m_P/2 E_\nu)^{3/2}$) for neutrino emission before (after) radiation-matter equality time, so that
\begin{eqnarray}
   \frac{d N_\nu}{dE_\nu}&=& 4\pi \frac{d \phi_\nu}{dE_\nu}=\frac{16 \,\Omega_P^0 \,\rho_{\text{crit}}^0  E_\nu}{m_P^3}\frac{t_r}{\tau_P} e^{- 4 \frac{t_r}{\tau_P} \frac{E_\nu^2}{m_P^2}} \,\Theta\left(\frac{m_P}{2} - E_\nu\right)\quad\quad\quad\quad\quad \hbox{(radiation)},
   \label{dnudERadrad}\\ 
   \frac{d N_\nu}{dE_\nu}&=& 4\pi \frac{d \phi_\nu}{dE_\nu}=\frac{6 \sqrt{2} \,\Omega_P^0 \,\rho_{\text{crit}}^0  \sqrt{E_\nu}}{m_P^{5/2}}\frac{t_m}{\tau_P} e^{- 2 \sqrt{2} \frac{t_m}{\tau_P} \frac{E_\nu^{3/2}}{m_P^{3/2}}} \,\Theta\left(\frac{m_P}{2} - E_\nu\right)\quad\quad \hbox{(matter).
   }\label{dnudERadmat}
\end{eqnarray}
Figure~\ref{fig:neutrino_flux} shows $\frac{d \phi_\nu}{dE_\nu}$ for an example set of values of the 3 inputs $\Omega_P^0$, $m_P$ and $\tau_P$ corresponding to an emission during the radiation dominated era. 
From now on, in the main body of this article, we will consider only emission during this era. 
Corresponding plots and related formulas for the matter dominated case can be found in appendix~\ref{app:matter-dominated}. 
We will not consider the intermediate case where the decays occur largely both before and after the matter-radiation equality time $t_{\text{eq}}$ (i.e.~for $\tau_P\simeq t_{\text{eq}})$. It can be obtained from combining both sets of formulas. 

Coming back to the radiation dominated era, from the equations above, it is easy to obtain the energy for which the flux is at its maximum $E_\nu^{\text{max}}$, as well as at the averaged neutrino energy
$\bar{E_\nu}$. Analytically, they are:
\begin{equation} \label{eq:enu_max_rad}
E_\nu^{\text{max}}= \frac{m_P}{2} \sqrt{\frac{\tau_P}{2 t_r}}\,,  \quad\quad\quad\quad\,\,  \bar E_\nu  = \frac{m_P}{2}\sqrt{\frac{\pi \tau_P}{4 t_r}}\,.
\end{equation}

\begin{figure}[t!]
    \centering
    
    \includegraphics[width=0.49\textwidth]{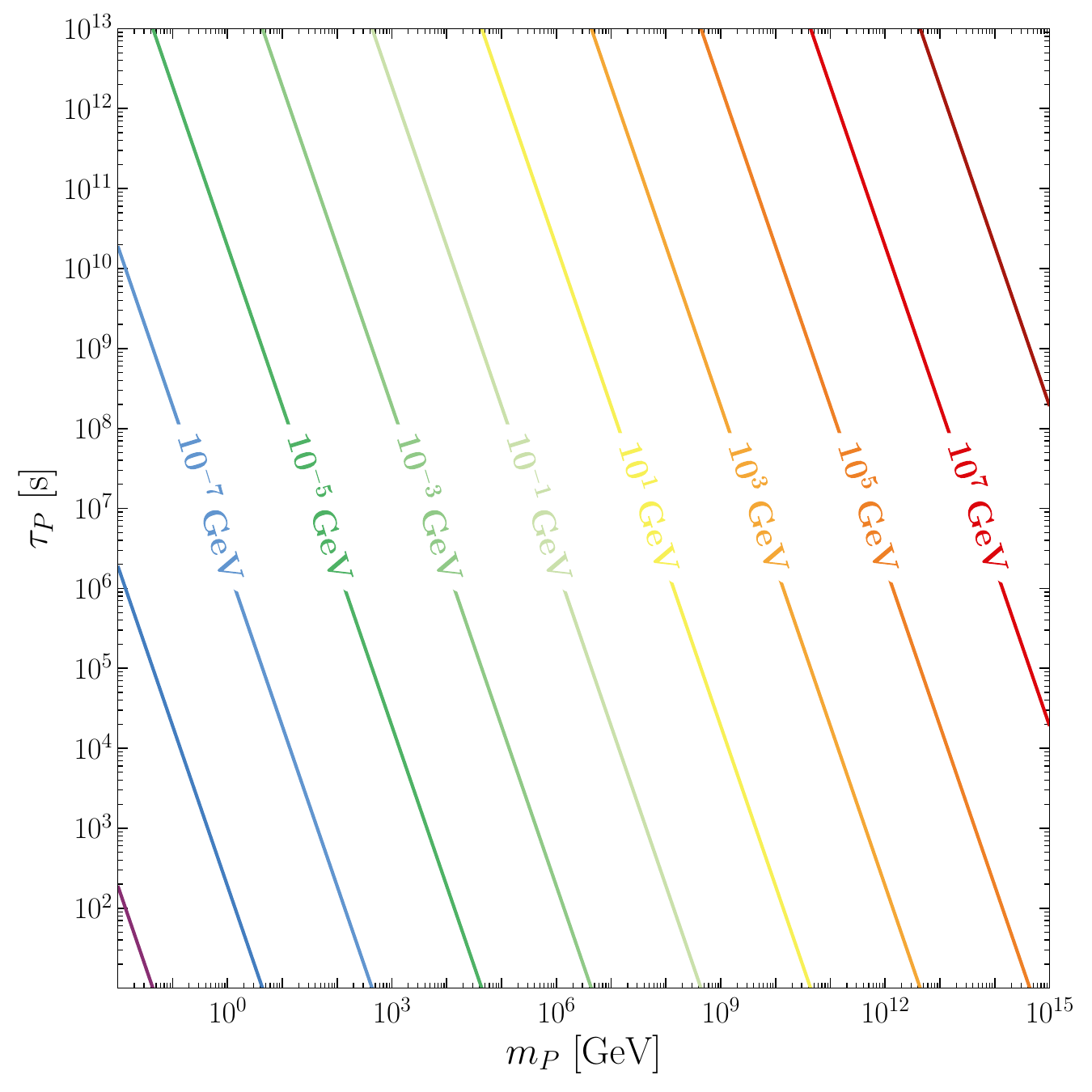}
    \includegraphics[width=0.49\textwidth]{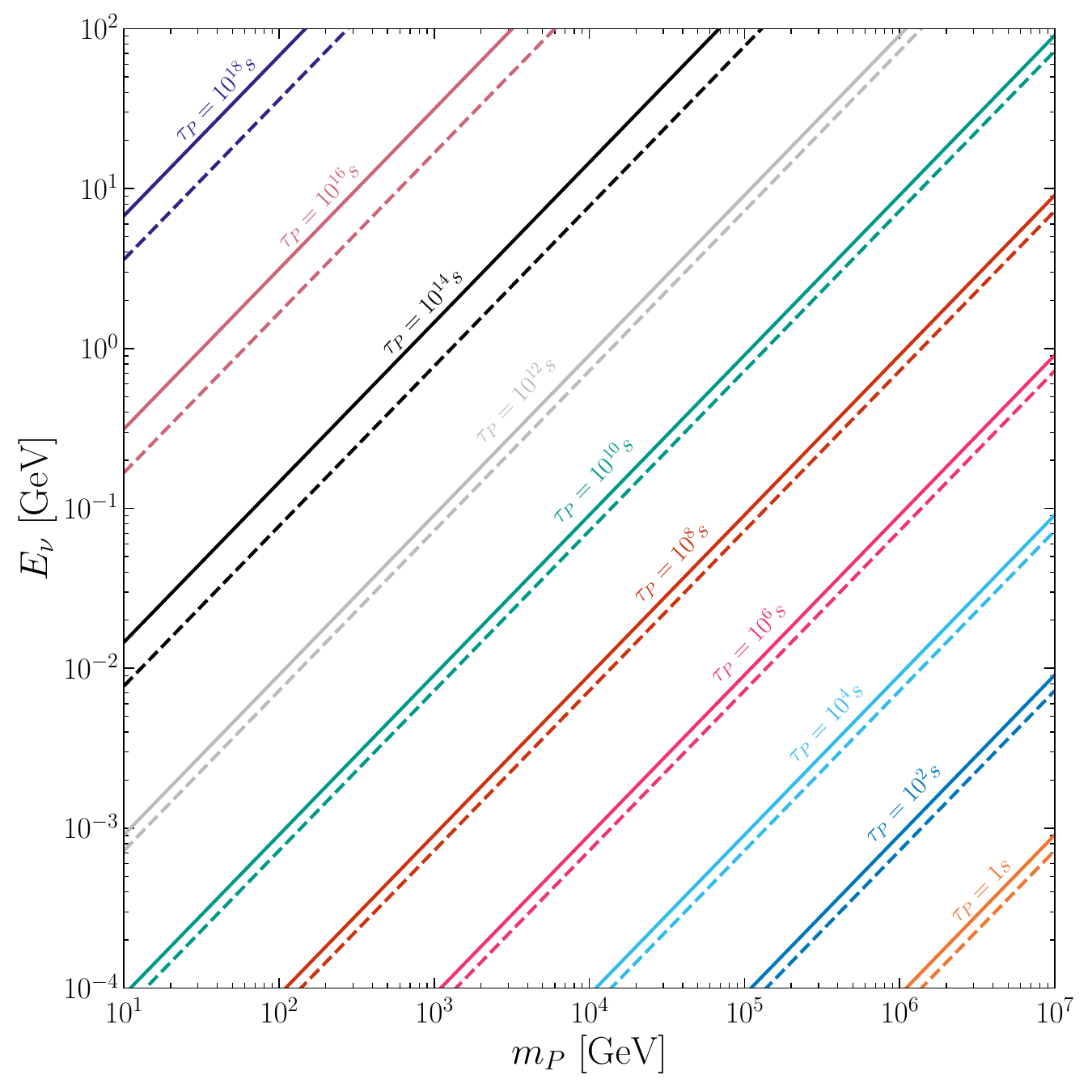}
    \caption{Left: The energy of today's neutrino at the maximum of the flux in the $m_P$-$\tau_P$ plane. Right: Mean energy (solid) and energy at the maximum (dashed) of today's neutrino flux as a function of the source particle mass $m_P$ for various values of the lifetime $\tau_P$ for a 2-body decay. }
    \label{fig:meanmax_energy}
\end{figure}

Thus, the energy at the maximum of the flux (average energy) is a factor $\sqrt{2}$ ($\sqrt{\pi}/2$) 
smaller than the naive estimate obtained if we assume that all neutrinos have been produced at $t_\text{inj}=\tau_P$, which gives $E_\nu^{\text{naive}}=\frac{m_P}{2}(\tau_P/t_{r})^{\frac{1}{2}}$. This is because a neutrino at the peak of the spectrum, or with an energy equal to the average energy, is not emitted at $t=\tau_P$ but at
\begin{equation} \label{eq:t_max_rad}
t^{\text{max}}= \frac{\tau_P}{2}  \quad\quad \bar{t} =  \tau_P \frac{\pi}{4}\,.
\end{equation}
Figure~\ref{fig:meanmax_energy} shows $E_\nu^{\text{max}}$ and $\bar{E}_\nu$ as a function of $m_P$ for various values of the lifetime.

To know more quantitatively by how much the redshift effect spreads the neutrino spectrum, that is to say up to which level the signal remains a sharp spectral feature, it is useful to see what is the interval of energies within which $50\%$, $67\%$, $90\%$ or $95\%$ of the events lie, taking a same values of $d N_\nu/d E_\nu$ at both extremities of the interval, see figure~\ref{fig:ratiospectrum} in appendix~\ref{app:sharpness} (i.e.~retaining in this figure only the events above an horizontal line).
Within these intervals, the energies sit within a factor equal to 2.5, 3.8, 11.9, and 22.6 respectively.
This means that the signal remains quite sharp. Note that, as expected, these ratios are independent of the source particle lifetime $\tau_P$ (as long as the decay predominantly occurs within the radiation dominated era, i.e.~for a lifetime of order of a few times smaller than $t_{eq}$ or smaller, see appendix~\ref{app:matter-dominated} for the matter dominated case).
Finally, note importantly that the general shape of the spectrum is independent of the $m_P$, $\tau_P$ (and obviously $f_P$) input parameters: if we consider eq.~(\ref{dnudERadrad}) for 2 different sets of parameters, the ratio
of the differential fluxes is a constant up to an appropriate constant rescaling of the energies
between both sets,
\begin{equation}
    \frac{d N_\nu}{dE_\nu}\Big|_{E_{\nu,2}}=\frac{f_{P,2}}{f_{P,1}}\cdot \frac{d N_\nu}{dE_\nu}\Big|_{E_{\nu,1}}\cdot \frac{m_{P,1}^2}{m_{P,2}^2}\sqrt{\frac{\tau_{P,1}}{\tau_{P,2}}}\,,
\end{equation}
with the energy rescaling as follows: $E_{\nu,2}/E_{\nu,1}=(m_{P,2}/m_{P,1})(\tau_{P,2}/\tau_{P,1})^{1/2}$. This formula follows directly from the energy dependence of the differential flux $\frac{d N_\nu}{dE_\nu}=A E_\nu e^{-E_\nu^2/B}$, as given from eq.~(\ref{dnudERadrad}). 
The energy rescaling to be made to get the proportionality of the spectra can be obtained for example by taking the ratio of the energies at the maximum of the flux, eq.~(\ref{eq:enu_max_rad}), $E^{\text{max}}_{\nu,2}/E^{\text{max}}_{\nu,1}=(m_{P,2}/m_{P,1})(\tau_{P_{2,2}}/\tau_{P,1})^{1/2}$. 
Finally, note that
if the source particle does not decay into a pair of neutrinos but into a neutrino and another particle ``X'', the neutrino spectrum induced in the detector is the same, modulo an obvious factor 2 suppression 
and the fact that an injected neutrino has an energy $(m_P^2-m_X^2)/(2m_P)$ rather than $m_P/2$.

\subsection{\texorpdfstring{Three body decay: $P \rightarrow \nu P' P''$}{Three body decay: P → ν P' P''}}

It is known that the 3-body decay of DM today into gamma rays or neutrinos leads to a gamma/neu\-trino energy spectrum which is quite sharp, even if not monochromatic. It is interesting to see, after the further spreading of the spectrum by the redshift effect, whether the signal remains relatively sharp.
With respect to the 2-body decay, the difference is that the decay at rest produces neutrinos with a momentum distribution $f_\nu$ which is not a delta function any more.
As a result, neutrinos of a given energy today are no longer produced all at the same time.
Therefore, at this point, it is useful to give the general formula which holds for non monochromatic neutrino production. 
The fundamental quantity that matters at production is ${d^2(a^3 n_\nu)}/(dE_\nu dt_\text{inj})$, the number of neutrinos produced in a comoving volume at a given time $t_\text{inj}$ per unit time within an energy interval which today gives neutrino energy within an interval $E_\nu$ and $E_\nu + dE_\nu$. 
For a given value of $E_\nu$, the neutrino number density per 
energy interval today is the time integral over it, or equivalently redshift integral,
\begin{equation}
    \frac{d N_\nu}{dE_\nu}= 4 \pi \frac{d \phi_\nu}{dE_\nu} = \int_0^{t_0}\frac{d^2(a^3 n_\nu)}{dE_\nu dt_\text{inj}} dt_\text{inj} = \int_{0}^{\infty}\frac{d^2(a^3 n_\nu)}{dE_\nu dt_\text{inj}}\frac{1}{(1+z_\text{inj}) H_\text{inj}}dz_\text{inj}\,.
    \label{dNdEgeneral}
\end{equation}
For a decay, by multiplying eq.~(\ref{dnudtlinecomoving}) by $df_\nu/dE_{\text{inj}}$, the fraction of neutrinos that are produced with energy within an energy interval $dE_{\text{inj}}$ (i.e.~the integral of $df_\nu/dE_{\text{inj}}$ over $E_\text{inj}$ is normalised to unity) and taking into account that $E_{\text{inj}}= E_\nu(1+z_\text{inj})$, one has at a given injection time (or corresponding redshift) and for a given $E_\nu$
\begin{equation}
    \frac{d^2(a^3 n_\nu)}{dE_\nu dt_\text{inj}} = \frac{\Delta n_\nu\Omega^0_P \rho^0_{crit}}{m_P \tau_P}e^{-t_\text{inj}/\tau_P} (1+z_\text{inj}) \frac{df_\nu}{dE_{\text{inj}}}\Big|_{E_{\text{inj}}=E_\nu(1+z_\text{inj})}\,,
    \label{d2n}
\end{equation}
where $\Delta n_\nu$ is the number of neutrinos produced per decay.
This gives
\begin{equation}
    \frac{d N_\nu}{dE_\nu} = 4 \pi \frac{d \phi_\nu}{dE_\nu}
    = \frac{\Delta n_\nu \Omega^0_P \rho^0_{crit}}{m_P \tau_P}\int_0^\infty dz_\text{inj}\frac{e^{-t_{\text{inj}}/\tau_P}}{H_\text{inj}} \left.\frac{df_\nu}{dE_{\text{inj}}}\right|_{E_{\text{inj}}=E_\nu(1+z_\text{inj})}\,.
    \label{dNdEgeneraldecay}
\end{equation}
For a decay into 2 neutrinos, $\Delta n_\nu=2$, $df_\nu/dE_{\text{inj}} = \delta(E_{\text{inj}} - m_P/2)$, and we recover the result of eq.~(\ref{dnudERad}).

As an example of 3-body decay, we will consider a fermion decaying 
into one neutrino (i.e.~$\Delta n_\nu=1$) and 2 scalars $\phi_{1,2}$ (with $m_{\phi_{1,2}}\lll  m_P$), as induced by a $\bar{\psi} \nu \phi_1 \phi_2$ local interaction.
The energy distribution, $df_\nu/dE_{\text{inj}}$, it gives at production is equal to $\Theta(m_P/2-E_{\text{inj}})\cdot 24 E_{\text{inj}}^2/m_P^3$. Plugging it in eq.~(\ref{dNdEgeneraldecay}), the spectrum obtained in the detector is 
\begin{equation}
    \frac{dN_\nu}{dE_\nu} = \frac{24 \Omega_P^0 \rho^0_{crit}}{m_P^4 \tau_P E_\nu}\int_{E_\nu}^{\frac{m_P}{2}} dE_{\text{inj}} E_{\text{inj}}^2 \frac{e^{-t_{\text{inj}}/\tau_P}}{H_\text{inj}}\,,
\end{equation}
where $(1+z_\text{inj})=E_{\text{inj}}/E_\nu$. This equation is valid independently of when the decay occurs, before or after the matter-radiation equality time, but as for the 2-body decay case, it will lead to different spectra in each case due to the different dependence of $H_\text{inj}$ and $t_\text{inj}$ in redshift. In the case of a decay within radiation dominated era, the integral can be expressed in terms of the error function as:

\begin{equation}
    \begin{split}
     \frac{dN_\nu}{dE_\nu} &= \frac{24 \Omega_P^0 \rho^0_{crit} t_r E_\nu}{m_P^3 \tau_P}\left[e^{-\frac{4 E_\nu^2 t_r}{m_P^2 \tau_P}} - \frac{2E_\nu}{m_P}\left(e^{-\frac{t_r}{\tau_P}} + \sqrt{\frac{\pi t_r}{\tau_P}}\left(\erf\left(\sqrt{\frac{t_r}{\tau_P}}\right)-\erf\left(\frac{2E_\nu}{m_P}\sqrt{\frac{t_r}{\tau_P}}\right)\right)\right)\right]   \\
                            & \approx \frac{24 \Omega_P^0 \rho^0_{crit} t_r E_\nu}{m_P^3 \tau_P}\left[e^{-\frac{4 E_\nu^2 t_r}{m_P^2 \tau_P}} - \frac{2E_\nu}{m_P}\left(e^{-\frac{t_r}{\tau_P}} + \sqrt{\frac{\pi t_r}{\tau_P}}\left(1-\erf\left(\frac{2E_\nu}{m_P}\sqrt{\frac{t_r}{\tau_P}}\right)\right)\right)\right] \,.
     \end{split}
\label{eq:flux_3body_rad}     
\end{equation}

Figure~\ref{fig:neutrino_flux} shows the spectrum this gives. Proceeding as for the 2-body decay, we get a transcendental equation for the energy at the maximum and a closed form solution for the mean energy
\begin{equation}
E_\nu^{\text{max}}= e^{\frac{-4(E_\nu^{\text{max}})^2}{m_P^2}}\left(\frac{m_P}{2} - \frac{2 (E_\nu^{\text{max}})^2 t_r}{m_P \tau_P}\right)\,,  \quad\quad \bar E_\nu  = \frac{m_P}{2} \sqrt{\frac{\pi \tau_P}{4 t_r}}\frac{3 }{4}\,.\footnote{$\erf(x)\to 1$ for $x\gtrsim 2$ which is verified in the case of a decay in the radiation era, i.e.\ $\tau_P \ll  t_r$, so that the mean energy reduces to this simpler expression.}
\label{Emax3body_rad}
\end{equation}

If the decay occurs at $a\ll 1$, which is the case for a decay during the radiation era, the transcendental equation simplifies (as $E_\nu^{\text{max}} \ll  m_P/2$), so that we can expand it at second order in $E_\nu^{\text{max}}/m_P$ and obtain:

\begin{equation}
E_\nu^{\text{max}}= \frac{m_P}{1 + \sqrt{5 + 4 \frac{t_r}{\tau_P}}}\,,  \quad\quad \bar E_\nu  = \frac{m_P}{2} \sqrt{\frac{\pi \tau_P}{4 t_r}}\frac{3}{4}\,.
\label{Emax3body_rad_bis}
\end{equation}

\begin{figure}[t!]
    \centering
    
    \includegraphics[width=0.49\textwidth]
    {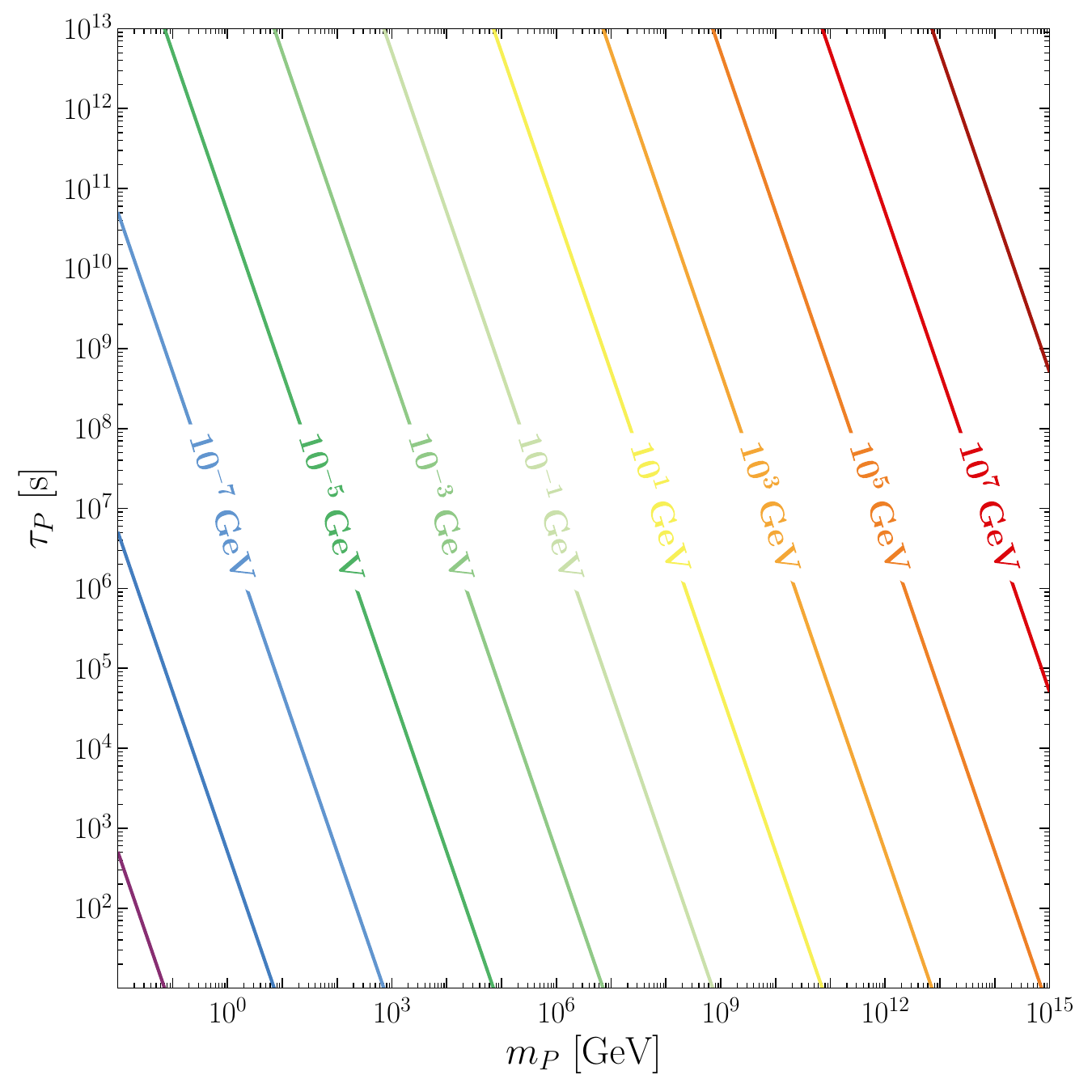}
    \includegraphics[width=0.49\textwidth]{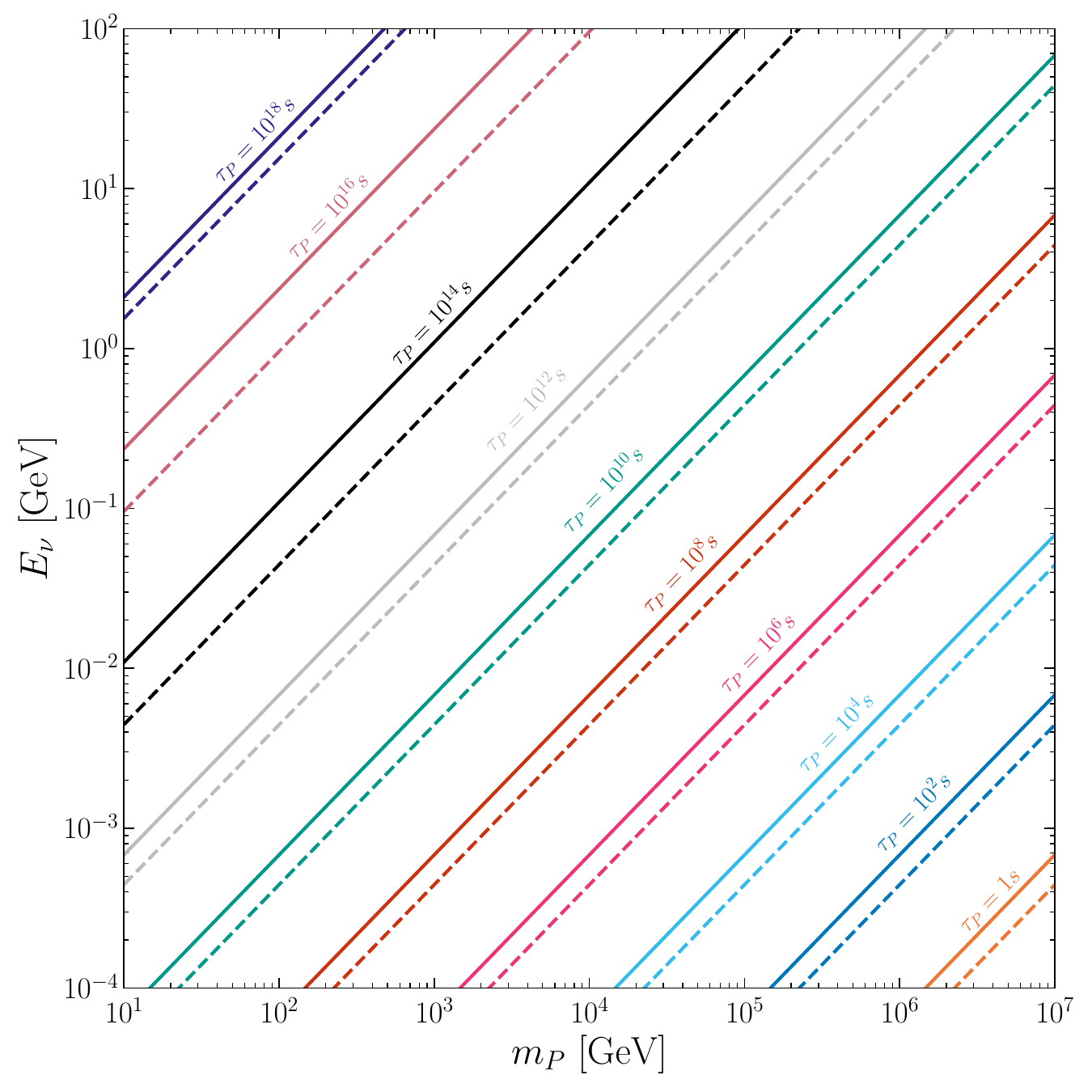}
    \caption{Same as figure~\ref{fig:meanmax_energy}, for a 3-body decay.
    }
    \label{fig:3body_lines_meanmax_energy}
\end{figure}

Thus, for $\tau_P\ll  t_r$, the maximum (averaged) energy is a factor ${3/2}$ ($\frac{9\sqrt{\pi}}{16}$)
smaller than the naive estimate obtained if we assume that all neutrinos have been produced at $t=\tau_P$, which gives $E_\nu^{\text{naive}}=\frac{m_P}{3}(\tau_P/t_{r})^{\frac{1}{2}}$. 
Figure~\ref{fig:3body_lines_meanmax_energy} shows $E_\nu^{\text{max}}$ and $\bar{E}_\nu$ as a function of $m_P$ for various values of the lifetime.

Figure~\ref{fig:ratiospectrum} in appendix~\ref{app:sharpness} shows the energy interval between which $50\%$, $67\%$, $90\%$ or $95\%$ of the events lie. 
Within these intervals, the energies lie within a factor equal to 3.2, 5.6, 22.5, and 47.4 respectively.
This means that the signal remains quite sharp, as for the 2-body decay, see also figure~\ref{fig:neutrino_flux}. 
The signal is hardly more spread than for the monochromatic 2-body decay case, this is due to the redshift induced spreading being somewhat larger than the difference in spread between the 2- and 3-body decay cases at production.

\subsection{\texorpdfstring{Box-shaped spectrum: $P\rightarrow P' P'\rightarrow \nu \bar{\nu} \nu \bar{\nu}$}{Box-shaped spectrum: P→P′P′→νν̄νν̄}}

The decay of a particle into two other particles which both subsequently undergo 2-body decays into neutrinos leads to a box-shaped spectrum. This spectrum is sharp in the sense that it involves a sharp infrared threshold and a sharp UV cut-off, with a flat spectrum in between. 
The energy distribution of the resulting neutrinos at the source is \cite{Ibarra:2012dw} 
\begin{equation}
\frac{df_\nu}{dE_{\text{inj}}} = \frac{2}{m_P\sqrt{1 - \Delta^2}} \Theta(E_{\text{inj}}-E_-)\Theta(E_+-E_{\text{inj}})\,,\quad\quad E_\pm = \frac{m_P}{4}\left( 1 \pm  \sqrt{1-\Delta^2}\right)\;,
\end{equation}
where the width of the box depends on the ratio of the masses of the mother and intermediate particles P and P': $\Delta = 2m_{P'}/m_P$.
One can check that we recover the two-body spectrum for $\Delta\to 1$, for a decaying particle twice lighter.
Thus, using eqs.~(\ref{dNdEgeneral}) and (\ref{d2n}), the produced spectrum, for $E_\nu < E_+$ is given by
\begin{equation}
    \frac{dN_\nu}{dE_\nu} = \frac{8 \Omega_0^P \rho_{\text{crit}}^0}{m_P^2 \tau_P E_\nu \sqrt{1- \Delta^2}}\int_{\max{(E_\nu, E_-)}}^{E_+}
                            dE_{\text{inj}}\,\frac{e^{-t_{inj}/\tau_P}}{H_\text{inj}}\;,
\end{equation}
with $(1+z_\text{inj})=E_{\text{inj}}/E_\nu$.
For a decay within the radiation dominated era, this integrates to
\begin{equation}
     \frac{dN_\nu}{dE_\nu} = \frac{8 \Omega_0^P \rho_{\text{crit}}^0}{m_P^2 \sqrt{1- \Delta^2}} \sqrt{\frac{\pi t_r}{\tau_P}}\left[\text{erf}\left(\frac{E_\nu}{\max(E_\nu, E_-)}\sqrt{\frac{t_r}{\tau_P}}\right) - \text{erf}\left(\frac{E_\nu}{E_+}\sqrt{\frac{t_r}{\tau_P}}\right)\right] \;.      
\end{equation}
Figure~\ref{fig:neutrino_flux} shows the energy spectrum one gets for an example value of $\Delta=0.2$. 
As expected, the sharp threshold and cut-off are smoothed, but still the raise and fall to and from the plateau remain sharp enough for a signal of this type to stand out of a neutrino background.

\subsection{\texorpdfstring{Out-of-equilibrium annihilation: $p \bar{p} \rightarrow \nu \bar{\nu}$}{Out-of-equilibrium annihilation: p p̅ → ν ν̅}}

If neutrinos are produced from a pair annihilation of particles $p$ and $\bar{p}$, for instance from $p\bar{p}\rightarrow \nu \bar{\nu}$, one could also get a sharp spectral feature of PHENU today. This requires that the process is not in thermal equilibrium, otherwise the sharp spectral feature would be erased (back into the thermal distribution).
Such an out-of-equilibrium annihilation could occur if the source particles are produced in the thermal bath at a certain epoch or if the annihilation, for various reasons, could start only at a given time.
Here we will assume the latter case, considering that the annihilation starts instantaneously as a result of a phase transition (that we assume to proceed quasi instantaneously at a time $t_\star$) with equal number of particles $p$ and antiparticles $\bar{p}$ after this phase transition.\footnote{An example is when the source particles are purely asymmetric before the phase transition (with, for instance, only $p$ particles). In this case the phase transition can quasi instantaneously lead to particle-antiparticle oscillations, leading to particle-antiparticle annihilations, see e.g.~\cite{Servant:2013uwa,Dhen:2015wra}.}
In this case, if the particles are non-relativistic at the phase transition time, the neutrino energy mass spectrum at the source is monochromatic as for a 2-body decay.
The number of particles annihilating, and thus of produced neutrinos, per unit time is
\begin{equation}
\frac{d n_P}{dt}+3H n_P=-\frac{2}{\Delta n_\nu}\frac{d n_\nu}{dt}= -\frac{1}{2} n_P^2 \langle \sigma v \rangle\,,
\label{dnudtscatt}
\end{equation}
where $\Delta n_\nu$ is the number of neutrinos produced per annihilation, $n_P$ is the sum of the number densities of particles $p$ and antiparticles $\bar{p}$ as a function of time, and $\langle \sigma v\rangle$ is the annihilation cross-section. This gives \cite{Dhen:2015wra}
\begin{equation}
    n_Pa^3= \frac{n_{P\star} a_\star^3}{1+\frac{\langle \sigma v\rangle}{2H}\frac{a_\star^3}{a^3}\Big(\frac{a}{a_\star}-1\Big)n_{P\star}}\,,
    \label{noversscatt}
\end{equation}
where the ``$\star$'' refers to the time $t_\star$.
Proceeding as for the 2-body decay, i.e.~taking $d n_\nu/dt$ from (\ref{dnudtscatt}), with $n_P$ from (\ref{noversscatt}), and multiplying by $a^3$ and $dt= dE_\nu/(H E_\nu)$ gives 
the differential neutrino flux in the detector today
\begin{equation}
\frac{d N_\nu}{dE_\nu}= \frac{1}{4\pi}\frac{d \phi_\nu}{d E_\nu}= \frac{\Delta n_\nu}{4}\frac{(\Omega^0_P \rho_{\text{crit}}^0)^2 m_P \langle \sigma v\rangle}{E_\nu^4 H_\text{inj}}\frac{1}{\Big(1+\Omega^0_P \rho_{\text{crit}}^0\frac{\langle \sigma v\rangle}{2H_\text{inj}}\frac{m_P^2}{E_\nu^3}\Big(\frac{E_\nu}{m_Pa_\star}-1\Big)\Big)^2}\,,
\label{dnudERadscatt}
\end{equation}
with $(1+z_\text{inj})=m_P/E_\nu$.
For the process to be out-of-equilibrium, one needs to check that the interaction rate for an equilibrium population of annihilating particles P is smaller than the Hubble rate at $t_\star$ (i.e.\ $(\langle \sigma v \rangle n_p^{\text{eq}}/  H)|_{t_\star}<1$). Assuming that the annihilation takes place in the radiation dominated era, this condition on the thermally averaged cross-section reads
\begin{equation}
    \langle \sigma v \rangle < \left(\frac{2\pi}{m_P T_\star} \right)^{3/2}\frac{e^{m_P/T_{\star}}}{2t_\star g_i}\;,
\end{equation}
where $t_\star=0.301 \frac{1}{\sqrt{g_\star}} \frac{m_{\text{Planck}}}{T_\star^2}$ with $g_\star$ the number of relativistic degrees of freedom at time $t_\star$
and $g_i$ the number of degrees of freedom of the annihilating particle $p$. This constraint is satisfied as soon as $T_\star$ is sufficiently lower than $m_P$, given the exponential suppression of $n_P^{\text{eq}}$, which is always the case for masses larger than the neutrino decoupling temperature $T_\text{dec}^\nu\simeq 1$~MeV.

Similarly, for the reaction to occur at all, we impose that the interaction rate, when considering the non-equilibrium abundance of the particle $P$, is larger than the Hubble rate at the time of annihilation (i.e.~$(\langle \sigma v \rangle n_p /H)|_{t_\star}>1$):
\begin{equation}\label{eq:sigma_min}
    \langle \sigma v \rangle > \frac{m_P}{ \Omega_P^0 \rho^0_{crit}}\sqrt{\frac{t_\star}{t_r^3}} = 2.3 \cdot 10^{-23} \text{ cm}^2\left(\frac{1}{f_P}\right)\left(\frac{m_P}{1 \text{ TeV}}\right)\left(\frac{t_\star}{10^8 \text{ s}}\right)^{1/2}\,.
\end{equation}
The energy spectrum this process gives, eq.~(\ref{dnudERadscatt}), is shown in figure~\ref{fig:neutrino_flux}. It shows a sharp
step function threshold followed by a tail scaling as $\sim 1/(E_\nu^4 H)\overset{\sim}{\propto} E_\nu^{-2}$ in the detector, totally different from the decay cases.
This difference stems from the facts that obviously an annihilation does not scale as an exponential decay law and that the rate is proportional to $n_P^2$ rather than to $n_P$, which brings an extra factor of $a^{-3}=(m_P/E_\nu)^3$. 
The possibility of this quite unique sharp spectral feature has, to our knowledge, not been considered before.

\subsection{Neutrino oscillations\label{sec:oscillation}}

So far we did not consider that PHENUs can have 3 different flavours, $\nu_e$, $\nu_\mu$ and $\nu_{\tau}$, as the sharp spectral features discussed above do not depend on the neutrino flavour considered.
Some of the results below will nevertheless depend on the injected flavour composition. However, a comparison of the characteristic time scales 
shows that before they reach the detector, or even before they could interact with the medium, the PHENUs have plenty of time to oscillate (see e.g.~\cite{Bianco:2025boy}). 
Thus, for any physical consideration the flavour composition of the flux is not far from a \nicefrac{1}{3}-\nicefrac{1}{3}-\nicefrac{1}{3} democratic composition.
For definiteness, we will consider everywhere below 
a $\nu_e\bar{\nu}_e$ injected flux.
From the oscillation probabilities encoded in the PMNS matrix
\begin{align}\label{eq:osc}
    (\mathbf{M}_{\text{osc}})_{\alpha\beta}=\sum_{i=1}^3 |U_{\alpha i}|^2|U_{\beta i}|^2\approx\begin{pmatrix} 0.55&0.17&0.28\\0.17&0.45&0.37\\0.28&0.37&0.35\end{pmatrix}\;,
\end{align}
this means that the PHENU flux is a quantum superposition of
55\%, 17\% and 28\% of $\nu_{e,\mu,\tau}$ respectively.

\subsection{Effects of radiative corrections\label{sec:FSR}}

\begin{figure}[t!]
    \centering
    
    \includegraphics[width=0.45\textwidth]{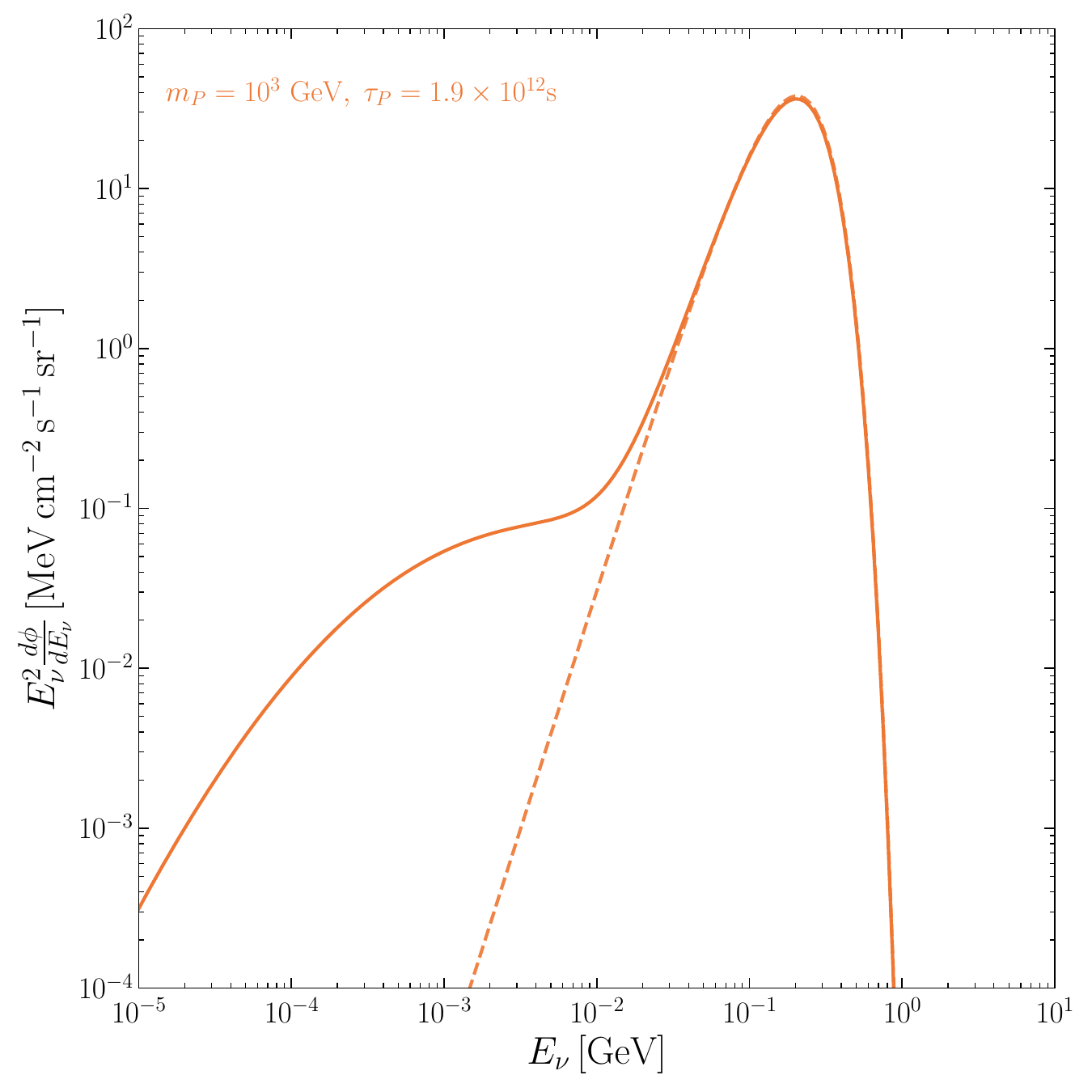}
    \includegraphics[width=0.45\textwidth]
    {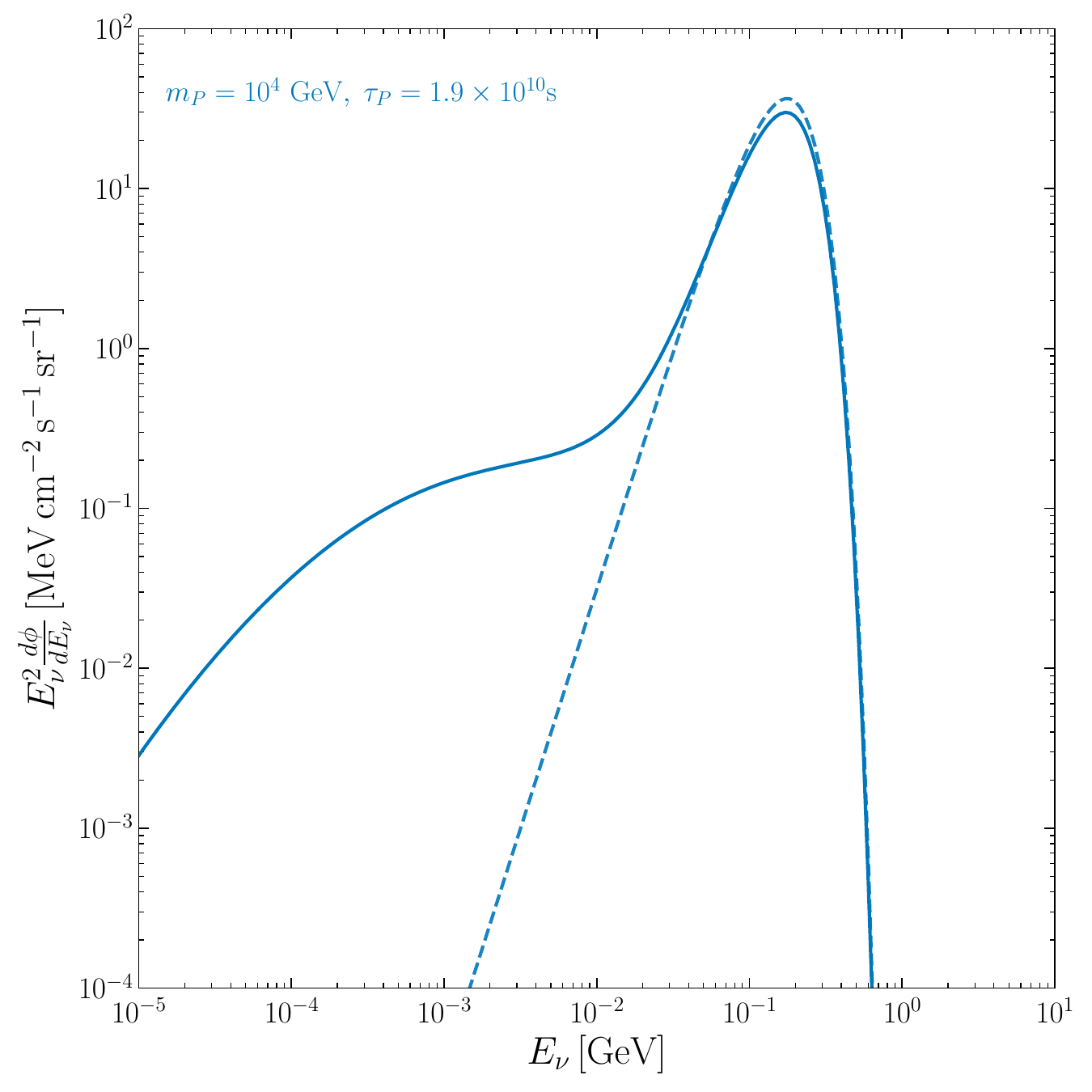}    
    \includegraphics[width=0.45\textwidth]{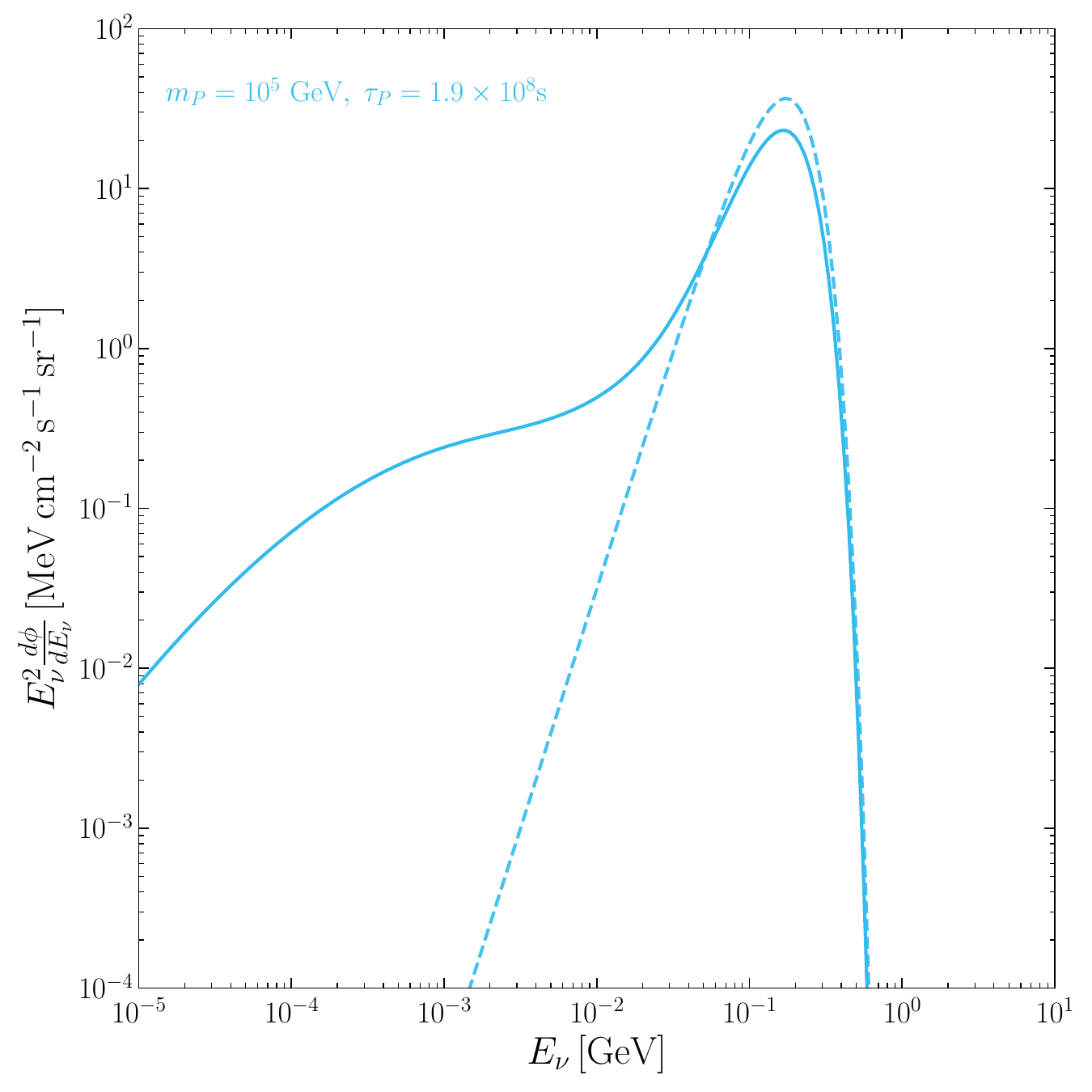}
    \includegraphics[width=0.45\textwidth]
    {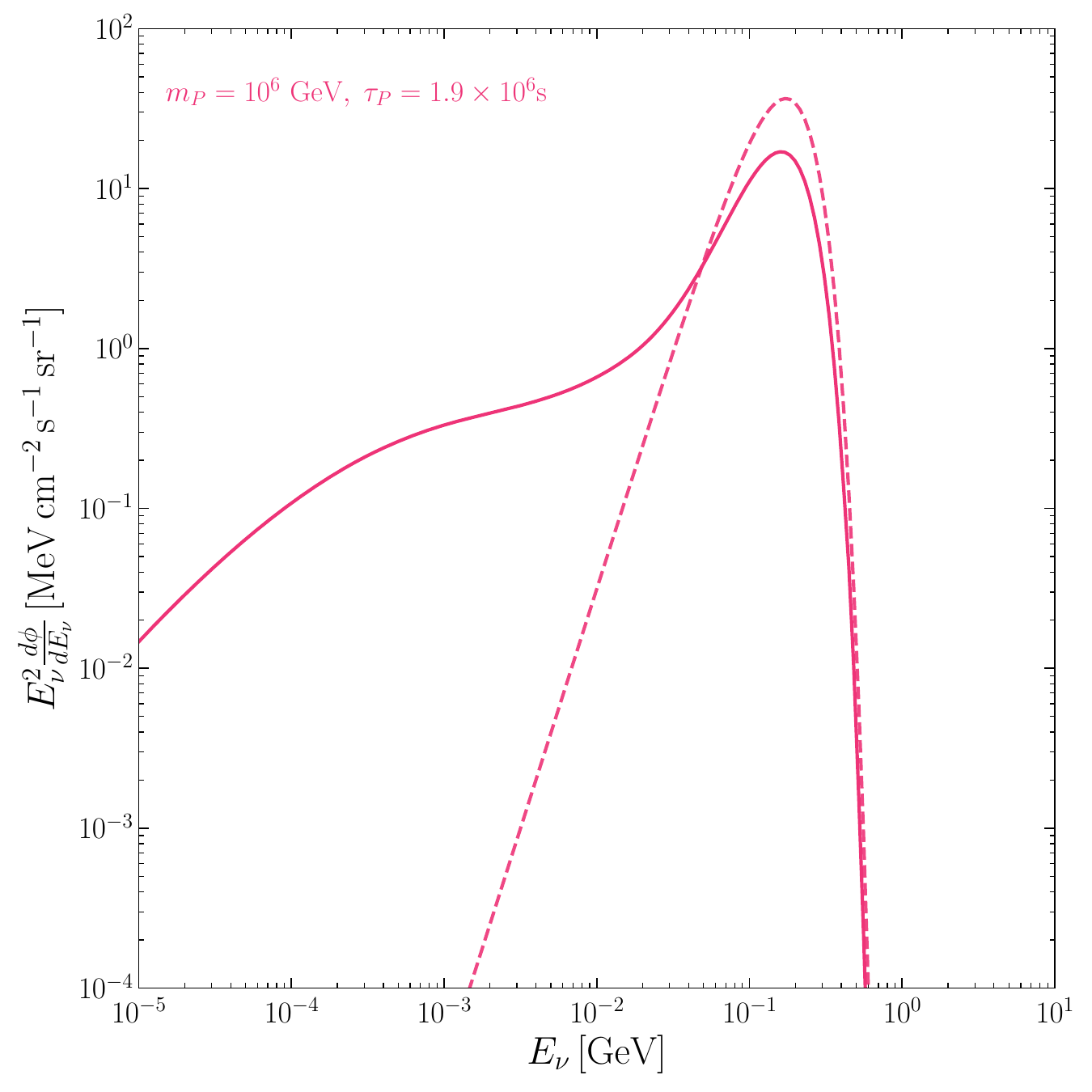}
    \caption{Energy spectrum with (solid) and without (dashed) final state radiation for a two-body decay of source particles with masses $m_P=1, 10, 100, 1000$~TeV.}
    \label{fig:FSR}
\end{figure}

The spectra we obtained so far do not take into account the effects of radiative corrections, i.e.~``final state radiation'' (FSR). 
Their effects depend on the energy of the injected neutrinos. 
Below the electroweak scale (EW), they are Fermi suppressed and thus quickly become negligible as we lower the mass of the source particle. 
Above it, they are in general relevant.
In this case, these corrections have the triple effect of smoothing the high energy part of the spectrum of the produced PHENU, producing secondary neutrinos of lower energies, and producing other particles that can have cosmological effects (mainly on BBN and CMB, see section~\ref{sec:BBN-CMB} below). 
The spreading of the sharp spectral feature they induce turns out to be  subdominant with respect to the one due to the redshift, and is relevant only for energies far above TeV. Using \textsc{Pythia}~8.3 \cite{Bierlich:2022pfr} to compute these corrections, for a source particle with 1, 10, 100, 1000~TeV mass decaying into 2 neutrinos (i.e.\ PHENUs produced with an energy equal to half this value), figure~\ref{fig:FSR} shows the spectrum obtained in the detector with and without radiative corrections.  For 1 TeV, the radiative corrections produce a non-negligible amount of lower energy neutrinos but leave the higher energy sharp spectral feature basically unchanged. Instead, for $m_P=1$~PeV, they reduce the energy spectrum at the maximum of the sharp spectral feature by a factor of 2.15, but still leave the spectral feature well sharp. 

\section{Determination of the lifetime/mass regions where the sharp spectral feature is unaffected by in-flight neutrino interactions}
\label{sec:scattering}

Once the PHENUs have been produced, they can undergo essentially 2 types of potentially relevant interactions while travelling through the universe:  scattering with a (C$\nu$B) neutrino,  and
``self"-scatterings of 2 PHENUs.\footnote{Scatterings on any other SM particle is suppressed. The scattering rate on baryons is suppressed by the small baryon to C$\nu$B neutrino ratio, whereas scattering rate on CMB photons is suppressed by the tiny corresponding cross-section.} 
Both of them can be elastic, $\nu\bar{\nu} \rightarrow \nu \bar{\nu}$, or inelastic, $\nu \bar{\nu}\rightarrow e^+ e^-, e^+\mu^-, e^- \mu^+, \mu^+ \mu^-, \pi^+\pi^-, ..., W^+ W^-$.\footnote{Elastic scatterings can be of the form $\nu_i\bar{\nu}_i \rightarrow \nu_j \bar{\nu}_j$ (with $i=j$ or $i \neq j$) and $\nu_i\bar{\nu}_j \rightarrow \nu_i \bar{\nu}_j$ (with $i \neq j$).} 
The goal of this section is to determine the regions of lifetime, mass, and also possible abundance, of the source particle for which such scatterings do or do not have an effect. 
Determining in particular the regions for which such scatterings are not expected to occur will allow to determine the (broad) region of parameter space where, consequently, the sharp spectral feature is not altered by these scatterings. 
Would an unaltered sharp spectral feature be observed, would this determination allow to tell us essential informations on the mass, lifetime and abundance of the source particle. 
Some of the results in this section are independent of the type of spectrum considered, but some others do somewhat depend on it. For definiteness, in those situations we will consider a 2-body decay (which is also representative of what happens for a 3-body decay).

\subsection{\texorpdfstring{Scatterings on C$\nu$B neutrinos}{Scatterings on CνB neutrinos}}\label{sec:CnuB}

Scatterings of a PHENU with a background neutrino result in a distortion of the expected spectrum, on top of the redshift and FSR effects shown in section~\ref{sec-spectra}. 
The effect of elastic scatterings is to replace a neutrino of a given initial energy by two neutrinos, with energies summing up to the initial state energies. 
Thus, it somewhat lowers the average energy of the PHENU neutrinos but increases the number of high energy neutrinos.
The effect of inelastic scatterings, instead, is to lower the overall flux since the neutrinos, in this case, disappear.
Both inelastic and elastic effects are expected to affect more the low energy part of the sharp spectral feature than its high energy part, since low energy neutrinos have been emitted earlier, and thus had the chance to interact with background neutrinos for a longer time. 

\begin{figure}[t!]
    \centering
    \includegraphics[width=0.5\linewidth]{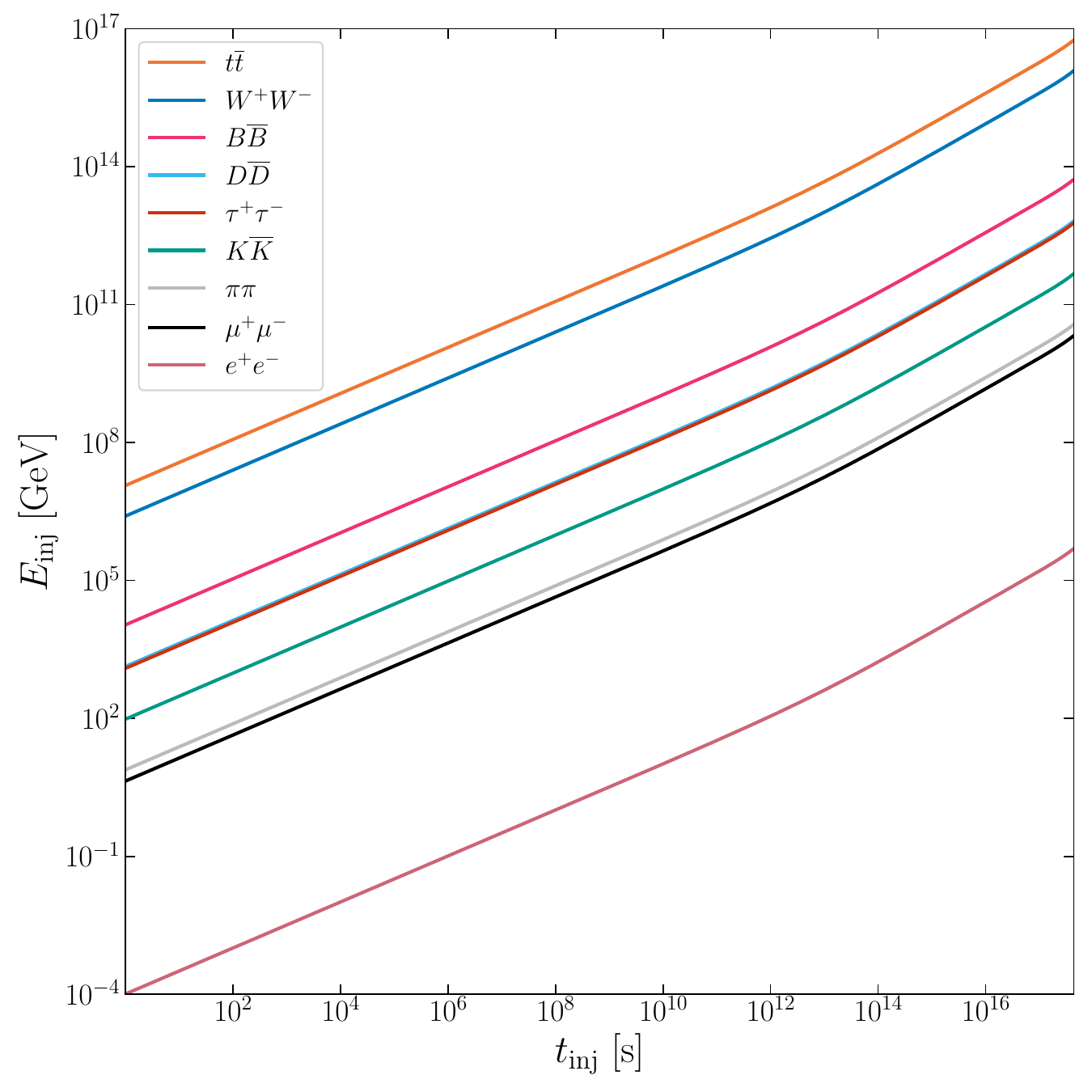}
    \caption{Minimum neutrino energy at injection $E_{\text{inj}}$ needed  to create the specified SM particle pair at a given injection time $t_{\text{inj}}$ through scattering over a thermal neutrino (considering that the thermal neutrino has an energy equal to its average energy).
    }
    \label{fig:thresholds}
\end{figure}

Since a background neutrino is (on average) much less energetic than an injected neutrino, the energy of the injected neutrino that is necessary to produce a pair of particles in an inelastic way must be significantly bigger than the mass of these produced particles.
For $e^+e^-$ pair production, i.e.~$\nu \nu_{\nu_{BG}}\rightarrow e^+ e^-$, the kinematic threshold for a scattering with background neutrinos at their mean energy is
\begin{equation}
    E_{\nu} \geq \frac{m_e^2}{\bar{E}_{\nu_{BG}}}\;,
\end{equation}
where $\bar{E}_{\nu_{BG}}$ is the mean background neutrino energy, $\bar{E}_{\nu_{BG}}=3.15 \,T_{\nu_{BG}}$, with $T_{\nu_{BG}}  = (1+z)(4/11)^{1/3}  T^0_{\text{CMB}}$ the temperature of the background neutrino at redshift $z$ (and similarly for other particle pair production, 
replacing $m_e$ by the corresponding masses). For a pair production at the mean relative velocity between neutrinos, one needs twice the energy of the threshold $E_\nu \ge 2m_e^2/\bar{E}_{\nu_BG}$.
Note that already for the next lightest particles, i.e.~$\mu^+ \mu^-$ or $\pi^+ \pi^-$ production, the threshold is significantly larger (by $\sim 5$ orders of magnitude). 
Figure~\ref{fig:thresholds} shows the minimum energy of the neutrino at injection $E_{\text{inj}}$ (i.e.~of $m_P/2$ for a 2-body decay) that is needed to enable the production of a given SM particle pair, for an injection taking place at a time $t_{\text{inj}}$, taking for the background neutrino its mean energy at this time. 
Let us recall here that for a given lifetime the peak of the 2-body injection occurs at $t_{\text{inj}}=\tau_P/2$.

The elastic cross-section, as well as the inelastic ($l^+l^-$) cross-section, are given in appendix~\ref{app:cross-section}.
In the following, we will consider a ``standard" neutrino background, where the neutrinos are democratically distributed between the 3 flavours with a negligible neutrino/anti-neutrino asymmetry. 
Note that the elastic cross-section profits from a larger combinatoric factor than the inelastic one. 
Starting from a $\nu_e \bar{\nu_e}$, including oscillations, and summing over all 3 cosmic neutrino flavours, and on the possible flavours in the final state for the elastic channels, the elastic to inelastic ($e^+e^-$) cross-section ratio is of about $13$ below the $Z$ resonance and $3.5$ above. 
Thus, one expects that the elastic type of distortion of the energy spectrum will be dominant.

To evaluate the region in which either one of, or both, the elastic and inelastic scatterings are important, or instead irrelevant, 
one can compute the average number of scatterings that a neutrino, injected at a given redshift $z_\text{inj}$, would undergo until it arrives to us (with an energy $E_\nu$ today, which we take as the energy it would have if it had not interacted),

\begin{equation}
    S_\nu (E_\nu,z_{\text{inj}}) = \int_0^{z_{\text{inj}}} \frac{dz}{(1+z) H(z)} \Gamma_\nu(E_\nu,z) \;,
    \label{Snuthermal}
\end{equation}
where $\Gamma_\nu(E_\nu,z)=\langle\sigma v\rangle n_{\nu_{BG}}$, with $n_{\nu_{BG}} = \frac{3}{4\pi^2}\zeta(3)(T^0_\nu)^3 (1+z)^3$ the number density of background neutrinos at a redshift $z$, and $\langle\sigma v\rangle$ the cross-section averaged over the thermal distribution of the background neutrinos, at a given redshift and energy. 
The scattering rate, i.e.~the number of scattering per unit time that a single injected neutrino undergoes is given by
\begin{equation}
    \Gamma_\nu(E_\nu,z) = \int\frac{d^3p_{\nu_{BG}}}{(2\pi)^3}\frac{1}{e^{E_{\nu_{BG}}/T_{\nu_{BG}}} + 1}\sigma(s)(1-\cos(\theta)) \,,
    \label{eq:Gammabackground}
\end{equation}
where $s=2 E E_{\nu_{{BG}}} (1-\cos\theta)$ is the centre of mass energy squared, $E = E_\nu (1+z)$ is 
the energy of the PHENU at redshift $z$, and $\theta$ is the angle between the PHENU and the background neutrino. 

\begin{figure}[t!]
    \centering
    \includegraphics[width=0.98\linewidth]{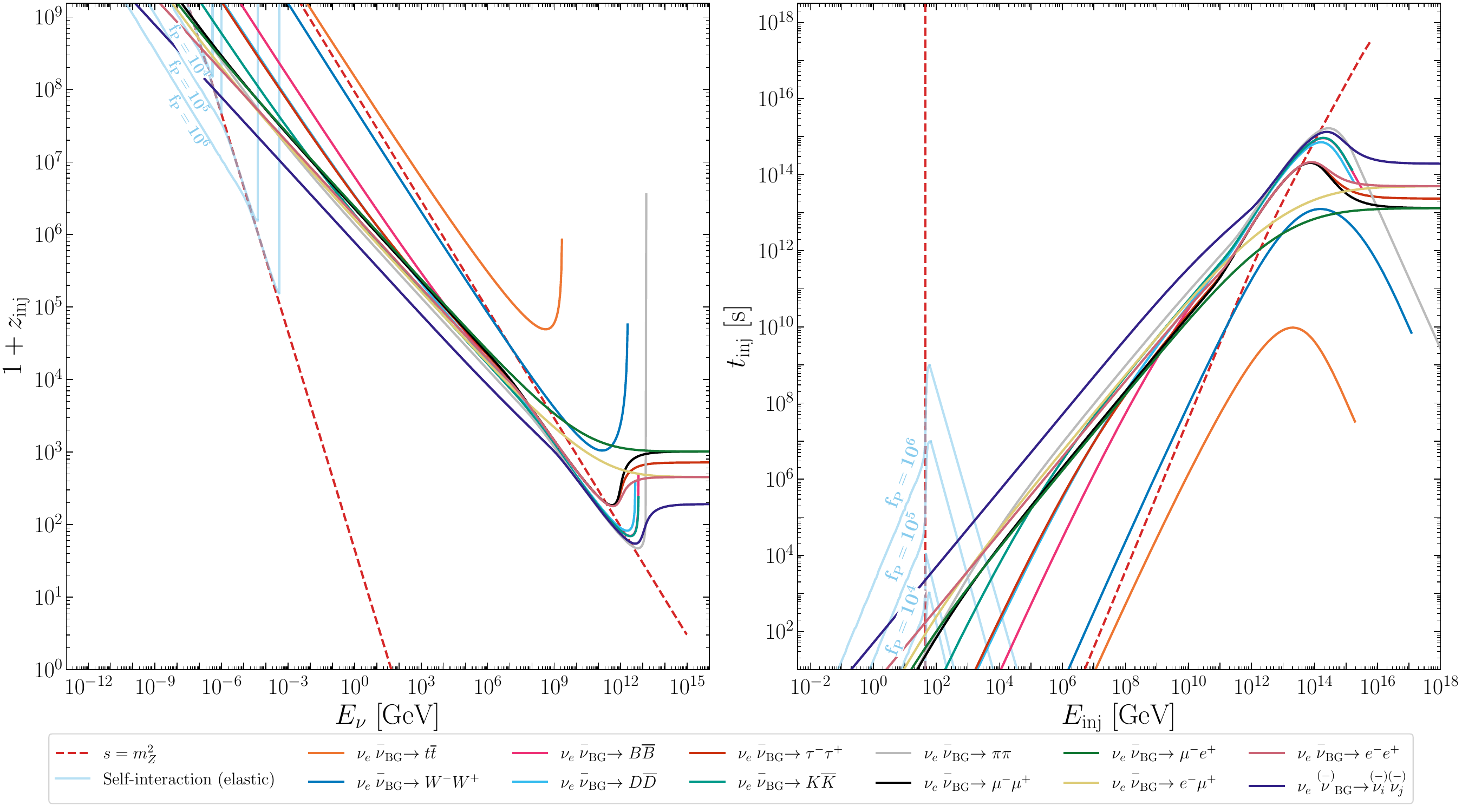}
    \caption{Left: values of the neutrino energy today and neutrino injection redshift leading, on average, to one scattering for the various possible elastic and inelastic scatterings, both from a collision on a background neutrino or on another PHENU. The dashed lines correspond to the resonance of the Z boson for both cases. The injected neutrinos are considered to be $\nu_e$, taking into account that, once produced, they oscillate into other flavours, see eq.~\eqref{eq:osc}. Right: same as the left panel, as a function of neutrino energy at production and time of injection. For a 2-body decay this right panel can also be interpreted as the number of reactions in the $m_P$-$\tau_P$ plane: the energy at production is simply $m_P/2$ and a neutrino emitted at the peak of the production spectrum, eq.~(\ref{eq:t_max_rad}), is injected at $t_{\text{inj}}=\tau_P/2$. For the scattering of 2 PHENUs the result depends on $f_P$, as indicated in the plot.}
    \label{fig:optical_depth}
\end{figure}

Integrating over the azimuthal angle and changing variable from $\theta$ to $s$, one gets (see also~\cite{Ema:2013nda}),
\begin{eqnarray}
    \Gamma_\nu(E_\nu, z) &=& \frac{1}{16\pi^2E_\nu^2(1+z)^2}\int dE_{{\nu_{BG}}} \frac{1}{e^{E_{\nu_{BG}}/T_{\nu_{BG}}} + 1} \int_0^{4 E E_{\nu_{BG}}}ds \,s\,\sigma(s)\;,\\
    \label{eq:average_scattering_rate}
    &=&  \frac{T_{\nu_{\text{BG}}}}{16 \pi^2 E_\nu^2(1+z)^2}\int_0^\infty ds s \ln(1+e^{-\frac{s}{4E_\nu(1+z)T_{\nu_{\text{BG}}}}})\sigma(s)\;.
\end{eqnarray}
where $T_{\nu_{BG}}$ is to be evaluated at the redshift considered.
Plugging this integral in eq.~(\ref{Snuthermal}), we first compute the redshift $z_{\text{max}}$ for which the average number of elastic and inelastic scatterings of the various types for a neutrino emitted at $z_{\text{max}}$ is equal to 1, left panel of figure~\ref{fig:optical_depth}.
For $z<z_{\text{max}}$ we can assume that the spectrum in the detector will be well described by the one from section~\ref{sec-spectra} (i.e.~the sharp spectral feature is not distorted), while for $z>z_{\text{max}}$ the effects of scatterings cannot be neglected.
The various lines for elastic and inelastic scatterings are shown in figure~\ref{fig:optical_depth}.
For inelastic scatterings we give the lines for $e^+e^-$ pair production as well as for heavier particle pair production.
Note that the smaller the mass of the final state, the sooner the neutrino lacks sufficient energy to create it, which explains the change of slope at lower energy for each of the inelastic processes. 
Thus, for a fixed value of $E_\nu$, 
one needs to go at higher redshift to have a neutrino energetic enough to be above this threshold.

For very high neutrino energy, the neutrino can scatter over the background neutrino with a centre of mass energy equal to the mass of the $Z$ boson.
This leads to a resonant enhancement of the cross-section (``Glashow resonance"), as can be seen on the figure.
The resonance requires that the neutrino energy at the time of the scattering is $m_Z^2/(2\bar{E}_{\nu_{BG}}[1-\cos(\theta)])$ where $\theta$ is the angle between the two neutrinos.
Thus, it can generically occur 
as long as its energy remains above $E_\nu^{\text{res}}\equiv m_Z^2/(4\cdot3.15 \cdot T_{\nu_{BG}})$ while redshifting.
This requires that the energy at source, $E_{\text{inj}}$ (equal to $m_P/2$ for a 2-body decay) is equal or larger than $E_\nu^{\text{res}}$.
This condition $E_{\text{inj}}=E_\nu^{\text{res}}$ gives the red dashed diagonal line in the figure.
Nevertheless, for $E_{\text{inj}} \gg E_\nu^{\text{res}}$ the occurrence of a resonant scattering just after injection is suppressed because the interval of (isotropic) scattering angles which leads to $\sqrt{s}=\sqrt{2 E_{\text{inj}} E_{\nu_{BG}}(1-\cos\theta)}$ within the interval $m_Z\pm\Gamma_Z/2$ is $\Delta \theta= \Gamma_Z/\sqrt{2 E_{\text{inj}} E_{\nu_{BG}}}$, i.e.~suppressed by a $1/\sqrt{E_{\text{inj}}}$ factor.
For $E_{\text{inj}} \gg E_\nu^{\text{res}}$ this suppression will be avoided when the neutrino has redshifted enough to have an energy of order $E_\nu^{\text{res}}$ but at this time the number of resonant scatterings is instead suppressed by the dilution of the neutrino number density.
This explains why in figure~\ref{fig:optical_depth}, to have one scattering on average, one must increase the injection redshift again when the injection energy goes above $E_\nu^{\text{res}}$ (i.e.~on the right side of the diagonal).
As a result, well above the resonance, the scattering rate is dominated again by non resonant contribution.
For an annihilation into charged leptons (which involves $W$ boson exchange diagrams), this leads to a flat behaviour of the one scattering line, since the cross-section scales as $s^0$.
For the other annihilation processes, instead, the cross-sections scale as $s^{-1}$, which imply that beyond a certain energy the average number of scatterings is always smaller than one, explaining that the one scattering line goes up vertically. 
Note also that in the left panel of figure~\ref{fig:optical_depth}, the reason why the $t\overline{t}$ and $W^+W^-$ channel lines are much above  the others lines is that they are never resonant (since $2m_t \gg m_Z$ and $2m_W > m_Z$).
These channels are also suppressed at high energy as the cross-section goes as $s^{-1}$ for $s$ much larger than $4\,m_W^2$ or $4\,m_t^2$.

To better understand when collisions with background neutrinos do matter, it is convenient to recast the results of the left panel of figure~\ref{fig:optical_depth} as a function of the injection parameters, $E_{\text{inj}}=E_\nu\cdot (1+z_\text{inj})$ and $t_{\text{inj}}$.
This is given in the right panel of figure~\ref{fig:optical_depth}. 
For a 2-body decay this plot can also be interpreted as the one scattering lines in the $m_P$-$\tau_P$ plane, as the energy at production is simply $m_P/2$ and a neutrino emitted at the peak of the production spectrum is injected at $t_{\text{inj}}=\tau_P/2$, eqs.~(\ref{eq:enu_max_rad}) and (\ref{eq:t_max_rad}).\footnote{
Alternatively, one could have considered the average number (over the whole energy spectrum) of scatterings a neutrino will undergo, by determining the number of scatterings each produced neutrino will have on average, integrating over all the produced neutrinos and dividing by the total number of produced neutrinos. 
 One can check that in this case, the transition lines corresponding to one scattering are also very close to the ones plotted in figure~\ref{fig:optical_depth}.}
This plot shows that the scattering occurrence increases for larger injected energies and lower emission times, i.e.~larger masses and lower lifetimes. 
Below the $Z$ resonance, higher masses give larger centre of mass energies and thus larger scattering rates since the cross-sections scale as $s$. 
Lower $\tau_P$ leads to earlier decay, increasing the neutrino travel distance along which the neutrinos can scatter.
Lower $\tau_P$ also means larger C$\nu$B temperature, implying larger $s$, i.e.~larger cross-sections.
For instance a neutrino emitted with $10^{16}$~GeV GUT scale energy will have its sharp spectral feature basically unaffected only if it has been emitted when the Universe was already $10^{15}$~seconds old. 
As another high energy scale example, for instance for a lifetime equal to the recombination time, the mass of the source particle must be smaller than $10^{11}$~GeV. And for lifetimes equal to, for example, $10^8$~sec ($10^5$~sec) one can still observe an unaltered sharp spectral feature for a source particle mass up to 10 TeV (1~GeV).
The $E_{\text{inj}}=E_\nu^{\text{res}}$ diagonal line is also indicated in this plot.

Note that what is relevant in figure~\ref{fig:optical_depth} is, above all, the elastic scattering line, because these scatterings are the fastest and thus have the larger distortion effect on the energy spectrum. 
Typically, given the factor $\sim 13$ between the respective cross-sections, when on average an inelastic scattering occurs at least once,  it will have already scattered elastically.  
This implies that the lines which give on average one inelastic scattering somewhat overestimate the effect of inelastic scattering because they assume that the incoming energy of the injected neutrino is the (redshifted) injected one.
However, in reality, this energy will be smaller due to elastic processes, suppressing the probability that an inelastic scattering occurs.
Thus, the inelastic scattering lines in figure~\ref{fig:optical_depth} have to be interpreted as conservative lines.

\begin{figure}[t!]
    \centering
    \includegraphics[width=0.5\linewidth]{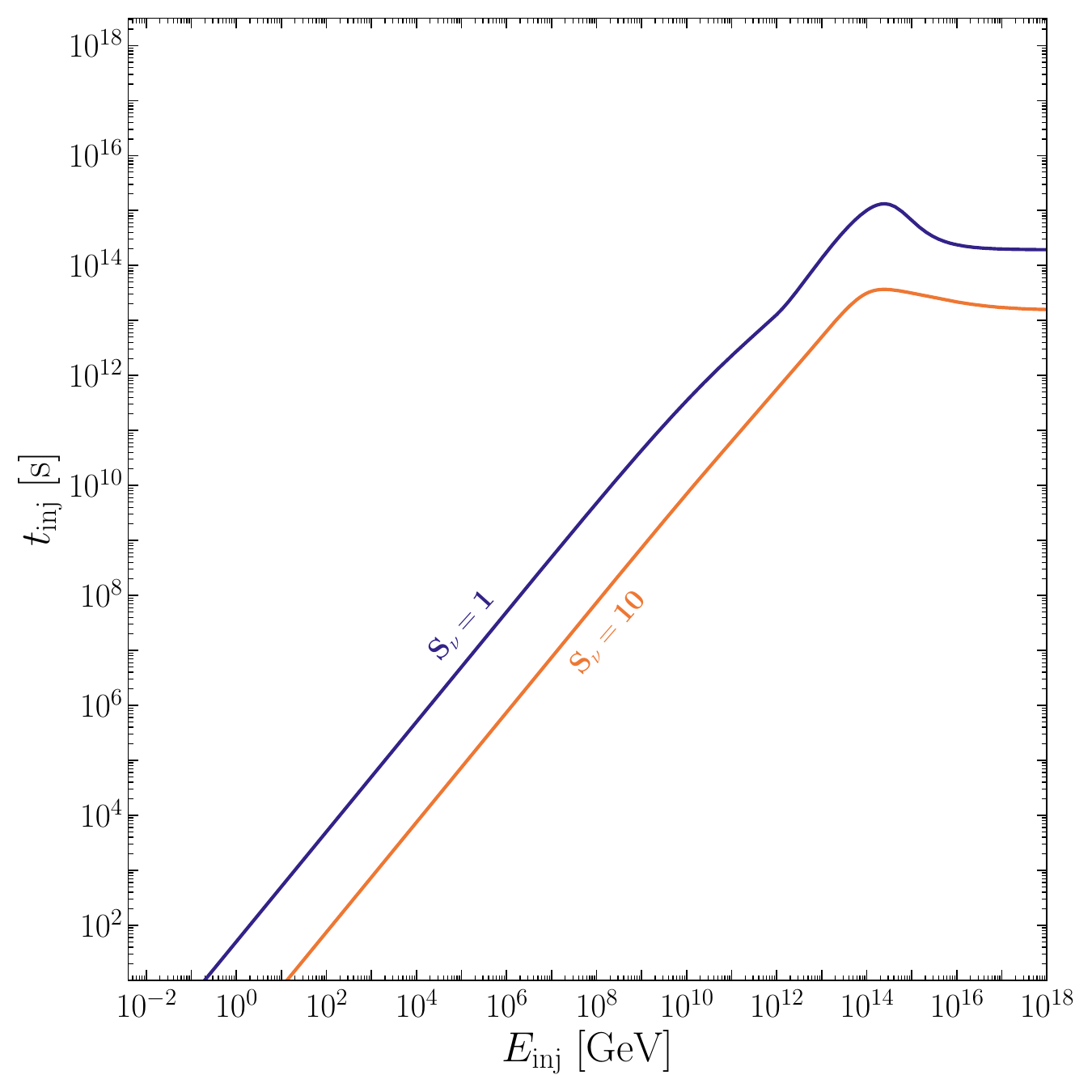}
    \caption{Values of $t_{\text{inj}}$ that lead to 10 elastic scatterings on average (orange line), as compared to the 1 scattering line (indigo line in the right panel of figure~\ref{fig:optical_depth}) as a function of the injection energy.}
    \label{fig:multi_scattering}
\end{figure}

To simply determine for which value of the parameters a PHENU undergoes on average one elastic scattering delimitates a region of parameter space where the spectrum will definitely not be significantly affected by these scatterings, but does not tell precisely where it will start to be really affected.
It can be checked that on average, a neutrino loses about $25\%$ of its energy each time it scatters elastically with a background neutrino (for energies well above the energy of the background neutrinos).
Thus, a single scattering clearly does not distort a sharp spectral feature much, but a few of them will already have an important effect.
Eq.~(\ref{eq:average_scattering_rate}) is adequate to determine when, on average, n scatterings will occur with $n=1$ but is not for $n>1$: 
to be precise, for multi-scattering, one should take into account that a fraction of the energy is lost after each scattering (as the cross-section depends on energy). 
A proper determination of when, on average, one has for instance 10 or 100 scatterings would typically require a Monte-Carlo simulation taking into account the dependence in the angle between both neutrinos entering in each collision and the resulting outgoing energies.
In this way, one could compute the distortion of the spectrum, which as said above is left for further work.
A rough estimate can nevertheless be performed using iteratively eq.~(\ref{eq:average_scattering_rate}) (to determine the time it takes to have the first collision on average, then the second one, etc.), assuming that after each collision, on average, the neutrino loses $25\%$ of its energy.
Proceeding as such, figure~\ref{fig:multi_scattering} shows, as a function of $E_{\text{inj}}$, what values of $t_{\text{inj}}$ lead to 10 elastic scatterings (i.e.~the fastest process).
Below the $Z$ resonance, the injected energy required to have 10 scatterings is approximately a factor of $\sim 30$ larger than that required for a single scattering. That this factor is sizeable is not surprising, since the energy of a neutrino after 10 elastic scatterings is only a few percent of its original energy (and since the cross-section grows with energy).
However, this factor is not very large either.
Thus, below the resonance, the region of parameters where the sharp spectral feature is affected, but in a mild way, is relatively narrow. Much larger are the parameter space regions where it is not affected at all or largely suppressed/flattened.
Around the resonance, however, the region is wider. 
Due to redshift and to the energy loss of the neutrino after each scattering, it will generally not undergo multiple resonant scatterings.

In Ref.~\cite{Gondolo:1991rn} the charged lepton pair production 
from scatterings on background neutrinos has been considered to estimate what is the suppression effect (mainly in the low energy part of the spectrum), but not the faster elastic scatterings. 
Following the detection of 2 PeV events by the Icecube collaboration \cite{IceCube:2013low}, Ref.~\cite{Ema:2013nda, Ema:2014ufa} explored the possibility that the high energy neutrino event could originate from the decay of a long-lived heavy relic. To this end, the authors evaluated the distortion effect of the neutrino spectrum due to elastic and inelastic scatterings over the background neutrinos by numerically solving the associated Boltzmann equations.
For PeV neutrinos (in the detector) the authors find that the (high energy part of the) spectrum is largely affected by scatterings on background neutrinos for injection redshifts above $10^4$ in agreement with what we expect from the left panel of figure~\ref{fig:optical_depth}.
A full Monte-Carlo simulation of the evolution of the spectrum would allow a systematic determination of this effect, based for instance on extensions of the codes developed in \cite{Ovchynnikov:2024xyd,Bianco:2025boy}.

\subsection{Self-scatterings of 2 PHENUs}\label{sec:selfscat}

At first sight one could believe, given that there are much less injected neutrinos than background ones, that the probability that 2 PHENUs self-scatter is much smaller than the one that a PHENU scatters over a background neutrino.
However, this must be checked explicitely because a PHENU is much more energetic than a background neutrino.
Thus, the centre-of-mass energy is much larger, and the cross-section much larger too (below the $Z$ resonance).
Moreover, this implies that the resonant scattering does not require as large PHENU energies as in the case of a scattering with a background neutrino.
A resonance may happen as soon as the sum of the energies of both PHENUs  at production is above $m_Z$.
As a result, the resonance requires much smaller PHENU energies at production than for a scattering on the background, and is therefore (potentially) relevant for lower energies today.
For energies well above $m_Z$, the resonant effect is nevertheless suppressed for reasons similar to those discussed above for scatterings with C$\nu$B neutrinos.

In order to quantify the above discussion, one can, as for the C$\nu$B scatterings above, compute the values of the parameters which lead on average to one scattering.
Unlike scatterings with a background neutrino, the number of self-scatterings that a neutrino undergoes depends not only on $z_\text{inj}$ and $E_\nu$ (or on $E_{\text{inj}}$ and $t_{\text{inj}}$) but also obviously on $f_P$, on the injected neutrino energy spectrum, and on the amount of distortion it experiences.
However, to compute the first scattering of a given injected neutrino one can consider that the other PHENU has still an unaffected spectrum.
For a two-body decay of a heavy particle $P$ of mass $m_P$ and lifetime $\tau_P$, the scattering rate can be obtained as for a scattering on a background neutrino, eqs.~(\ref{eq:Gammabackground}) and (\ref{eq:average_scattering_rate}), replacing the C$\nu$B momentum distribution, $\tilde{f}_{\nu_\text{BG}}=(e^{E_{\nu_\text{BG}}/T_{\nu_\text{BG}}}+1)^{-1}$ by the PHENU distribution at the redshift $z$ of the scattering. The latter can be straightforwardly obtained from the
$dN_\nu/dE$ distribution at this redshift which is just the distribution today, eq.~(\ref{dnudERad}), multiplied by $(1+z)^{3}$,
\begin{equation}
 \tilde{f_\nu}= \frac{E^2}{2\pi^2} \frac{dN_\nu}{dE}\Big|_z =\frac{\Omega_P \rho^0_{\text{crit}}E}{\pi^2m_P \tau_PH_\text{inj}}e^{-t_{\text{inj}}/\tau_P}(1+z)^3\,,
\end{equation}
where $E$ is the energy of the PHENU considered at the scattering redshift $z$, whereas $H_\text{inj}$ and $t_\text{inj}$ are evaluated at redshift $(1+z_\text{inj})=m_P/2E_\nu$. This gives
\begin{equation}
    \Gamma ( E_\nu, z) = \frac{\Omega_P^0 \rho_{\text{crit}}^0(1+z)}{4 E_\nu^2  m_P \tau_P} \int_0^{2E_\nu(1+z)m_P} ds \sigma(s) s \int_{\frac{s}{4 E_\nu (1+z)}}^{\frac{m_P}{2}} 
                                        \frac{dE}{E^3} \frac{e^{-t_{\text{inj}}/\tau_P}}{H_\text{inj}}\;.
\end{equation}
From the rate, the number of scatterings that a PHENU (of energy $E_{\nu}$ today, created at a redshift $z_\text{inj}$) has on average is equal to one if 
\begin{equation}
    S_\nu(z_\text{inj}, E_\nu) = \int_0^{z_\text{inj}} \frac{dz}{(1+z) H(z)} \Gamma(E_\nu (1+z), z)=1\;.
\end{equation}
In practice, one can approximate these integrals by considering that all PHENUs have been created at a redshift $z_\text{inj}$ with monochromatic energy $m_P/2$.
Making this approximation, figure~\ref{fig:optical_depth} shows the values of input parameters that fulfil this equation, for various values of the abundance $f_P$.
The resonance maximum occurs when $E_\text{inj}\simeq m_Z/2$ (red dashed lines), i.e.~for much smaller injected neutrino energy than for scatterings on background neutrinos, as expected.
Also, the larger $f_P$, the larger is the probability of scattering, and the larger the injection time must be to have only one scattering on average. For $E_{\text{inj}}$ below the resonance line or not too far above it, one observes that for $f_P$ larger than $\sim 10^{4}-10^5$ the self-interactions occur more often than the scatterings on background neutrinos, so that their effect cannot be neglected. 
However, beyond the fact that this consequently concerns only very large values of $f_P$ and relatively low lifetime, this is relevant only for energies today in the detector that are very low, below MeV or even far below (see left panel of the figure), too low to be interesting experimentally, see section~\ref{experiments}.\footnote{It can also be checked that for most of this region where self-scatterings could be relevant, the abundance that is needed is so large that it would lead to an early matter domination era. However, as just stated, this is anyway not relevant for a possible observation of a PHENU flux.} 

\section{BBN and CMB constraints}\label{sec:BBN-CMB}

\subsection{\texorpdfstring{$\Delta N_{\text{eff}}$-CMB constraint}{ΔNeff-CMB constraint}}\label{sec:delta_neff}

\begin{figure}[t!]
    \centering
    \includegraphics[width=0.49\textwidth]{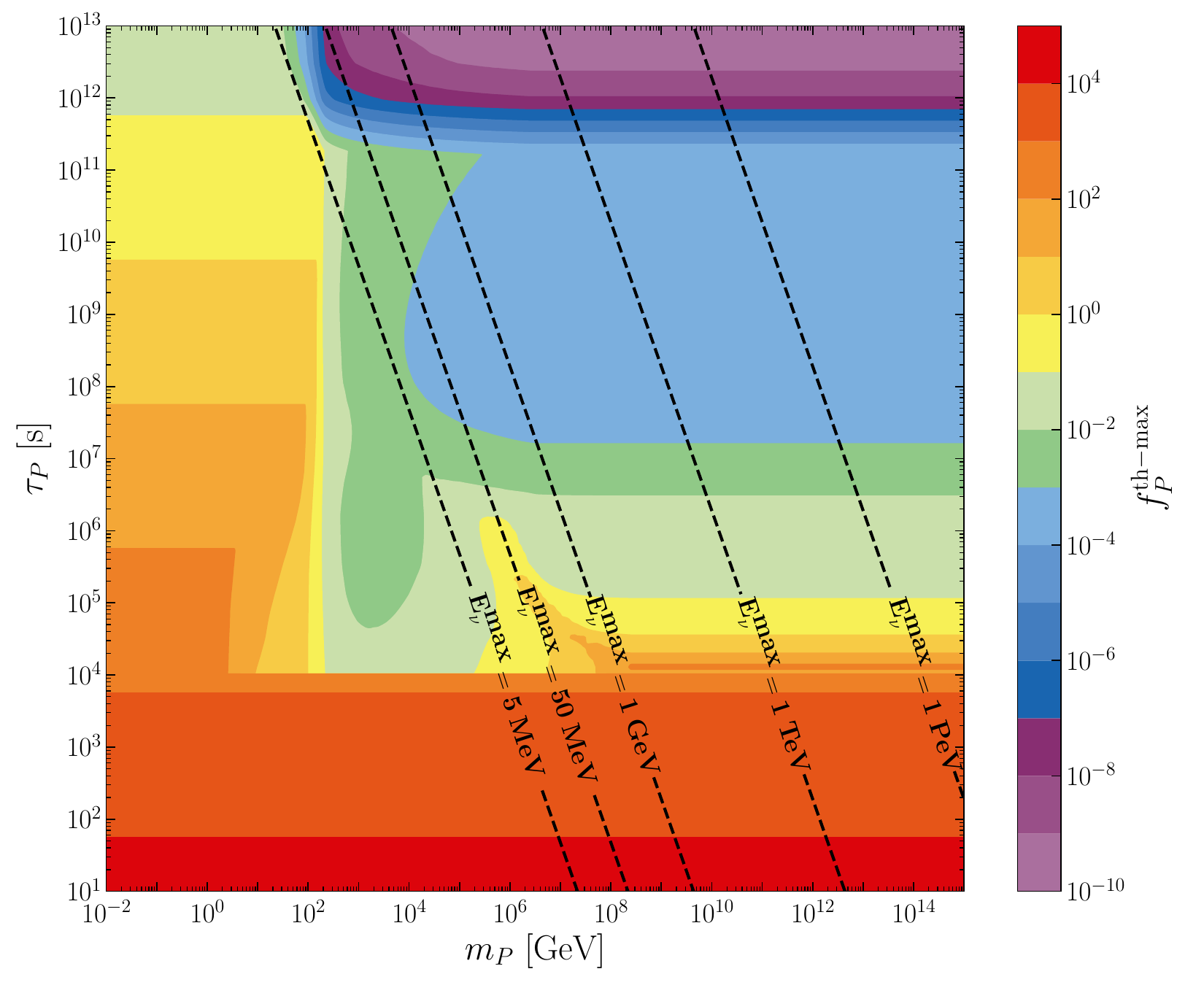}
    \includegraphics[width=0.49\textwidth]{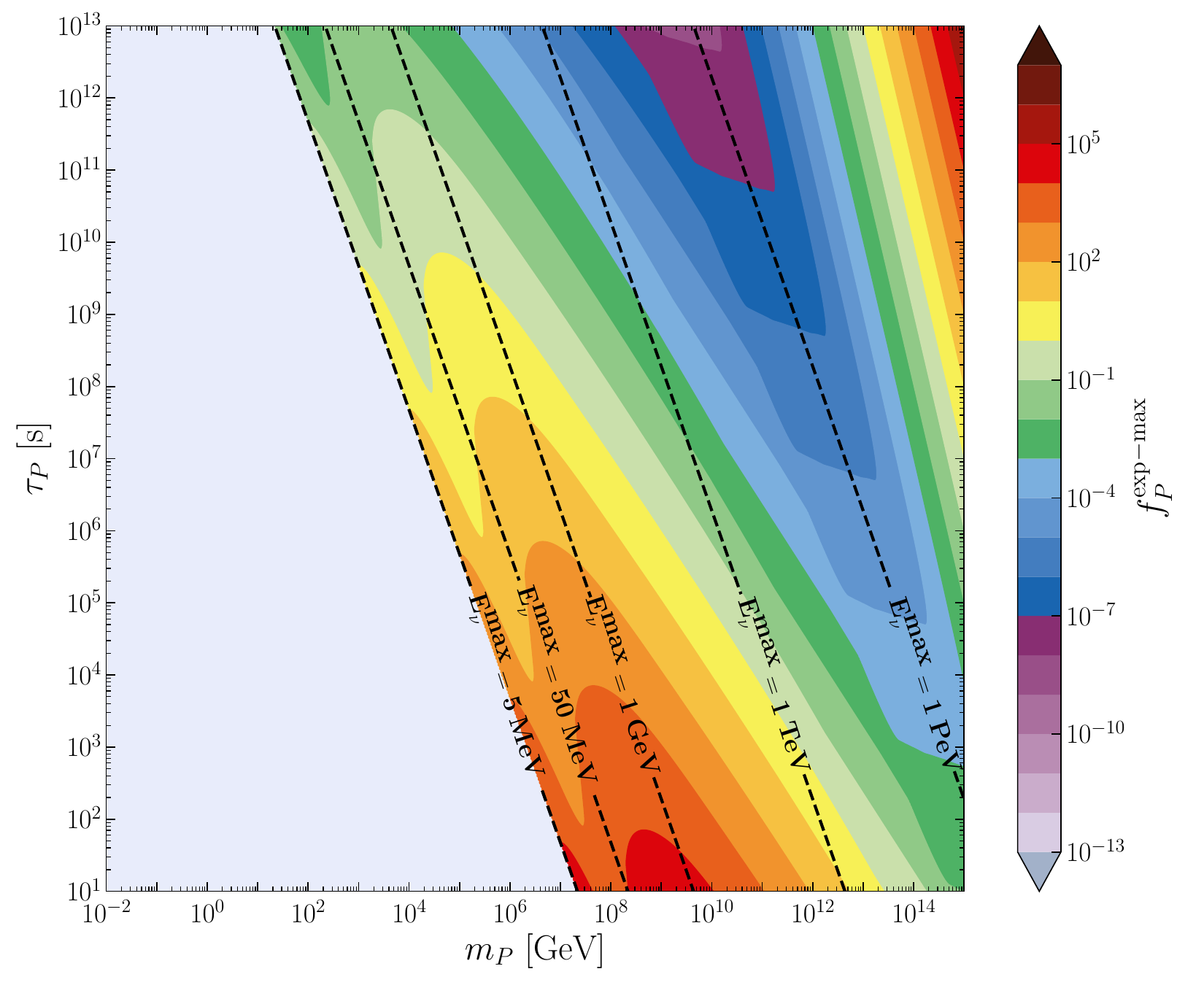}
    \caption{Upper bounds on $f_P$ as a function of $m_P$ and $\tau_P$: theoretical constraints (left panel) and experimental constraints (right panel). In the grey region the energy at the maximum of the spectrum is below 5 MeV and thus difficult to probe experimentally.\protect\footnotemark\ The black dashed lines indicate the corresponding energy at the maximum of the spectrum in the detector.}
    \label{fig:f_p_upper_limits}
\end{figure}

The abundance of the particle decaying into primordial neutrinos is constrained by CMB observations, as the injected radiation counts for $\Delta N_{\text{eff}}$, defined as:

\begin{equation}\label{eq:delta_neff}
    \Delta N_{\text{eff}} = \frac{\rho_{\nu_{\text{inj}}}(t_{\text{rec}})}{2\frac{7}{8}\frac{\pi^2}{30}T^4_{\nu_{BG}}(t_{\text{rec}})}
    =\frac{8}{7}\left( \frac{11}{4} \right)^{(4/3)} \frac{\rho_{\nu_{\text{inj}}}(t_{\text{rec}})}{\rho_\gamma(t_{\text{rec}})}\;,
\end{equation}
where $\rho_{\nu_{\text{inj}}}$ and $\rho_\gamma$ are the PHENU and photon energy densities, with $t_{\text{rec}}$ the recombination time.
The energy injected as neutrinos by the decay of the source particle at a given time is equal to the number of P particle decay per unit time multiplied by the number of neutrino produced by each decay $\Delta n_\nu$, e.g.\ right hand side of eq.~(\ref{dnudtline}) for a two-body decay, times the average energy injected per neutrino, e.g.~$m_P/2$ for the two-body decay.
Subsequently, the neutrino energy is redshifted by a factor $a_\text{inj}/a_{\text{rec}}$ (where $a$ is to be taken at injection time $t_\text{inj}$) and their number is diluted by an $(a_\text{inj}/a_{\text{rec}})^3$ volume factor so that
\begin{equation}
\rho_{\nu_{\text{inj}}}(t_{\text{rec}})=\int_0^{t_{\text{rec}}} dt_{\text{inj}} \frac{\Omega_P \rho_{\text{crit}}^0 a_\text{inj}}{\tau_P a^4_\text{rec}}\frac{2 \bar{E}_\text{inj}}{m_P}\frac{\Delta n_\nu}{2}e^{-t_{\text{inj}}/\tau_P}\;.
\end{equation}
In this equation $\bar{E}_\text{inj}$ is the average energy of the neutrino at injection. It is equal to $m_P/2$, $3m_P/8$ and $m_P/4$ for the 2-body, 3-body and box-shaped cases respectively.
Assuming that all decays take place during the radiation dominated era, this integral gives
\begin{equation}
 \rho_{\nu_{\text{inj}}}(t_{\text{rec}})=  \frac{\Omega_P \rho_{\text{crit}}^0}{2 a^4_\text{rec}}\sqrt{\frac{\pi \tau_P}{t_r}} \frac{\Delta n_\nu}{2}\cdot A
\end{equation}
where $A=1, 3/4, 1/2$ for the 3 cases respectively.  
Thus, using eq.~\eqref{eq:delta_neff} one gets
\begin{equation}
    f_P =  \left(\frac{\Delta N_{\text{eff}}}{0.29}\right) \Big(\frac{5.6 \cdot 10^9 s}{\tau_P}\Big)^{1/2} \left(\frac{2}{\Delta n_\nu}\right) \cdot A^{-1} \,.
    \label{DeltaNeffbound}
\end{equation}
Taking as upper bound the value 0.29 \cite{Planck:2018vyg}, this directly gives an upper bound on the abundance $f_P$ as a function of the lifetime $\tau_P$, independently of the mass $m_P$. For the 2-body case (with $\Delta n_\nu=2$) it is given by the horizontal bands in the left panel of figure~\ref{fig:f_p_upper_limits} and constitutes the first of the several theoretical upper bounds that hold, all given in this figure.

\begin{figure}[t!]
    \centering
    \includegraphics[width=0.6\textwidth]{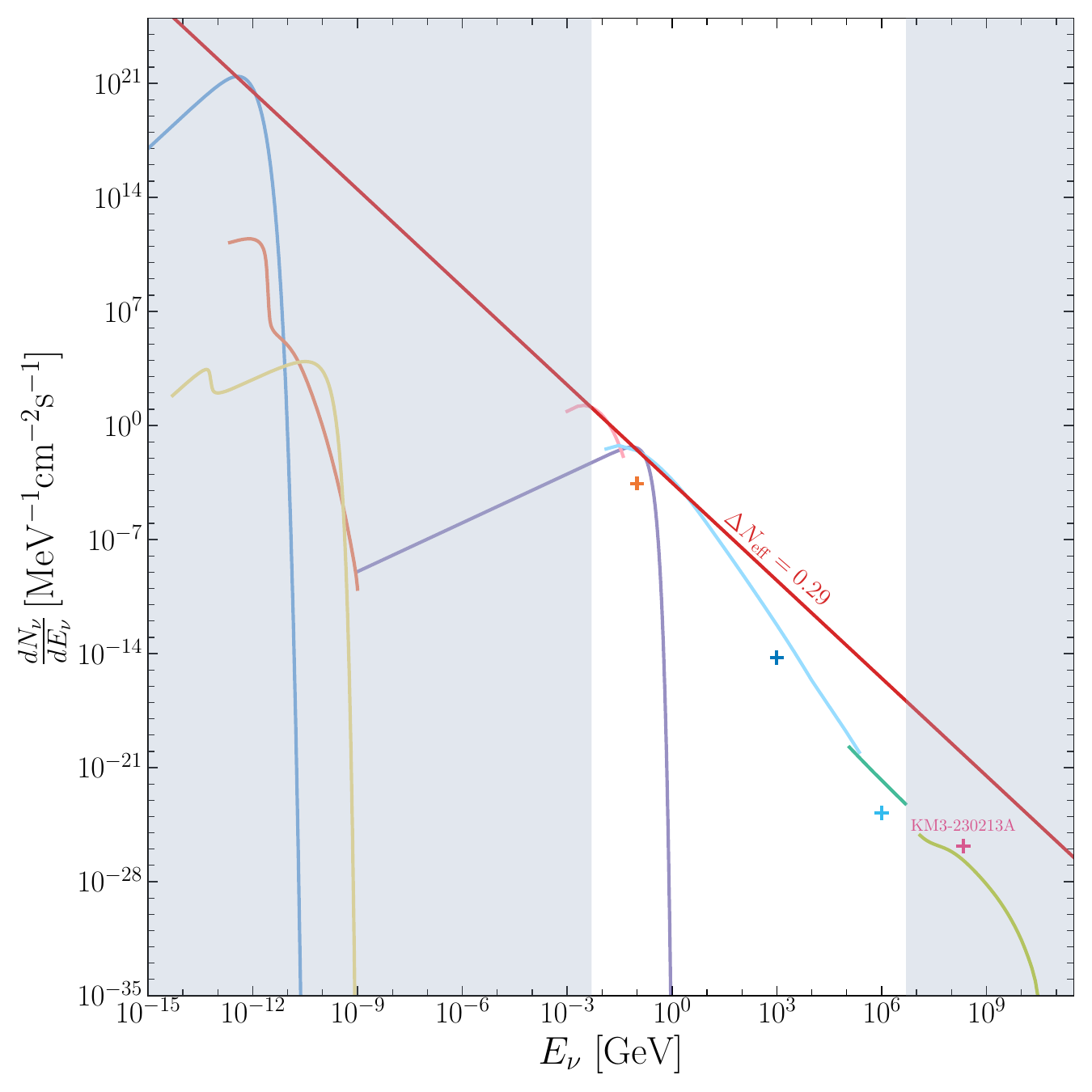}
    \caption{The upper bound on the flux from 2-body decays into neutrinos at its maximum from the $\Delta N_{\text{eff}}<0.29$ constraint (red line). Also shown are the various observed (atmospheric and astrophysical) and expected isotropic flux (cosmic neutrino background, BBN neutrinos, DSNB  and cosmogenic neutrinos), from \cite{Vitagliano:2019yzm}.
    The blank region (from 5~MeV to 5~PeV) is the one we consider to set experimental bounds on the source particle abundance. An example of 2-body decay sharp spectral feature, that saturates the $\Delta N_{\text{eff}}$ constraint, is also given in indigo (see also the related horizontal $\Delta N_{\text{eff}}$ upper bound in the left panel of figure~\ref{fig:f_p_upper_limits}). The four coloured points correspond to the benchmark values used for figure~\ref{fig:on-off}. In particular, the ``KM3-230213A" labelled point corresponds to the recent ultra-high energy neutrino event from KM3NeT, using the central values of figure~5 from Ref.~\cite{KM3NeT:2025npi}.}
    \label{fig:neff_bound}
\end{figure}

\footnotetext{One has checked that for all these upper bounds, at $t\sim\tau_P$ the source particle energy density is subleading with respect to the SM particles radiation energy density, so that the Universe is radiation dominated at this time (as assumed to obtain these bounds).}
The bound on $\Delta N_{\text{eff}}$ can be translated to an upper bound on the flux produced as a function of $m_P$ and $\tau_P$.
In particular, it gives an upper bound on the differential flux at its maximum $E_\nu^{\text{max}}$. 
For a 2-body decay this upper bound remarkably depends only on $E_\nu^{\text{max}}$.
Plugging eq.~(\ref{DeltaNeffbound}) in eq.~(\ref{dnudERad}) taken at $E_\nu=E_\nu^{\text{max}}$, and using eq.~(\ref{eq:enu_max_rad}), the bound is (see also \cite{McKeen:2018xyz})
\begin{equation}
    \frac{d \Phi_\nu^{\text{max}}}{dE_\nu} < 1.71\cdot 10^{-7}\frac{\rho_{\text{crit}}^0}{(E_\nu^{\text{max}})^2}\left(\frac{\Delta N_{\text{eff}}}{0.29}\right)\;.
\end{equation}
This gives a simple diagonal upper bound in the neutrino energy versus flux today plane, scaling as $(E_\nu^{\text{max}})^{-2}$, as given in figure~\ref{fig:neff_bound}.
This upper bound on the flux can be compared with observed isotropic fluxes, also given in this figure, as will be discussed in more details in section~\ref{experiments}.\footnote{
Note that, as the bound scales as $E_\nu^2$, the energy for which the spectrum hits the limit is not the peak energy of the spectrum but the peak energy of $E_\nu^2 \frac{d \phi_\nu }{d E_\nu}$. This difference is nevertheless small.}
Figure~\ref{fig:neff_bound} also shows an example of 2-body PHENU flux that just saturates the $\Delta N_{\text{eff}}$ upper bound.

For the annihilation case, one obtains similarly
\begin{equation}
    \rho_{\nu_\text{inj}}(t_\text{rec}) = \int_{t_\star}^{t_\text{rec}}dt_\text{inj} \frac{(\Omega^0_P \rho^0_\text{crit})^2\langle\sigma v\rangle t_r^3}{2 m_P t_\text{rec}^2 t_\text{inj}}\frac{1}{\left[1 + \frac{\langle \sigma v \rangle \Omega^0_P \rho^0_\text{crit}}{m_P}\sqrt{\frac{t_r^3}{t_\text{inj}}}\left(\sqrt{\frac{t_\text{inj}}{t_\star}} -1\right)\right]^2}\;.
\end{equation}
This can be integrated analytically (which leads to a long expression we will not give) and put in eq.~(\ref{eq:delta_neff}) to obtain the corresponding upper bound on $f_P$.
Using the lower bound on the cross-section from eq.~(\ref{eq:sigma_min}), one can also get a general conservative bound
that is
\begin{equation}
    \rho_{\nu_\text{inj}} > \frac{\Omega^0_P \rho^0_\text{crit}}{8\left(1-\sqrt{\frac{t_\star}{4t_\text{rec}}}\right)}\sqrt{\frac{t_r^3 t_\star^2}{t^5_\text{rec}}}\left(2\sqrt{\frac{t_\text{rec}}{t_\star}} - 2 + \left(2\sqrt{\frac{t_\text{rec}}{t_\star}}- 1\right)\ln\left[2\sqrt{\frac{t_\text{rec}}{t_\star}} - 1\right]\right)\,,
\end{equation}
which, using eq.~\eqref{eq:delta_neff}, turns into the upper bound
\begin{equation}
    f_P < \left(\frac{\Delta N_\text{eff}}{0.29}\right) \left(\frac{1.8\cdot 10^{12}s}{t_\star}\right)\frac{1-\sqrt{\frac{t_\star}{4t_\text{rec}}}}{2\sqrt{\frac{t_\text{rec}}{t_\star}} - 2 + \left(2\sqrt{\frac{t_\text{rec}}{t_\star}}- 1\right)\ln\left[2\sqrt{\frac{t_\text{rec}}{t_\star}} - 1\right]}\,,
\end{equation}
where the fraction on the right hand side of the inequality is $\simeq 1$ for $t_\star \ll t_\text{rec}$, so that in this limit we have
\begin{equation}
    f_P \lesssim 320 \left(\frac{\Delta N_\text{eff}}{0.29}\right) \left(\frac{5.6 \cdot 10^9 s}{t_\star}\right)\,.
\end{equation}

\subsection{BBN photo/hadro-disintegrations and CMB anisotropies/distortions}

Via radiative corrections, high-energy neutrinos produce SM particles that, unlike neutrinos, feel the electromagnetic interaction (charged leptons, $W$ bosons, ...) and possibly also the strong interaction (hadrons via quark production).
These neutrino FSR processes become important for injected neutrino energy around the electroweak scale or beyond.
Production of electromagnetic and hadronic material can also result from the scattering of an injected neutrino with a background neutrino
or from the self-scattering of two injected neutrinos. 
Once produced, this
material can subsequently photo- or hadro-disintegrate nuclei that have been previously formed during the BBN epoch.
This could spoil the remarkable agreement of the SM with the observations of light-element abundances. 
This results in an upper bound on the abundance $f_P$ as a function of $m_P$ and $\tau_P$.
This has been considered in \cite{Hambye:2021moy} (for photo-disintegration effects from FSR only) and in much more detail in \cite{Bianco:2025boy}, adding in particular hadro-disintegration from FSR effects as well as photo- and hadro-disintegration from scatterings with a C$\nu$B neutrino and of 2 injected neutrinos, resulting in constraints that can be way more stringent for a large fraction of the parameter space (see also \cite{Kanzaki:2007pd} for an early work on some of the effects of neutrino production of particles through scatterings on background neutrinos).
Injected high-energy neutrinos can also modify the CMB anisotropies as well as distort the CMB energy spectrum. 

 We use the results of figure~10 of \cite{Bianco:2025boy}, that gives the photo- and hadro-disintegration upper bounds on $f_P$ for various masses as a function of the lifetime, and use the code developed in that reference to determine these bounds for more values of the mass.\footnote{In particular we also use this code for masses well beyond the $\sim 10$~TeV  value up to which it has been mostly considered in \cite{Bianco:2025boy}. This is expected to be safe, even if it could be subject to further investigation. For such energies, we only consider photo- and hadro-disintegration processes induced by FSR to establish the constraint as the scatterings (thermal and non-thermal) are subleading. At very high mass, the constraint are basically constant, 
 as a result of the saturation of the fraction of electromagnetic and hadronic material produced for very high masses, see \cite{Bianco:2025boy}.} 
 These constraints have been derived for $\tau_P>10^4$~sec.
The left panel of figure~\ref{fig:f_p_upper_limits} shows, as a function of $m_P$ and $\tau_P$, the upper bounds that hold on $f_P$, combining these BBN constraints obtained in \cite{Bianco:2025boy} with the CMB ones which we take in \cite{Hambye:2021moy} (the latter based mostly on \cite{Acharya:2019uba}, see also \cite{Poulin:2016nat,Chluba:2020oip,Fu:2020wkq}). 
This panel shows, as in \cite{Bianco:2025boy}, that the ``direct" BBN and CMB upper bounds on $f_P$, as a function of $m_P$ and $\tau_P$, are always more stringent than the ones from the $\Delta N_{\text{eff}}$ constraint as soon as the source particle mass $m_P$ is above a few GeV for short lifetime and above $\sim 100$~GeV for larger lifetime.
It would be important to determine the BBN constraints for shorter lifetimes, $\tau_P<10^4$~sec, where they are presumably stringent too.
In the absence of a precise determination of these constraints for such short lifetimes, we will only consider the $\Delta N_{\text{eff}}$ bound in this region, as shown in the left panel of figure~\ref{fig:f_p_upper_limits}. 
Also, a comparison of the left panel of figure~\ref{fig:f_p_upper_limits} with the right panel of figure~\ref{fig:optical_depth} shows, as expected, that a large part of this region where these direct bounds are relevant lies within the region in which one expects in-flight interactions to occur, but not only.\footnote{This is due to FSR which does not require any in-flight interaction and also to the fact that, even if on average a PHENU scatters less than once on a background neutrino, these scatterings can have an effect on BBN abundances (whereas to have a sizeable distortion of the sharp spectral feature one needs the average number of scatterings per PHENU to be substantially larger than one).}

Note, finally, that from these constraints one can also set an upper bound on the differential flux at the energy where the spectrum is at its maximum, $E_\nu^{\text{max}}$.
This bound does not depend only on $E_\nu^{\text{max}}$ (in contradistinction to the $\Delta N_{\text{eff}}$ constraint) but depends on both $m_P$ and $\tau_P$, or equivalently on both $m_P$ and $E_\nu^{\text{max}}$ (using eq.~\eqref{eq:enu_max_rad}).
As a result, we will not plot them, see sections~\ref{ratiosection} and \ref{reconstructionsection} for related discussions.

\section{Experimental upper bounds on neutrino fluxes\label{experiments}}

A fully detailed analysis of what actually are the experimental bounds on $f_P$, as a function of $m_P$ and $\tau_P$, would require several dedicated analysis, all the way from sub-eV energies to well above PeV energies, which is beyond the scope of this analysis.
One can nevertheless estimate these bounds in the following way.

First of all, for definiteness and simplicity, we consider that the measured and expected fluxes are those given by figure~1 of Ref.~\cite{Vitagliano:2019yzm}. These fluxes are reconstructed theoretical fluxes which reproduce the data well (wherever there is data). 
Out of these fluxes, we retain only the isotropic ones that have been measured: the atmospheric flux and the astrophysical IceCube flux.
Together, they cover energies from $\sim 50$~MeV \cite{PhysRevD.85.052007, Super-Kamiokande:2015qek} to a few PeV (and we take as maximum value $\sim 5$~PeV) \cite{IceCube:2017zho}.
We will also consider what we get if we go down to $\sim 5$~MeV, considering the expected dominant diffuse supernova neutrino background (DSNB), between $\sim 5$~MeV and $\sim 50$~MeV, as if it had already been observed.
 This is justified as it is not excluded that such a DSNB flux would be observed in a relatively near future in the Super-Kamiokande detector or by the HyperK and Juno experiments (and current upper bounds on it are not that far from the expected flux \cite{Super-Kamiokande:2021jaq,Harada:2023apz}).
 We do not consider any solar fluxes as they can be easily separated from a PHENU isotropic flux by cutting out an appropriate solid angle around the Sun from the experimental data.

In the right panel of figure~\ref{fig:f_p_upper_limits} we plot, for the 2-body decay case and as a function of $m_P$ and $\tau_P$, the upper bound on $f_P$ which holds under the requirement that the neutrino flux never exceeds the already observed fluxes (atmospheric and astrophysical from $\sim 50$~MeV to $\sim 5$~PeV) and also does not exceed the expected DSNB flux (between $\sim 5$~MeV and $\sim 50$~MeV). 
The lines corresponding to an energy at the maximum today equal to $5$~MeV and $50$~MeV delimitate this region. Also given are the lines along which this energy is equal to 1~GeV, 1~TeV and 1~PeV. 

In contrast to the theoretical constraints on $f_P$, which for most of the parameter space do not depend much on $m_P$, the experimental constraints have a more involved dependence on $m_P$. 
First of all, as figure~\ref{fig:neff_bound} shows, the experimentally observed flux essentially scales as $E_\nu^2$, which explains that the bound on $f_P$ gets more stringent for large $m_P$ (and fixed values of $\tau_P$). 
However, for very large values of $m_P$, they relax, first because the IceCube neutrino spectrum is less steep than the atmospheric one, and subsequently when the energy of the maximum of the spectrum goes beyond the largest experimental energy $\sim 5$~PeV. 
There is still an experimental constraint for $m_P$ beyond this threshold but it gets quickly weaker as it comes only from the low energy tail of the sharp spectral feature that must not go over the experimental flux at 5~PeV. 
Instead, for energies at the maximum below 5 MeV the constraint becomes irrelevant very quickly as the high energy tail of the 2-body spectrum is very steep, and thus for simplicity we shade in grey the region where the energy at the maximum is below 5~MeV. 
A PHENU flux observation at such energies would be clearly challenging (see nevertheless \cite{McKeen:2018xyz} for possibilities at $\sim$~eV energies by the Ptolemy experiment).
Slightly above this 5~MeV threshold, the reason for the up-and-down behaviour of the bounds stems from the atmospheric flux which flattens towards low energies (around 50-200~MeV energies) and from the DSNB flux that takes over at lower energies in a steeper way (between 5 and 35~MeV).

Obviously, better bounds on $f_P$ could be obtained from proper experimental analysis, taking into account that the observed flux is well fitted by the atmospheric background (and astrophysical background power law), so that there would be room for a PHENU flux basically only within the error bars. 
For instance, if the error bars allow for a PHENU flux only at the level of 10\% of the measured flux, the upper bounds on $f_P$ are the one given in the right panel of figure~\ref{fig:f_p_upper_limits}, divided by a factor 10.
As already stated above, we will not do this detailed analysis here, left for further publications. 
That would presumably improve the upper bounds of figure~\ref{fig:f_p_upper_limits} by $\sim$ 1 order of magnitude where the atmospheric flux is the most precisely measured.

\begin{figure}[t!]
    \centering
    \includegraphics[width=0.5\textwidth]{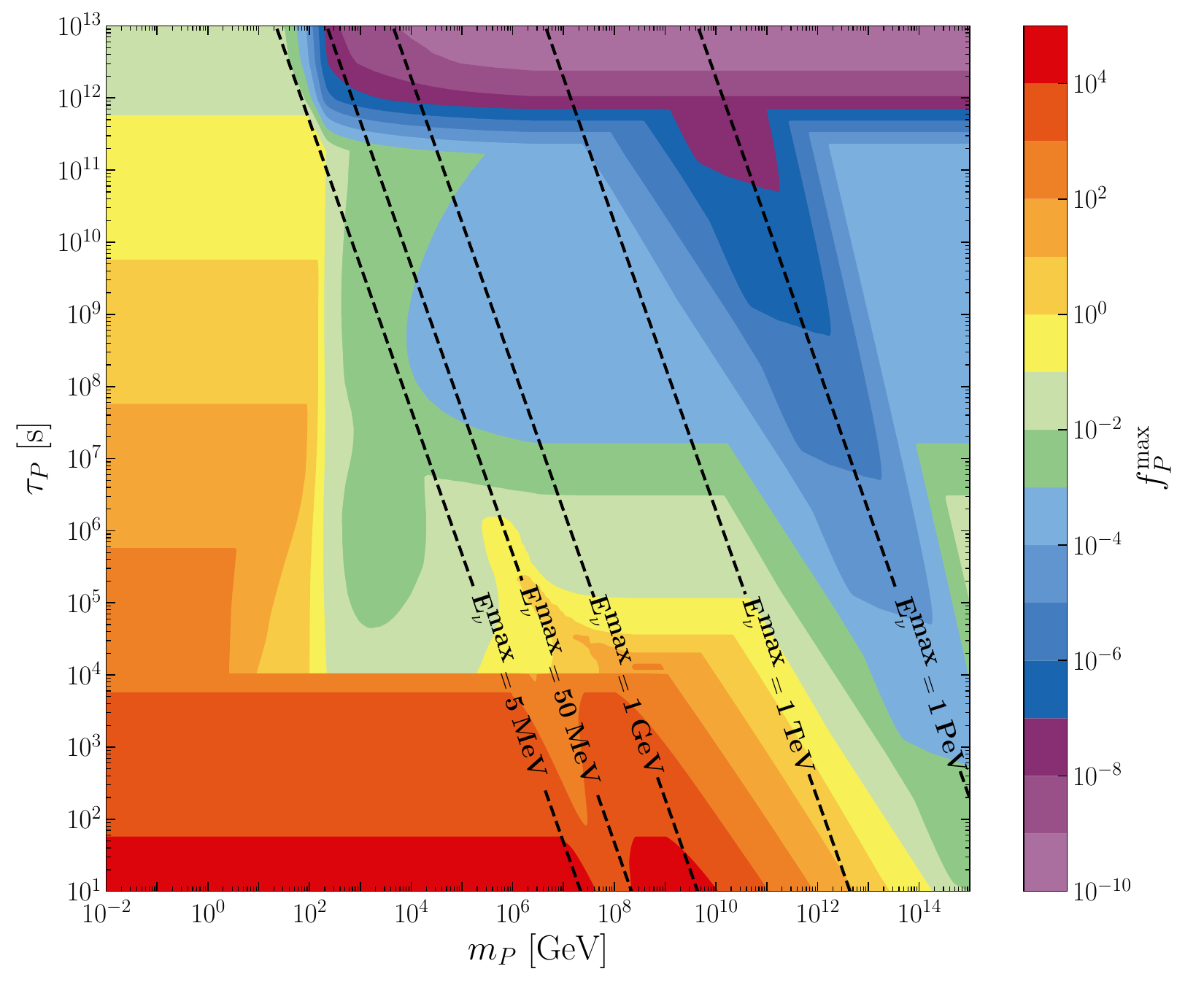}
 \caption{Upper bounds on $f_P$ combining all theoretical constraints (left panel of figure~\ref{fig:f_p_upper_limits}) with the experimental constraints (right panel of figure~\ref{fig:f_p_upper_limits}).} \label{fig:f_p_upper_limits_combined}
\end{figure}

In figure~\ref{fig:f_p_upper_limits_combined}, we show the upper bound on $f_P$ which results from combining all theoretical constraints and the current experimental constraints (i.e.~both panels of figure~\ref{fig:f_p_upper_limits}). 
This is the current global upper limit on $f_P$.

Finally, for curiosity, in the left panel of figure~\ref{fig:f_p_upper_limits_expected} in the appendix we draw the same as in the right panel of figure~\ref{fig:f_p_upper_limits}, but requiring that the flux never exceeds not only the already observed isotropic fluxes but also all other expected isotropic fluxes (C$\nu$B neutrinos, BBN neutrinos, cosmogenic neutrinos). 
This would hold for the case where the PHENU flux would not exceed expected fluxes and that the latter would be measured (which, of course, in some cases is very far from being conceivable in the near future). 
In other words this tells us what abundance of the source particle would give a PHENU flux at the level of the various expected fluxes. 
Further plots applying to this situation can be found in the appendix too.

\section{Mass and lifetime regions that could be probed experimentally in the light of the theoretical constraints\label{ratiosection}}

From both experimental and theoretical constraints above, one can now finally determine the region of parameters which could lead to an observable flux in agreement with the CMB and BBN constraints. 

First, figure~\ref{fig:neff_bound} allows us to compare the observed fluxes with the $\Delta N_{\text{eff}}$ upper bound that holds on them. 
The atmospheric, IceCube astrophysical and DSNB fluxes sit around or below the $\Delta N_{\text{eff}}$ constraint. 
Thus, an observation of a PHENU flux in the corresponding energy ranges is not forbidden by $\Delta N_{\text{eff}}$. 

\begin{figure}[t!]
    \centering
    \includegraphics[width=0.49\textwidth]{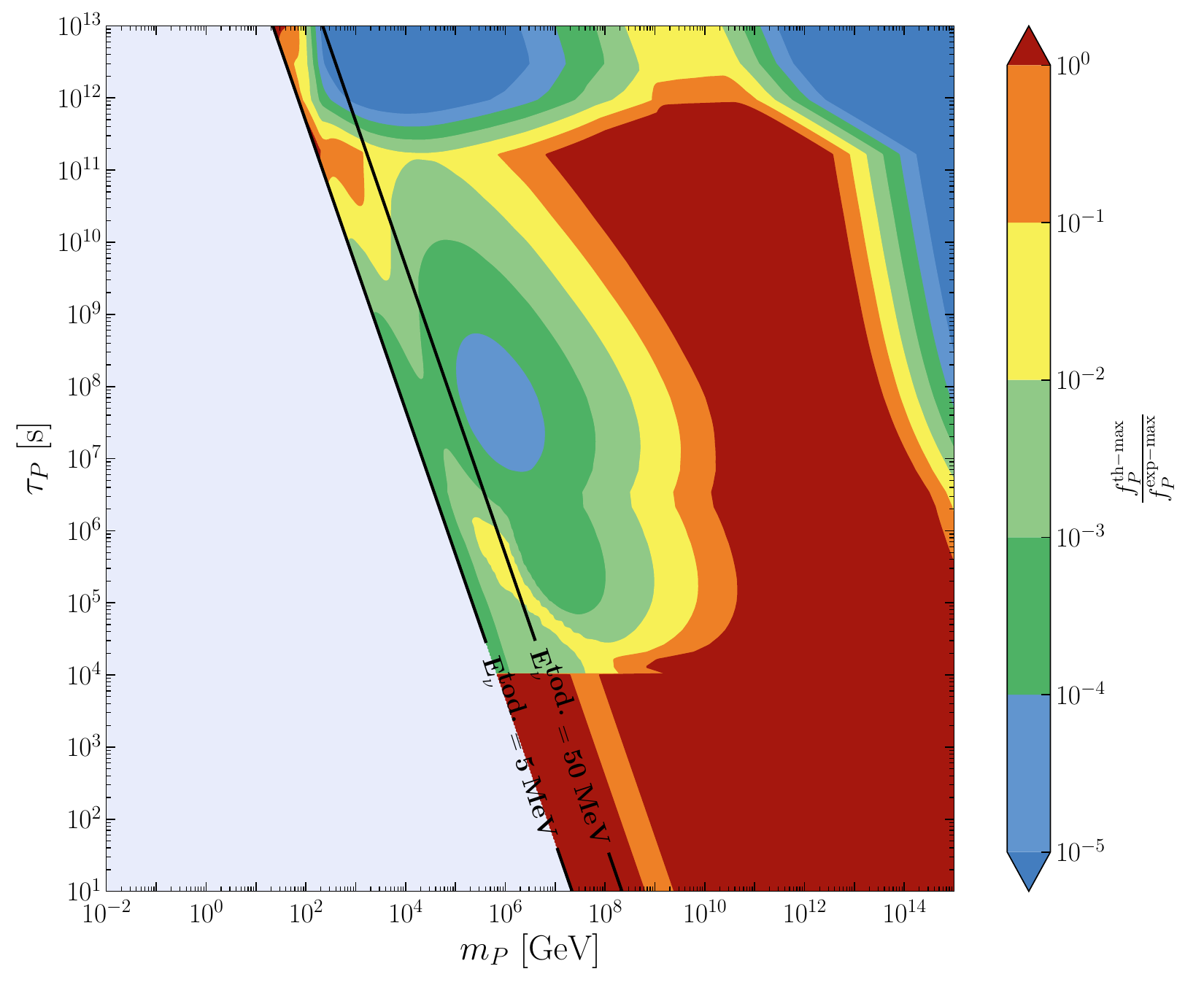}
    \includegraphics[width=0.49\textwidth]{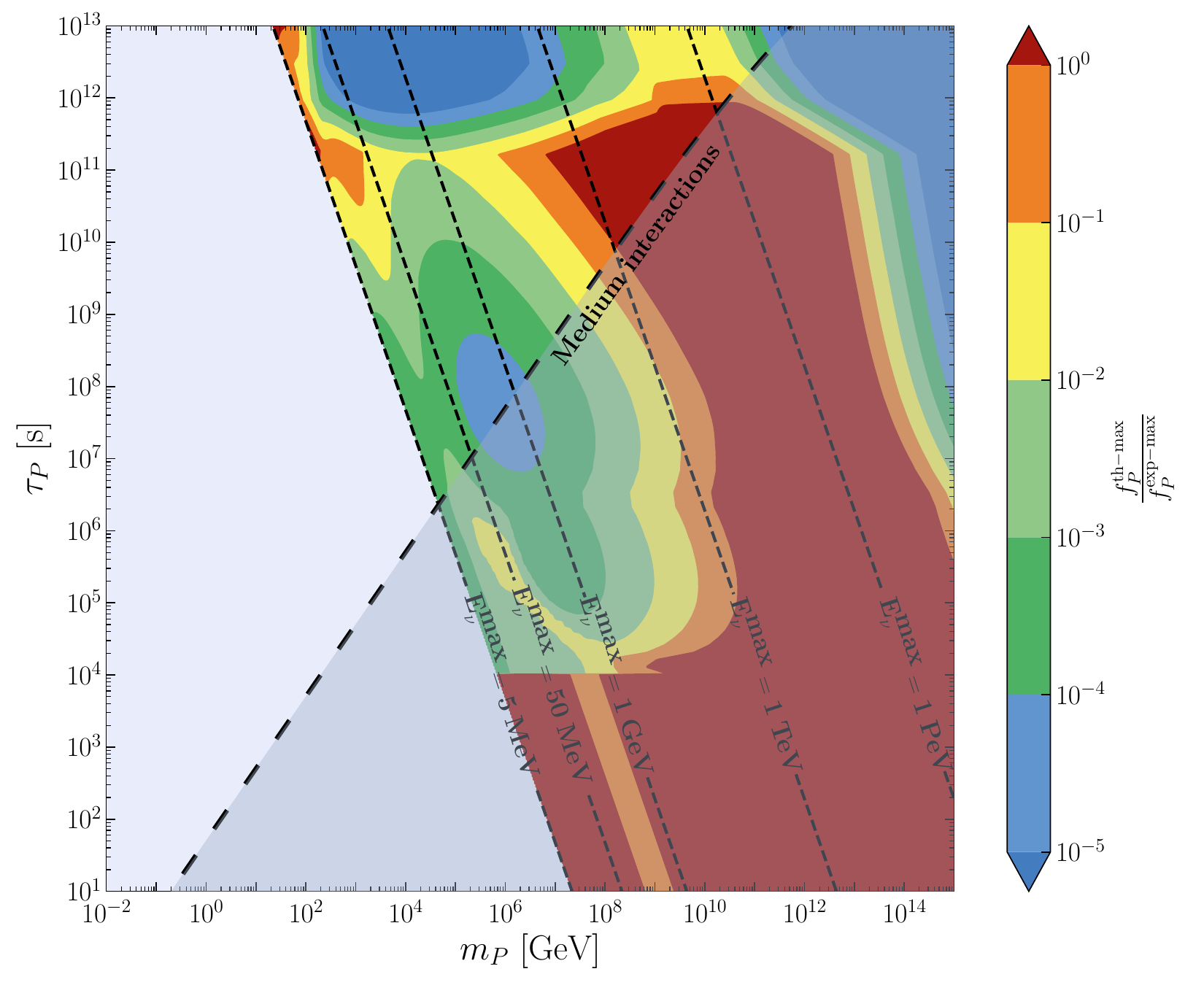}
    \caption{Left: The ratio of the maximal $f_P$ allowed by the CMB, BBN and $\Delta N_{\text{eff}}$ constraints (left panel of figure~\ref{fig:f_p_upper_limits}) with the one allowed by experimental constraints (right panel of figure~\ref{fig:f_p_upper_limits}). The light grey region corresponds to parameters which give $E_\nu^{\text{max}}<5$~MeV, difficult to probe experimentally.
    Right: same, adding the condition (dashed black line) that the produced PHENU undergoes on average less than one elastic scattering with a background neutrino on its way to the detector, so that the sharp spectral feature is safely undistorted (unshaded region).}
    \label{fig:ratio}
\end{figure}

As discussed above, the $\Delta N_{\text{eff}}$ constraint is not the only theoretical one and in general not the most stringent one. 
A better determination of what is allowed comes from a comparison of both
panels in figure~\ref{fig:f_p_upper_limits}. 
One must look for the regions where the theoretical limits are above, or not far below, the experimental ones.
One observes that the theoretical constraints allow for a flux of PHENUs that could be above the experimentally observed fluxes (or above a non-negligible fraction of them) in large regions of the parameter space but not everywhere. 
To better determine what are the possibilities of observing a PHENU flux it is convenient to plot instead, as given in the left panel of figure~\ref{fig:ratio}, the ratio of theoretical and experimental upper bounds that hold on the abundance, $f_P^{\text{th-max}}/f_P^{\text{exp-max}}$ (i.e.\ the ratio of the constraints given in both panels of figure~\ref{fig:f_p_upper_limits}).
Here as well, the light grey region corresponds to neutrino energies today (at the maximum of the spectrum) below 5 MeV. 
In the $E^{\text{max}}_\nu> 5$~MeV region, wherever the theory over experimental constraint ratio is larger than unity (dark red region) the theoretical constraints allow the PHENU flux to saturate the experimental flux (but does not necessarily need to, as it is allowed to be smaller too). 
The orange region is certainly experimentally accessible too, as it corresponds to a ratio within the $10^{-1}$-1 range. 
Similarly the yellow region ($10^{-2}$ to $10^{-1}$) is promising. 
The light green region gives a $10^{-3}$ to $10^{-2}$
isotropic PHENU excess out of the isotropic background fluxes, which is more challenging experimentally. 
Dark green and light blue regions give a ratio within the $10^{-3}$-$10^{-4}$ and $10^{-4}$-$10^{-5}$ ranges respectively.

In the right panel of figure~\ref{fig:ratio} we have added to the left panel the line which delimitates the region where one expects that the PHENU will undergo scatterings on the C$\nu$B neutrino from the region where we do not expect any such scatterings.
This line corresponds to the indigo line in figure~\ref{fig:optical_depth}, which gives on average one elastic scattering on the way to the detector (i.e.~the fastest process among all the processes considered in figure~\ref{fig:optical_depth}). 
It shows that the region where such scatterings do matter (shaded region in the right panel of figure~\ref{fig:ratio}) covers a large part of the experimentally accessible region shown in the left panel of that same figure. 
As discussed in section~\ref{sec:scattering}, in this region (or at least as soon as we consider a case sizeably inside it) the sharp spectral feature is expected to be largely affected by these scatterings.
But still a rather large region where such scatterings are not expected, and thus sharp spectral features are expected to remain unaffected, appears to be experimentally accessible (unshaded colourful region). 
It is spanned by source particles with masses ranging from a few tens of GeV to $\sim 10^{11}$~GeV. 
Thus, the new physics scales that can be probed can be many orders of magnitude above the electroweak scale, actually at scales that can be much closer to the GUT scale (without reaching it) than the electroweak scale! 
The corresponding relevant range of lifetimes $\tau_P$ spans from $10^9$~sec all the way to about the recombination time $\sim 10^{13}$~sec.\footnote{It is somewhat surprising that even for an emission at the time of recombination, an observable flux is also possible (given the strong CMB constraints that apply in this case). The reason is that for few TeV to few PeV energies in the detector today, the Icecube constraints on $f_P$ are event stronger or comparable to the CMB ones, see figure~\ref{fig:f_p_upper_limits}.} 
Thus, the observation of a sharp spectral feature in the near future would not allow us to probe epochs as early as around the BBN era, but it would still provide access to periods well before the recombination time and the corresponding CMB emission time.

The right panel of figure~\ref{fig:ratio} also shows contour lines corresponding to neutrino energies at the maximum of the spectrum in the detector equal to $5$~MeV, $50$~MeV, 1~GeV, 1~TeV and 1~PeV.
This shows that an unaffected sharp spectral feature is to be observed within energies from 5~MeV (or 50~MeV) to $\sim 10$~PeV. Thus, they have to be found mostly at energies relevant for the Super-Kamiokande, Juno, Hyper-Kamiokande, IceCube and KM3Net detectors. 
A ``double binning strategy", i.e.~binning the neutrino events both in energy and direction, would be necessary to observe the sharp spectral feature and check that the signal is isotropic. 
This is possible considering cascade events at neutrino telescopes because the energy resolution for such events is sufficiently accurate. 
This has been performed for the first time by the IceCube collaboration, for the search of a monochromatic flux of neutrinos (``$\nu$-line") that could have been emitted by the annihilation or decay of DM particles into 2 neutrinos in the Milky Way centre, see \cite{IceCube:2023ies} and also \cite{ElAisati:2015ugc}. 

Note that in figure~\ref{fig:ratio} we did not delimitate the (low energy and low mass) region where the self-scatterings would be important. 
This region depends on one more parameter (i.e.~$f_P$), 
but taking the upper bound on $f_P$ from figure~\ref{fig:f_p_upper_limits} and plugging it in figure~\ref{fig:optical_depth}, we find that the concerned energies in the detector are far too low to be observable (i.e.~below $\sim 10$ keV, sitting in the grey region in figure~\ref{fig:ratio}).

Finally, it would be important to further investigate the large shaded region to determine how far one could go within this region still obtaining a distinctive signal out of the background. 
As mentioned above, this would require a proper determination of the scattering effects, presumably best obtained from Monte Carlo simulations, in the spirit of the codes that have been developed in \cite{Ovchynnikov:2024xyd,Bianco:2025boy}.

\section{Reconstruction of the source particle properties from the observation of a neutrino flux\label{reconstructionsection}}

Besides establishing which regions of the parameter space could lead to an observable flux, one must also consider a more experimentally driven question, the other way around: suppose a flux of PHENUs is observed with a given intensity, a given peak energy, and a given sharp spectral feature (corresponding to one of the sharp features above), what would be the values of the 3 input parameters that could explain such an observed signal?

\begin{figure}[t!]
    \centering
    \includegraphics[width=0.49\linewidth]{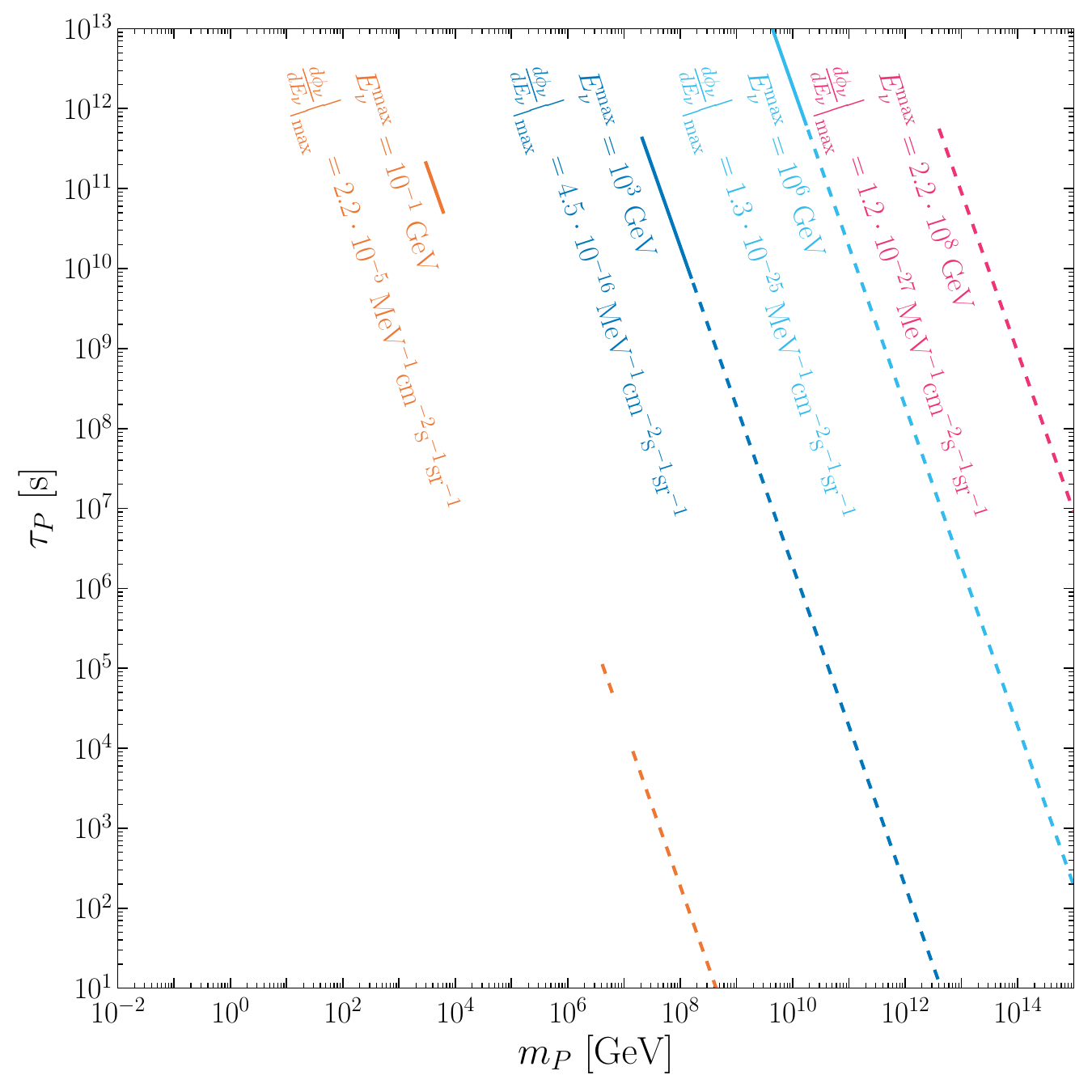}
    \includegraphics[width=0.49\linewidth]{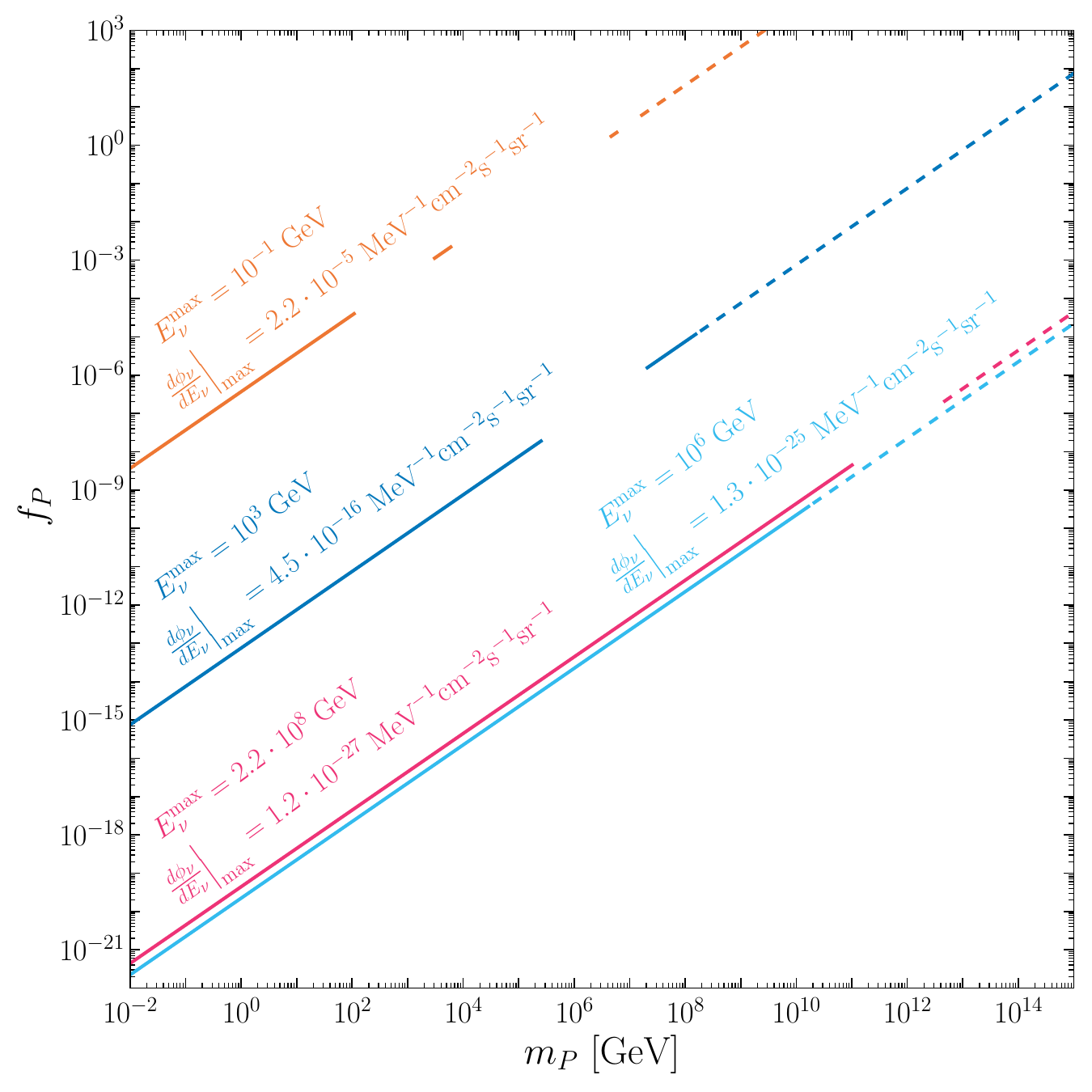}
 \caption{One dimensional parameter space allowed by theoretical constraints, for given values of energy and flux intensity at the maximum of the flux, in the $(\tau_P, m_P)$ (left) and $(f_p, m_P)$ (right) planes. The dashed parts of the lines correspond to sets of parameters for which interactions with the medium cannot be neglected. The benchmark points (orange, light and dark blue) used lead to fluxes which reach a percent of the isotropic experimental flux (at the maximum of the flux), while the recent KM3NeT ultra-high energy neutrino event (pink) is given by the central values of figure~5 from Ref.~\cite{KM3NeT:2025npi}. They are indicated by correspondingly coloured points in figure~\ref{fig:neff_bound}.}
    \label{fig:on-off}
\end{figure}

To address this question, here too we will consider the case 
of an observed 
2-body decay spectrum. 
If the spectrum is not observed precisely enough to see the spreading effect of the radiative corrections (that can be sizeable for large values of $m_P$, see section~\ref{sec:FSR}), 
the spectrum is essentially the one of figure~\ref{fig:neutrino_flux} and can be characterised by 2 experimental quantities: the energy at its maximum, $E_\nu^{\text{max}}$ and its flux intensity which we can take at the maximum too, $\left.d\phi_\nu /d E_\nu\right|_\text{max}$.
Given that there are 3 inputs but 2 experimental quantities, there is one free parameter left. 
For definiteness, we will take this free parameter to be $m_P$.
From the energy at the maximum one can determine $\tau_P$ as a function of $m_P$, as given in eq.~(\ref{eq:enu_max_rad}), which can be rewritten as
\begin{equation}
    \tau_P=8 t_r\Big(\frac{E_\nu^{\text{max}}}{m_P}\Big)^2\;.
    \label{tauPexp}
\end{equation}
Similarly, one can obtain $f_P$ as a function of both experimental inputs and $m_P$, via eqs.~(\ref{dnudERad}) and (\ref{eq:t_max_rad}), which give
\begin{equation}
    f_P=4\pi \left.\frac{d \phi_\nu}{dE_\nu}\right|_\text{max}\cdot\frac{m_P \,E_\nu^{\text{max}}}{2\,\Omega_{DM}^0\rho_{\text{crit}}^0}\;.
    \label{fPexp}
\end{equation}
Next, for both experimental inputs, via those equations, one can determine the range of values of $m_P$ which are allowed by the theoretical constraints. 
To this end, for a given value of $m_P$, and of the corresponding value of $\tau_P$, eq.~(\ref{tauPexp}), one can simply check if the corresponding value of $f_P$, eq.~(\ref{fPexp}), is below the maximum value of $f_P$ allowed by the theoretical constraints as given in the left panel of figure~\ref{fig:f_p_upper_limits}.
In figure~\ref{fig:on-off} we show, for benchmark values of both experimental inputs, what are the corresponding allowed values of $m_P$ and $\tau_P$ (left panel in close correspondence with figure~\ref{fig:ratio}) or of $m_P$ and $f_P$ (right panel). 
These values sit along one dimensional lines.
Thus, an experimental observation could considerably restrict the 
parameter space that might be at the origin of the observed signal, on top of identifying the type of process which is responsible for it (from the type of sharp spectral feature observed). 
Note that in figure~\ref{fig:on-off} a portion of the lines (dashed part) corresponds to a region of parameter space where we expect the sharp spectral feature to be affected by scatterings with C$\nu $B neutrinos (corresponding to the shaded region in the right panel of figure~\ref{fig:ratio}), so that they are partially or largely incompatible with the sharp spectral feature that would be observed. 
Thus, only the solid part of the lines fulfil this consistency check.\footnote{For curiosity, note that one of the lines in figure~\ref{fig:on-off} corresponds to a spectrum with a maximum at the (best-fit) energy and flux of the recently observed ultra-high energy neutrino event by KM3NeT as given in figure~5 of Ref.~\cite{KM3NeT:2025npi}. One observes that this falls in the region where multiple scatterings are expected. Therefore, determining whether this event could originate from a PHENU (while satisfying all constraints) would require computing the distorted spectrum and checking its compatibility with the KM3NeT and (lower energy) IceCube observations, which is beyond the scope of the present work.}
In appendix~\ref{app:plots}, we also give the regions of both experimental inputs that are allowed for $m_P$ equal to $10^{0,3,6,9,12,15}$~GeV, see figure~\ref{fig:reconstruction_emax_phi}.

In practice, for large values of $m_P$, one could even further lift the one dimensional degeneracy that is left so far, thanks to the radiative corrections. 
As discussed in section~\ref{sec:FSR}, final state radiation due to radiative corrections induces a spreading effect of the spectrum that depends on the source particle mass $m_P$, and that is sizeable for $m_P$ well above TeV, see figure~\ref{fig:FSR}.
Thus, in this case, if the spectrum is measured precisely enough, one could fully determine the values of all three source particle parameters.

\section{Case with democratic production of electrons and neutrinos}

A list of models which could lead to a relic particle decaying into neutrinos with branching ratio close to 100$\%$ can be found in \cite{Bianco:2025boy}. 
For these models the constraints above hold. 
Nevertheless, one might wonder what happens in a scenario where the decay would proceed into neutrinos and into charged lepton pairs channels with $\sim 50\%$ branching ratios each (as can happen in many different frameworks as a result of $SU(2)_L$ invariance) or more generally in any scenario where the primary produced particles are not only neutrinos.
For source particle mass of order the electroweak scale or below, the BBN and CMB photo- and hadro-disintegration constraints on the production of charged leptons are several order of magnitude more stringent than on the production of neutrinos, except for lifetime below $\sim 10^4$ to $10^5$~sec \cite{Bianco:2025boy}. 
However, for the bulk of the observable region (unshaded colourful region in the right panel of figure~\ref{fig:ratio}), which concerns much larger injected energies, this is not the case, as the theoretical constraints differ only by a factor of a few or less \cite{Kawasaki:2017bqm,Bianco:2025boy}. 
For such large masses, the photo- and hadro-disintegration constraints are similar because for the neutrino channel they come mainly from
final state radiation of electromagnetic and hadronic material by the neutrino, and the fraction of electromagnetic or hadronic energy produced by a neutrino is hardly smaller than the one produced by a charged lepton \cite{Kawasaki:2017bqm,Bianco:2025boy}. 
As for the $\Delta N_{\text{eff}}$ constraint it is also about the same, because for such energies the amount of radiation that the produced relativistic charged leptons release is not far from unity.
Thus, a PHENU observation does not require that the process producing them goes exclusively to neutrinos. For instance, the ``lepton democratic" scenarios could also lead to an observable flux.

\section{Conclusion}

In this work, we performed a general study of the possibility that a flux of primordial high energy neutrinos with characteristic sharp spectral features could be produced and observed. The observation of such a flux would allow not only to probe the early Universe before recombination but also to probe possible new particle physics at energy scales that may be many orders of magnitude above the electroweak scale. 

We identified four types of sharp spectral features. 
The three first arise from the decay of an unstable relic into two or three neutrinos, or into two intermediate particles decaying into neutrinos. 
These processes are known to produce a sharp spectral feature in the DM indirect detection context, for instance from very slow decay of non-relativistic DM particles in the centre of the Milky Way today. 
In this context, they produce a neutrino line, a sharp 3-body spectrum and a box-shaped spectrum, respectively.
However, the PHENU signal that one of these processes would induce would be very different from the DM decay case.
Beyond the fact that it would be produced by an unstable relic particle with a much shorter lifetime, and that its abundance is not fixed by any experimental constraint (unlike DM), the PHENU flux does not point towards the galactic centre but is isotropic. 
Additionally, the sharp spectral feature differs because the neutrinos produced by the decay are emitted at different times and thus experience different redshifts until today.
This redshift effect broadens the energy spectrum but still results in a quite sharp spectral feature at the detector.
The fourth sharp spectral feature we considered arises from an out-of-equilibrium annihilation occurring at a specific time in the early Universe, resulting in a completely different sharp spectral feature (distinct once again from the neutrino line produced by non-relativistic DM annihilations in the galactic centre today).

There are several theoretical upper bounds that apply on the abundance of an unstable relic decaying into neutrinos in the Early Universe, coming from constraints on $\Delta N_{\text{eff}}$, CMB spectral distortions and anisotropies, and the photo- and hadro-disintegration of light nuclei induced by neutrino production. These constraints are summarised in the left panel of figure~\ref{fig:f_p_upper_limits}. 
As for the experimental constraints, they are given in the right panel of figure~\ref{fig:f_p_upper_limits}. 
We obtained them by requiring that the PHENU flux does not exceed the already observed isotropic flux. 

Combining both theoretical and experimental constraints, we found that there is a wide region of relic particle mass, lifetime and abundance for which a PHENU flux could be observed in a relatively near future, as given by the left panel of figure~\ref{fig:ratio}, with source particle masses above a few hundreds GeV and lifetimes all the way from the neutrino decoupling time to the recombination time.
Nevertheless, a sharp spectral feature would reach us unaffected
only if the neutrinos would not interact with C$\nu B$ neutrinos, or self interact, on their way to the detector.
Requiring that, on average, a PHENU would have less than one such scattering we find that a large fraction of the above region does not satisfy this criterion, but a relatively wide region still remains.
This observable and unaffected region spans masses between a few tens of GeV to $10^{11}$~GeV (with the most promising region between $10^6$~GeV and $10^{11}$~GeV), showing that in this way we could indeed test new particle physics many orders of magnitude above the electroweak scale. 
This range is mainly determined by the lower energy threshold at a few MeV and the upper energy threshold at a few PeV of the experiments today. 
The associated lifetimes sit between $\sim 10^9$~sec and the recombination time $\sim10^{13}$~sec. 
Thus, the observation of a PHENU flux with an unaffected sharp spectral feature allows testing early times well before recombination even though not all the way to the BBN epoch. 
The reason is that for lifetimes shorter than $10^9$~sec, interactions with the C$\nu$B neutrinos become frequent at such large neutrino energies. Instead, the requirement of having no self-interactions turns out to be much less relevant, as it concerns only low source particle mass leading to neutrino energies today well below
the few MeV low energy experimental thresholds (on top of concerning only very large source particle abundances). The most promising detector energies to look for a PHENU flux sit between $\sim10$~GeV and a few PeV (even if there are possibilities down to a few MeV), a range which is fully relevant for experiments such as for instance Hyper-Kamiokande, JUNO, IceCube and KM3NeT, looking for events with good energy resolution (such as cascade events in neutrino telescopes, crucial to observe the smoking-gun sharp spectral feature out of a much broader neutrino background).

It has to be noted also that the analysis we performed does not apply only to scenarios where the relic particle decay exclusively into neutrinos. Most of the observable and unaffected region of parameter space that we found is expected to be valid also, for instance, for a scenario where the decay would proceed equally to neutrinos and charged leptons, because for the high masses that this region spans, the fraction of electromagnetic/hadronic energy that a neutrino injects is not far from unity, as from charged leptons. 

The general analysis we performed could be improved in many ways. The photo- and hadro-disintegration bounds that we considered, from Ref.~\cite{Bianco:2025boy}, should be determined for lifetimes below $10^4$ seconds. A precise determination of the experimental constraints on a PHENU flux would require several dedicated analysis directly from the data, at various energies all the way from a few MeV to beyond PeV. A precise determination of the energy spectrum distortion that interactions along the propagation of the neutrinos would induce, as a function of the input parameters, would also be very interesting, to determine in particular how much of the shaded region in the right panel of figure~\ref{fig:ratio} would still lead to an observable distinctive feature.

\acknowledgments{
We thank S. Bianco, N. Esser, G. Facchinetti, J. Frerick, S. Palomares-Ruiz, K. Schmidt-Hoberg and especially M. Hufnagel for useful discussions. This work is supported by the Belgian IISN convention 4.4503.15 and by the Brussels laboratory of the Universe - BLU-ULB. The work of NGY is further supported by the Communauté française de Belgique through a FRIA doctoral grant. Computational resources have been provided by the Consortium des Équipements de Calcul Intensif (CÉCI), funded by the Fonds de la Recherche Scientifique de Belgique (F.R.S.-FNRS) under Grant No. 2.5020.11 and by the Walloon Region.
}

\appendix

\section{Emission during the matter dominated era} \label{app:matter-dominated}

\begin{figure}[t!]
    \centering
    \includegraphics[width = 0.49\textwidth]{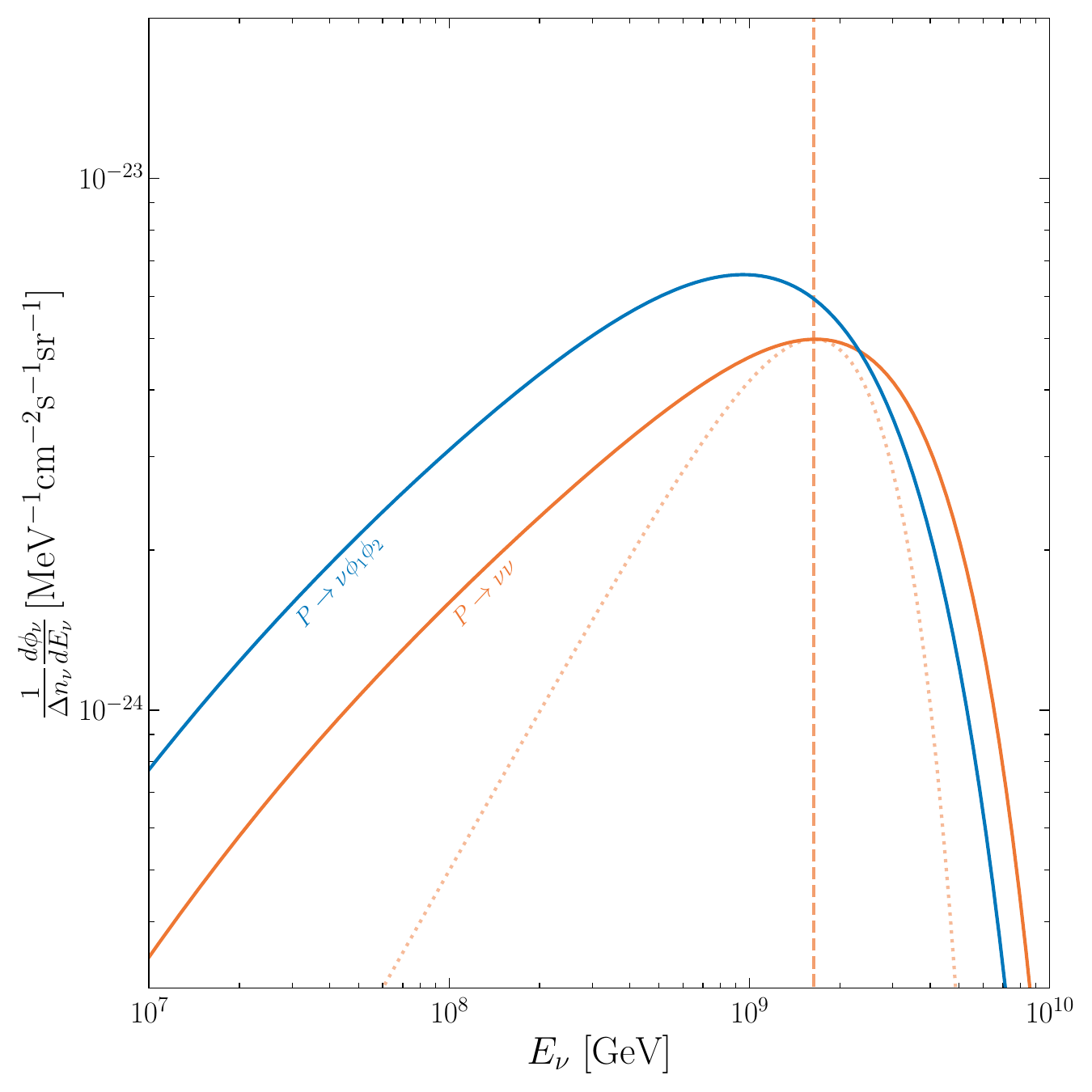}
    \includegraphics[width=0.49\linewidth]{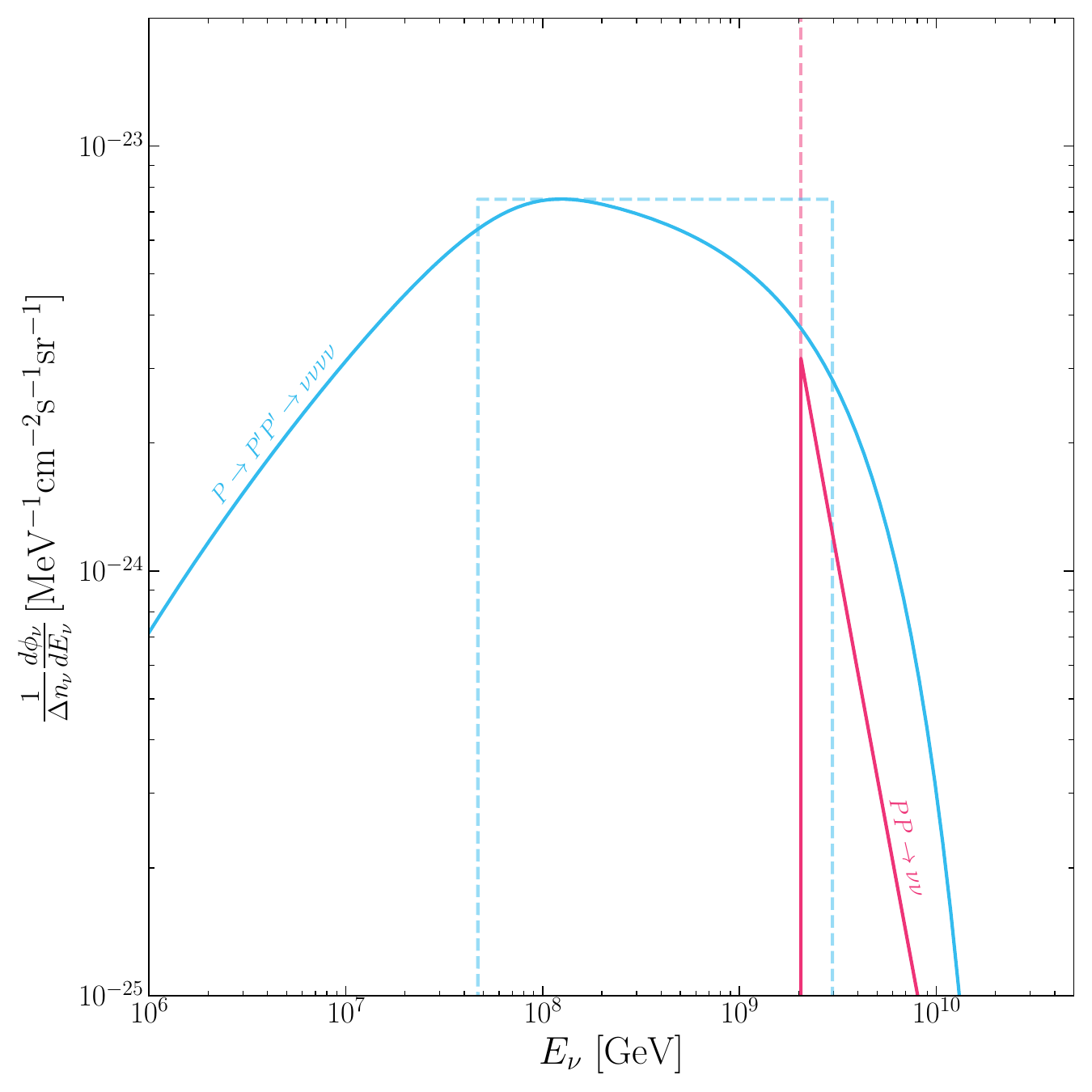}
    \caption{Same as figure~\ref{fig:neutrino_flux} but for an emission during the matter dominated era, $m_P = 10^{11}$ GeV and lifetime $\tau_P = 10^{16}$~s. The orange dotted line on the left panel corresponds to the (slightly sharper) 2-body flux emitted during the radiation era, as given in figure~\ref{fig:neutrino_flux} (rescaled for comparison).}
    \label{fig:neutrino_flux_mat}
\end{figure}

For an emission during the matter dominated era, the 4 types of sharp spectral features considered above are not the same as for an emission during the radiation dominated era, figure~\ref{fig:neutrino_flux}. They are given in figure~\ref{fig:neutrino_flux_mat}. The difference between both cases is sizeable. 
As can be seen in this figure the 2-body spectrum for the matter dominated case, eq.~\eqref{dnudERadmat}, is about a factor $1.3$ wider than the 2-body spectrum for the radiation dominated case, eq.~\eqref{dnudERadrad}. As already mentioned above, just above these equations, the difference is due to a different time dependence of the Hubble constant in eq.~\eqref{dnudERad}. For such a 2-body decay, the energy at the maximum of the spectrum, the average energy, and the corresponding emission times become
\begin{equation} \label{eq:enu_max_mat}
E_\nu^{\text{max}}= \frac{m_P}{2} \left(\frac{{\tau_P}}{3t_m}\right)^{2/3}\,,  \quad\quad \bar E_\nu  = \frac{m_P}{2}\Gamma(5/3)
\left(\frac{\tau_P}{t_m}\right)^{2/3}\,,
\end{equation}
\begin{equation} \label{eq:t_max_mat}
t^{\text{max}}= \frac{\tau_P}{3}\,, \quad\quad  \bar{t}  = \tau_P\, \Gamma^{3/2}(5/3)\,,
\end{equation}
to be compared with eqs.~(\ref{eq:enu_max_rad}) and (\ref{eq:t_max_rad}).

As for the 3-body decay flux in the matter dominated era, it is given approximately by
\begin{equation}
    \frac{dN_\nu}{dE_\nu} = \frac{6 \Omega_P^0 \rho_{\text{crit}}^0 \sqrt{E\nu}}{m_P^4}\frac{t_m}{\tau_P}\left[\sqrt{2} m_P^{3/2} E_2\left(2\sqrt{2} \left(\frac{E_\nu^{\text{max}}}{m_P}\right)^{3/2}\frac{t_m}{\tau_P}\right) - 4 E_\nu^{3/2} E_2\left(\frac{t_m}{\tau_P}\right)\right]\,,
\end{equation}
to be compared with eq.~(\ref{eq:flux_3body_rad}). 
In this equation $E_1$ and $E_2$ are the exponential integral function $E_n(z)$. 

For a 3-body decay the energy at the maximum and average energy are
\begin{equation}
\begin{split}
(E_\nu^{\text{max}})^{3/2}&=\frac{m_P^{3/2} \tau_P e^{-2\sqrt{2} \left(\frac{E_\nu^{\text{max}}}{m_P}\right)^{3/2}\frac{t_m}{\tau_P}}}{8\sqrt{2}}\left[t_m E_1\left(2\sqrt{2} \left(\frac{E_\nu^{\text{max}}}{m_P}\right)^{3/2}\frac{t_m}{\tau_P}\right) + \tau_P E_2\left(\frac{t_m}{\tau_P}\right)\right]^{-1}\,, \\
\bar E_\nu  &= \frac{9 m_P \left[t_m^2 (3t_m + 8 \tau_P)e^{-t_m/\tau_P} + \left(\frac{t_m}{\tau_P}\right)^{1/3}\tau_P^3(8\Gamma(8/3) - 3\Gamma(11/3, t_m/\tau_P))\right]}{320 t_m \tau_P^2 (1 - e^{-\frac{t_m}{\tau_P}})}\,,
\label{Emax3body_mat}
\end{split}
\end{equation}
where $\Gamma(s, x)$ is the incomplete gamma function. For $\tau_P \ll t_m$, i.e.\ for decays that happen well before today, the average energy simplifies further to:

\begin{equation}
    \bar E_\nu = \frac{9 \Gamma(8/3)}{40}m_P\left( \frac{\tau_P}{t_m}\right)^{2/3}\;.
\end{equation}

\section{Sharpness of the signal in the radiation dominated case}\label{app:sharpness}

\begin{figure}[t!]
    \centering
    \includegraphics[width=0.49\linewidth]{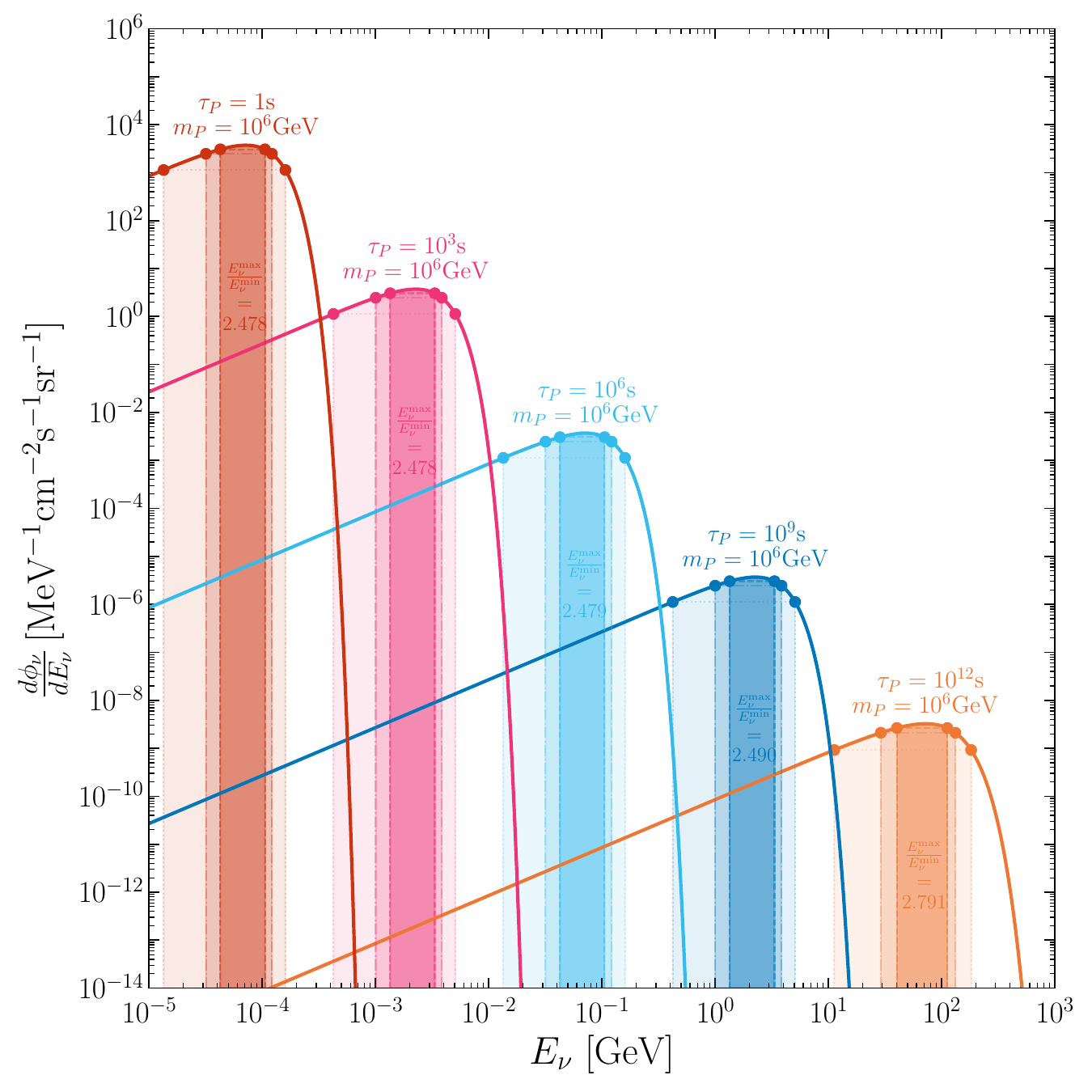}
    \includegraphics[width=0.49\linewidth]{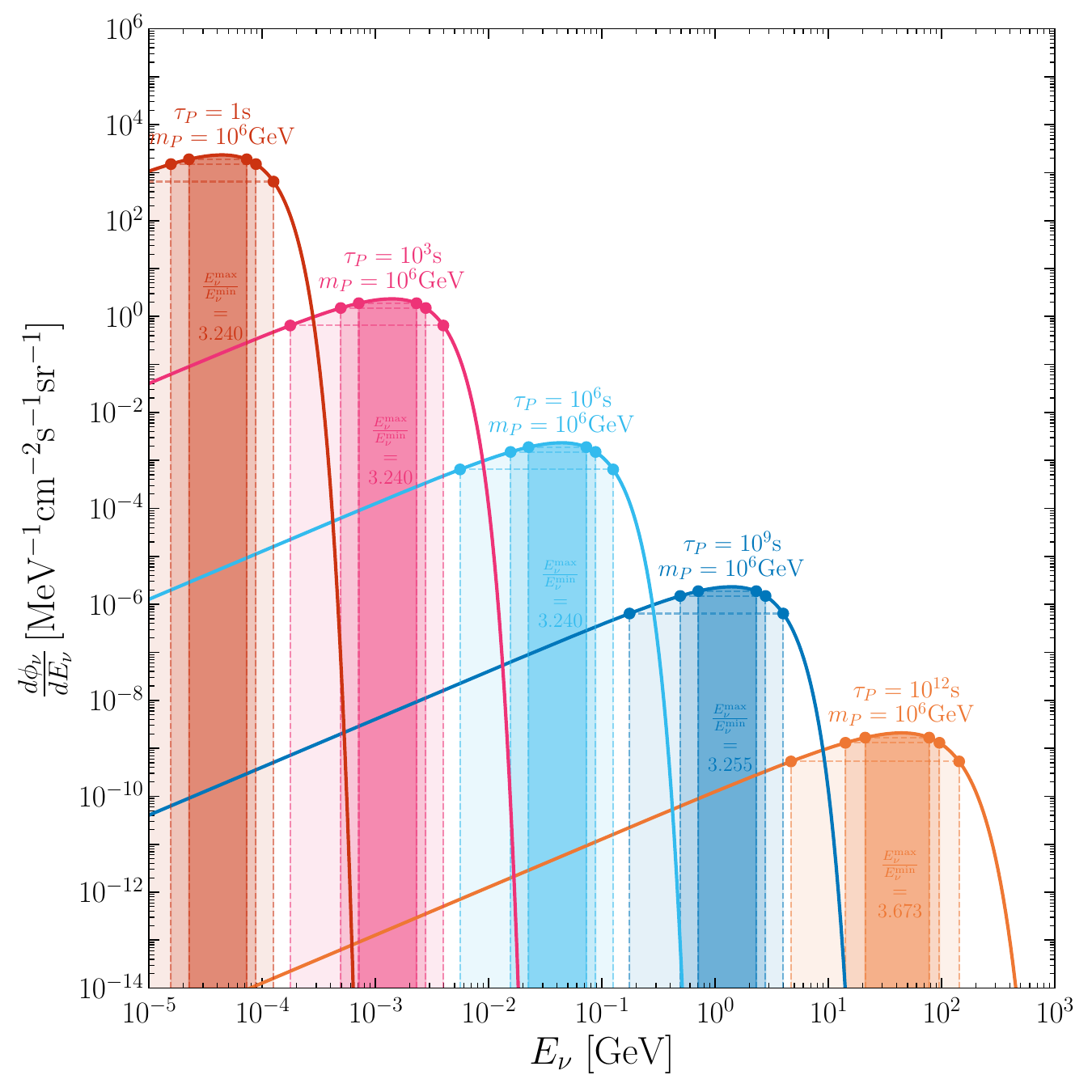}
    \caption{Energy intervals for which 50\%, 67\% and 90\% of the neutrinos are contained for a two-body decay (left) and a three-body decay (right), for various values of lifetime $\tau_P$. The darker shadowed region corresponds to the 50\% interval while the lighter one corresponds to the 90\% one.}
    \label{fig:ratiospectrum}
\end{figure}

To quantify the spreading of the line spectrum due to the redshift effect in the 2-body and 3-body cases, it can be useful to plot, as we do in figure~\ref{fig:ratiospectrum}, the energy interval within which, for an emission during the radiation dominated epoch,
50\%, 67\% and 90\% of the neutrinos of the flux are contained. 
The energy intervals shown on figure~\ref{fig:ratiospectrum} are obtained by requiring that the flux has the same value at their lower and upper bounds.

\section{Cross-sections}\label{app:cross-section}

In this appendix, we give the analytic expression for the scattering cross-sections used in section~\ref{sec:scattering}, computed using \textsc{FeynCalc}~10 \cite{MERTIG1991345, Shtabovenko:2023idz}. The elastic cross-sections are given by
\begin{equation}\label{eq:elastic_cross_section}
    \frac{d\sigma}{dt} = \frac{g^4}{64\pi \cos(\theta_W)^4}\frac{C(s, t, m_Z)}{(t-m_Z^2)^2}\;,
\end{equation}
where $t$ and $s$ are the Mandelstam variables, $\theta_W$ is the Weinberg angle, $g$ is the $SU(2)_L$ coupling constant and the $C$ factor is given in table~\ref{tab:neutrino-cross-section}.

\renewcommand{\arraystretch}{2}
\begin{table}
    \centering
    \begin{tabular}{|c|c|}
        \hline
        Process & $C(s, t, m_Z)$ \\ 
        \hline
        $\nu_l \overline{\nu}_l \to \nu_l \overline{\nu}_l$ & $\frac{(s+t)^2 (s+t - 2 m_Z^2)^2}{s^2 (s-m_Z^2)^2}$ \\
        $\nu_l \overline{\nu}_l \to \nu_{l'} \overline{\nu}_{l'}$ & $\frac{(s+t)^2 (t-m_Z^2)^2}{s^2 (s-m_Z^2)^2}$\\
        $\nu_l \nu_l \to \nu_l \nu_l$ & $\frac{(s+2m_Z^2)^2}{(s+t+m_Z^2)^2}$ \\
        $\nu_l \overline{\nu}_{l'} \to \nu_l \overline{\nu}_{l'}$ & $\frac{(s+t)^2}{s^2}$ \\
        $\nu_l \nu_{l'} \to \nu_l \nu_{l'}$ & 1 \\
        \hline
    \end{tabular}
    \caption{C factors for the elastic scattering cross-sections in eq.~\eqref{eq:elastic_cross_section}.}
    \label{tab:neutrino-cross-section}
\end{table}

For an inelastic scattering into a charged lepton pair $\nu_l \overline{\nu}_l \to l^- l^+$, the cross-section is
\begin{equation}
\frac{d\sigma}{dt} = \frac{g^4}{64\pi s^2 (s-m_Z^2) (t-m_W^2)} \Bigg\{K_1
+ \frac{s-m_Z^2}{t-m_W^2}K_2 + \frac{(t-m_W^2)\tan(\theta_W)^4}{s-m_Z^2} K_3\Bigg\}\;,
\end{equation}
with
\begin{align*}
    K_1 &= (2\sin(\theta_W)^2 -1)\left(s\frac{m_l^4}{2 m_W^2} + (s+t-m_l^2)^2\right)
+ 2 \sin(\theta_W)^2 m_l^2 \left(s + \frac{(m_l^2 -t)^2}{2 m_W^2}\right)\;, \\
    K_2 &= (s+t)(s+t+2m_l^2) + m_l^4\left(1 + \frac{s}{m_W^2} + \frac{(m_l^2 - t)^2}{4 m_W^4}\right)\;, \\
    K_3 &= s^2 + 2(m_l^4 -2m_l^2 t + t(s+t))
- \frac{m_l^2 s + (s+t-m_l^2)^2}{\sin(\theta_W)^2} + \left(\frac{s+t-m_l^2}{2\sin(\theta_W)^2}\right)^2\;.
\end{align*}

For an inelastic scattering into a different lepton flavour $\nu_l \overline{\nu}_l \to l'^+ l'^-$, the cross-section is:

\begin{equation}
   \frac{d\sigma}{dt} = \frac{g^4 \tan(\theta_W)^4}{64 \pi s^2 (s - m_Z^2)^2} K_3\;.
\end{equation}
All cross-sections have also been computed using \textsc{MadGraph}5 \cite{Alwall:2014hca}.
\section{Additional plots}\label{app:plots}

\begin{figure}[t!]
    \centering
    \includegraphics[width=0.49\linewidth]{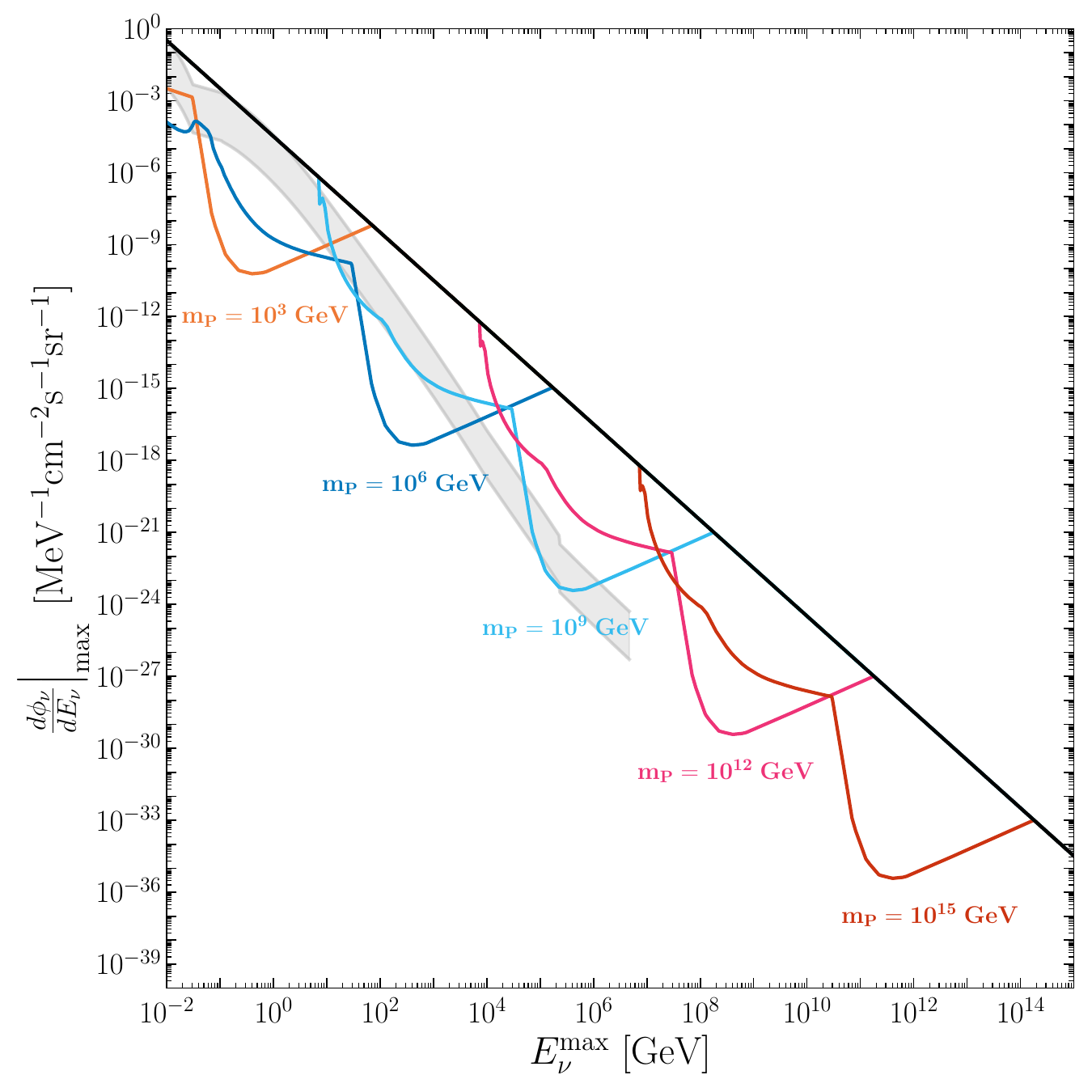}
    \caption{Theoretical constraints in the plane of the experimental inputs $E_\nu^{\text{max}}$ and $\left.d\phi_\nu /d E_\nu\right|_\text{max}$, the energy and the intensity of the flux at the maximum of the spectrum. Each line corresponds to the BBN + CMB constraint for a given mass $m_P$, and the black line corresponds to the limit from $\Delta N_{\text{eff}}<0.29$. Also shown is the region, in grey, for which the flux is contained within $f_P^{\text{exp-max}}$ (i.e.~the constraint on the abundance from experimental observations) and $10^{-2}f_P^{\text{exp-max}}$. For $m_P = 1$~GeV, the constraint from $\Delta N_{\text{eff}}$ (in black) is the most stringent. }
    \label{fig:reconstruction_emax_phi}
\end{figure}

Figure~\ref{fig:reconstruction_emax_phi} gives the regions of experimental input values $E_\nu^{\text{max}}$ and $\left.\frac{d\phi_\nu}{d E_\nu}\right|_\text{max}$ that are allowed by theoretical and experimental constraints for various values of $m_P$, see the discussion in section~\ref{reconstructionsection}.

\begin{figure}[t!]
    \centering
    \includegraphics[width=0.49\textwidth]{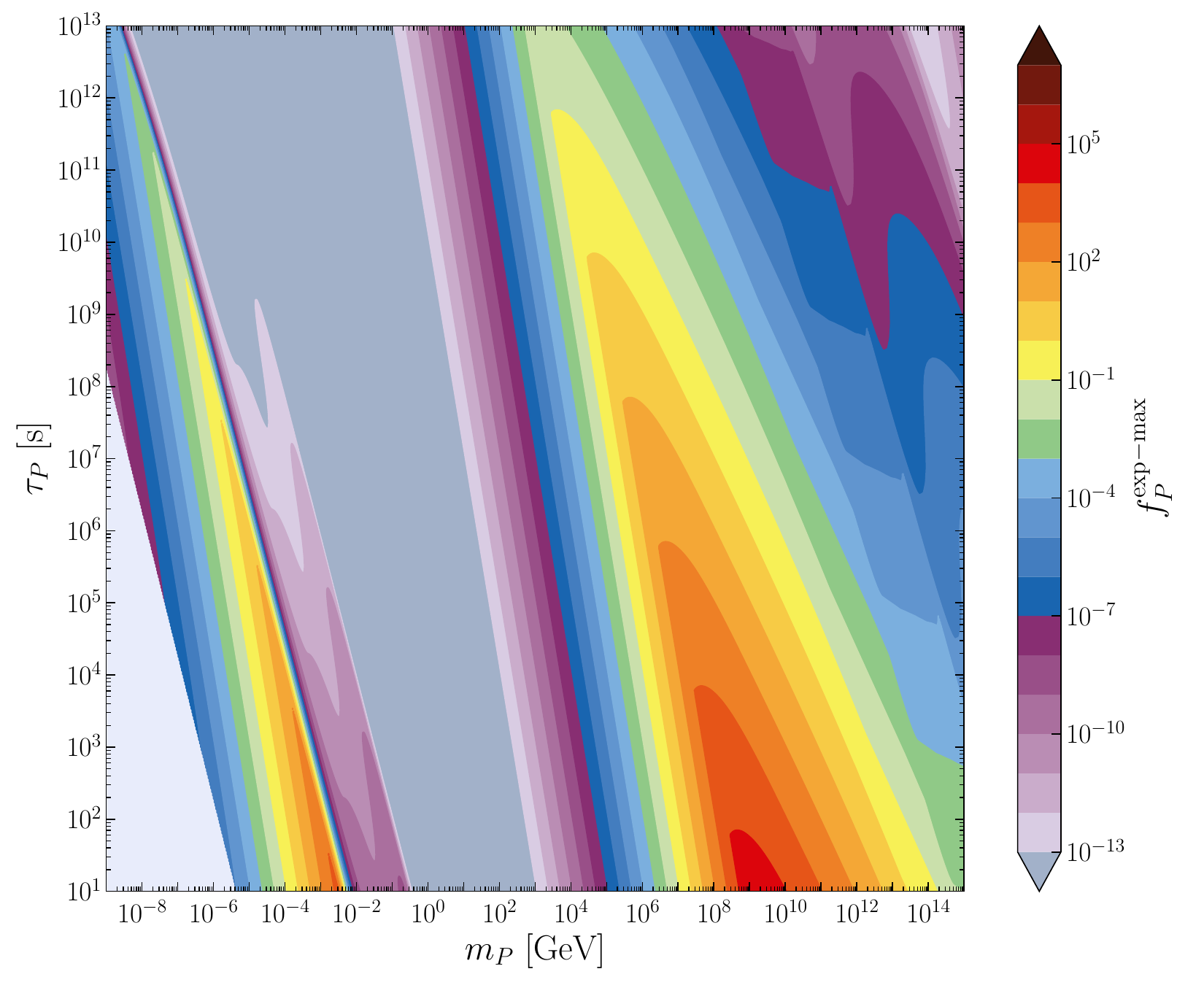}
    \includegraphics[width=0.49\textwidth]{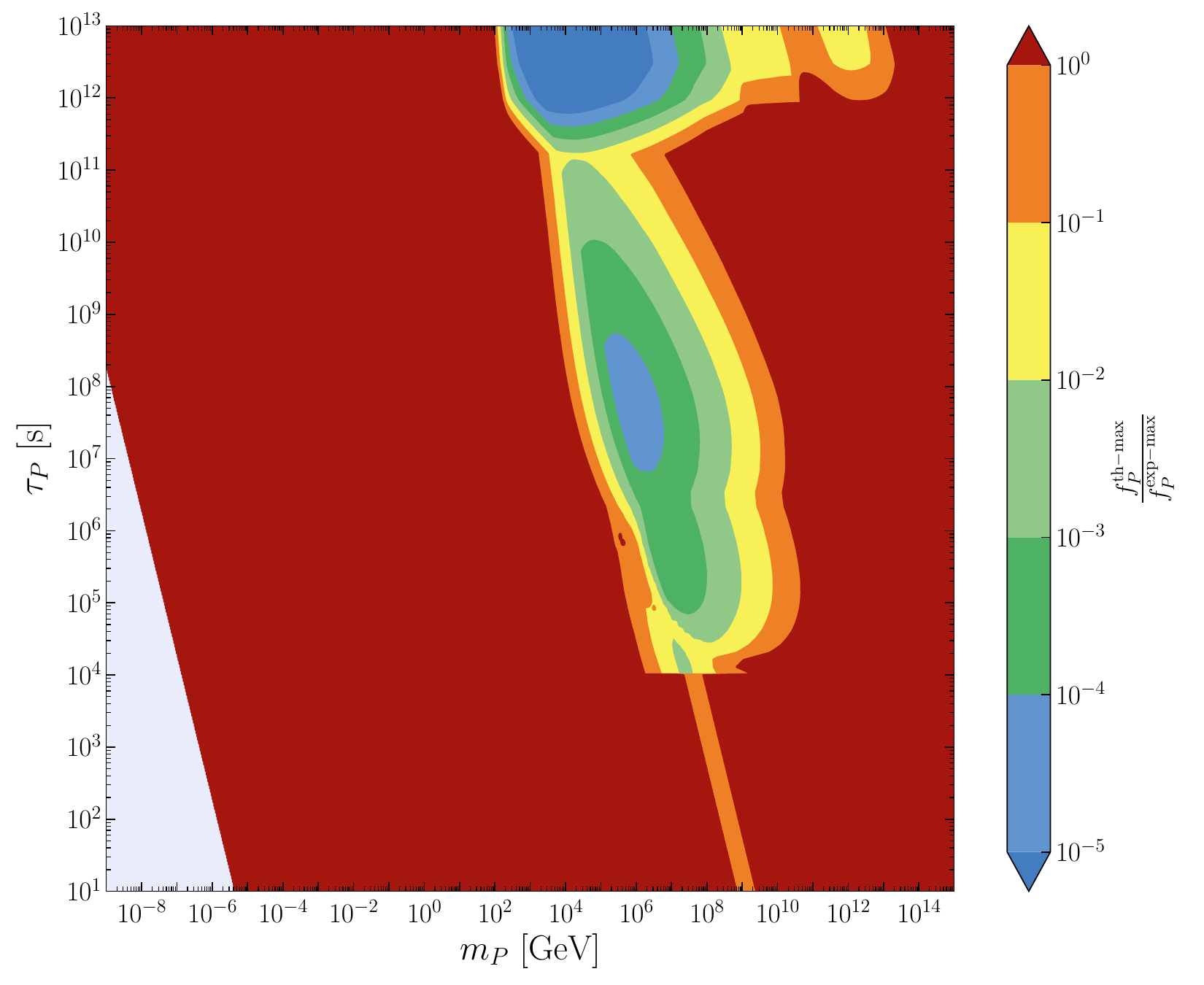}
    \caption{Same as the right panel of figure~\ref{fig:f_p_upper_limits} (left panel) and same as the left panel of figure~\ref{fig:ratio} (right panel),  assuming that the PHENU flux is nowhere larger than the measured and expected isotropic fluxes (i.e.~all fluxes shown in figure~\ref{fig:neff_bound}).}
    \label{fig:f_p_upper_limits_expected}
\end{figure}

Finally, out of curiosity, if instead of considering only the observed isotropic fluxes, one were to assume that all expected but not yet measured isotropic fluxes (C$\nu$B, BBN neutrinos, cosmogenic) were also observed as expected, then the experimental upper bound on $f_P$ shown in the right panel of figure~\ref{fig:f_p_upper_limits} would be replaced by the one in the left panel of figure~\ref{fig:f_p_upper_limits_expected}. In this case the left panel of figure~\ref{fig:ratio} would be replaced by the right panel of figure~\ref{fig:f_p_upper_limits_expected}.
This is a completely unrealistic situation but this sets the source particle abundance which would give a signal at the same level as the observed and expected isotropic fluxes.

\bibliographystyle{JHEP}
\bibliography{biblio}

\providecommand{\href}[2]{#2}\begingroup\raggedright\begin{thebibliography}{10}

\bibitem{Frampton:1980rs}
P.H.~Frampton and S.L.~Glashow, \emph{{Unstable heavy particles}}, \href{https://doi.org/10.1103/PhysRevLett.44.1481}{\emph{Phys. Rev. Lett.} {\bfseries 44} (1980) 1481}.

\bibitem{Gondolo:1991rn}
P.~Gondolo, G.~Gelmini and S.~Sarkar, \emph{{Cosmic neutrinos from unstable relic particles}}, \href{https://doi.org/10.1016/0550-3213(93)90199-Y}{\emph{Nucl. Phys. B} {\bfseries 392} (1993) 111} [\href{https://arxiv.org/abs/hep-ph/9209236}{{\ttfamily hep-ph/9209236}}].

\bibitem{Kanzaki:2007pd}
T.~Kanzaki, M.~Kawasaki, K.~Kohri and T.~Moroi, \emph{{Cosmological Constraints on Neutrino Injection}}, \href{https://doi.org/10.1103/PhysRevD.76.105017}{\emph{Phys. Rev. D} {\bfseries 76} (2007) 105017} [\href{https://arxiv.org/abs/0705.1200}{{\ttfamily 0705.1200}}].

\bibitem{Ema:2013nda}
Y.~Ema, R.~Jinno and T.~Moroi, \emph{{Cosmic-Ray Neutrinos from the Decay of Long-Lived Particle and the Recent IceCube Result}}, \href{https://doi.org/10.1016/j.physletb.2014.04.021}{\emph{Phys. Lett. B} {\bfseries 733} (2014) 120} [\href{https://arxiv.org/abs/1312.3501}{{\ttfamily 1312.3501}}].

\bibitem{Ema:2014ufa}
Y.~Ema, R.~Jinno and T.~Moroi, \emph{{Cosmological Implications of High-Energy Neutrino Emission from the Decay of Long-Lived Particle}}, \href{https://doi.org/10.1007/JHEP10(2014)150}{\emph{JHEP} {\bfseries 10} (2014) 150} [\href{https://arxiv.org/abs/1408.1745}{{\ttfamily 1408.1745}}].

\bibitem{McKeen:2018xyz}
D.~McKeen, \emph{{Cosmic neutrino background search experiments as decaying dark matter detectors}}, \href{https://doi.org/10.1103/PhysRevD.100.015028}{\emph{Phys. Rev. D} {\bfseries 100} (2019) 015028} [\href{https://arxiv.org/abs/1812.08178}{{\ttfamily 1812.08178}}].

\bibitem{Jaeckel:2020oet}
J.~Jaeckel and W.~Yin, \emph{{Boosted Neutrinos and Relativistic Dark Particles as Messengers from Reheating}}, \href{https://doi.org/10.1088/1475-7516/2021/02/044}{\emph{JCAP} {\bfseries 02} (2021) 044} [\href{https://arxiv.org/abs/2007.15006}{{\ttfamily 2007.15006}}].

\bibitem{IceCube:2013low}
{\scshape IceCube} collaboration, \emph{{Evidence for High-Energy Extraterrestrial Neutrinos at the IceCube Detector}}, \href{https://doi.org/10.1126/science.1242856}{\emph{Science} {\bfseries 342} (2013) 1242856} [\href{https://arxiv.org/abs/1311.5238}{{\ttfamily 1311.5238}}].

\bibitem{Halzen:1995hu}
F.~Halzen, B.~Keszthelyi and E.~Zas, \emph{{Neutrinos from primordial black holes}}, \href{https://doi.org/10.1103/PhysRevD.52.3239}{\emph{Phys. Rev. D} {\bfseries 52} (1995) 3239} [\href{https://arxiv.org/abs/hep-ph/9502268}{{\ttfamily hep-ph/9502268}}].

\bibitem{Lunardini:2019zob}
C.~Lunardini and Y.F.~Perez-Gonzalez, \emph{{Dirac and Majorana neutrino signatures of primordial black holes}}, \href{https://doi.org/10.1088/1475-7516/2020/08/014}{\emph{JCAP} {\bfseries 08} (2020) 014} [\href{https://arxiv.org/abs/1910.07864}{{\ttfamily 1910.07864}}].

\bibitem{Wu:2024uxa}
Q.-f.~Wu and X.-J.~Xu, \emph{{High-energy and ultra-high-energy neutrinos from Primordial Black Holes}}, \href{https://doi.org/10.1088/1475-7516/2025/02/059}{\emph{JCAP} {\bfseries 02} (2025) 059} [\href{https://arxiv.org/abs/2409.09468}{{\ttfamily 2409.09468}}].

\bibitem{Planck:2018vyg}
{\scshape Planck} collaboration, \emph{{Planck 2018 results. VI. Cosmological parameters}}, \href{https://doi.org/10.1051/0004-6361/201833910}{\emph{Astron. Astrophys.} {\bfseries 641} (2020) A6} [\href{https://arxiv.org/abs/1807.06209}{{\ttfamily 1807.06209}}].

\bibitem{Ibarra:2012dw}
A.~Ibarra, S.~Lopez~Gehler and M.~Pato, \emph{{Dark matter constraints from box-shaped gamma-ray features}}, \href{https://doi.org/10.1088/1475-7516/2012/07/043}{\emph{JCAP} {\bfseries 07} (2012) 043} [\href{https://arxiv.org/abs/1205.0007}{{\ttfamily 1205.0007}}].

\bibitem{Servant:2013uwa}
G.~Servant and S.~Tulin, \emph{{Baryogenesis and Dark Matter through a Higgs Asymmetry}}, \href{https://doi.org/10.1103/PhysRevLett.111.151601}{\emph{Phys. Rev. Lett.} {\bfseries 111} (2013) 151601} [\href{https://arxiv.org/abs/1304.3464}{{\ttfamily 1304.3464}}].

\bibitem{Dhen:2015wra}
M.~Dhen and T.~Hambye, \emph{{Inert Scalar Doublet Asymmetry as Origin of Dark Matter}}, \href{https://doi.org/10.1103/PhysRevD.92.075013}{\emph{Phys. Rev. D} {\bfseries 92} (2015) 075013} [\href{https://arxiv.org/abs/1503.03444}{{\ttfamily 1503.03444}}].

\bibitem{Bianco:2025boy}
S.~Bianco, P.F.~Depta, J.~Frerick, T.~Hambye, M.~Hufnagel and K.~Schmidt-Hoberg, \emph{{Photo- and Hadrodisintegration constraints on massive relics decaying into neutrinos}},  \href{https://arxiv.org/abs/2505.01492}{{\ttfamily 2505.01492}}.

\bibitem{Bierlich:2022pfr}
C.~Bierlich et~al., \emph{{A comprehensive guide to the physics and usage of PYTHIA 8.3}}, \href{https://doi.org/10.21468/SciPostPhysCodeb.8}{\emph{SciPost Phys. Codeb.} {\bfseries 2022} (2022) 8} [\href{https://arxiv.org/abs/2203.11601}{{\ttfamily 2203.11601}}].

\bibitem{Ovchynnikov:2024xyd}
M.~Ovchynnikov and V.~Syvolap, \emph{{Primordial Neutrinos and New Physics: Novel Approach to Solving the Neutrino Boltzmann Equation}}, \href{https://doi.org/10.1103/PhysRevLett.134.101003}{\emph{Phys. Rev. Lett.} {\bfseries 134} (2025) 101003} [\href{https://arxiv.org/abs/2409.15129}{{\ttfamily 2409.15129}}].

\bibitem{Vitagliano:2019yzm}
E.~Vitagliano, I.~Tamborra and G.~Raffelt, \emph{{Grand Unified Neutrino Spectrum at Earth: Sources and Spectral Components}}, \href{https://doi.org/10.1103/RevModPhys.92.045006}{\emph{Rev. Mod. Phys.} {\bfseries 92} (2020) 45006} [\href{https://arxiv.org/abs/1910.11878}{{\ttfamily 1910.11878}}].

\bibitem{KM3NeT:2025npi}
{\scshape KM3NeT} collaboration, \emph{{Observation of an ultra-high-energy cosmic neutrino with KM3NeT}}, \href{https://doi.org/10.1038/s41586-024-08543-1}{\emph{Nature} {\bfseries 638} (2025) 376}.

\bibitem{Hambye:2021moy}
T.~Hambye, M.~Hufnagel and M.~Lucca, \emph{{Cosmological constraints on the decay of heavy relics into neutrinos}}, \href{https://doi.org/10.1088/1475-7516/2022/05/033}{\emph{JCAP} {\bfseries 05} (2022) 033} [\href{https://arxiv.org/abs/2112.09137}{{\ttfamily 2112.09137}}].

\bibitem{Acharya:2019uba}
S.K.~Acharya and R.~Khatri, \emph{{CMB anisotropy and BBN constraints on pre-recombination decay of dark matter to visible particles}}, \href{https://doi.org/10.1088/1475-7516/2019/12/046}{\emph{JCAP} {\bfseries 12} (2019) 046} [\href{https://arxiv.org/abs/1910.06272}{{\ttfamily 1910.06272}}].

\bibitem{Poulin:2016nat}
V.~Poulin, P.D.~Serpico and J.~Lesgourgues, \emph{{A fresh look at linear cosmological constraints on a decaying dark matter component}}, \href{https://doi.org/10.1088/1475-7516/2016/08/036}{\emph{JCAP} {\bfseries 08} (2016) 036} [\href{https://arxiv.org/abs/1606.02073}{{\ttfamily 1606.02073}}].

\bibitem{Chluba:2020oip}
J.~Chluba, A.~Ravenni and S.K.~Acharya, \emph{{Thermalization of large energy release in the early Universe}}, \href{https://doi.org/10.1093/mnras/staa2131}{\emph{Mon. Not. Roy. Astron. Soc.} {\bfseries 498} (2020) 959} [\href{https://arxiv.org/abs/2005.11325}{{\ttfamily 2005.11325}}].

\bibitem{Fu:2020wkq}
H.~Fu, M.~Lucca, S.~Galli, E.S.~Battistelli, D.C.~Hooper, J.~Lesgourgues et~al., \emph{{Unlocking the synergy between CMB spectral distortions and anisotropies}}, \href{https://doi.org/10.1088/1475-7516/2021/12/050}{\emph{JCAP} {\bfseries 12} (2021) 050} [\href{https://arxiv.org/abs/2006.12886}{{\ttfamily 2006.12886}}].

\bibitem{PhysRevD.85.052007}
{\scshape Super-Kamiokande Collaboration} collaboration, \emph{Supernova relic neutrino search at super-kamiokande}, \href{https://doi.org/10.1103/PhysRevD.85.052007}{\emph{Phys. Rev. D} {\bfseries 85} (2012) 052007}.

\bibitem{Super-Kamiokande:2015qek}
{\scshape Super-Kamiokande} collaboration, \emph{{Measurements of the atmospheric neutrino flux by Super-Kamiokande: energy spectra, geomagnetic effects, and solar modulation}}, \href{https://doi.org/10.1103/PhysRevD.94.052001}{\emph{Phys. Rev. D} {\bfseries 94} (2016) 052001} [\href{https://arxiv.org/abs/1510.08127}{{\ttfamily 1510.08127}}].

\bibitem{IceCube:2017zho}
{\scshape IceCube} collaboration, \emph{{The IceCube Neutrino Observatory - Contributions to ICRC 2017 Part II: Properties of the Atmospheric and Astrophysical Neutrino Flux}},  \href{https://arxiv.org/abs/1710.01191}{{\ttfamily 1710.01191}}.

\bibitem{Super-Kamiokande:2021jaq}
{\scshape Super-Kamiokande} collaboration, \emph{{Diffuse supernova neutrino background search at Super-Kamiokande}}, \href{https://doi.org/10.1103/PhysRevD.104.122002}{\emph{Phys. Rev. D} {\bfseries 104} (2021) 122002} [\href{https://arxiv.org/abs/2109.11174}{{\ttfamily 2109.11174}}].

\bibitem{Harada:2023apz}
{\scshape Super-Kamiokande} collaboration, \emph{{First result of a search for Diffuse Supernova Neutrino Background in SK-Gd experiment}}, \href{https://doi.org/10.22323/1.444.1173}{\emph{PoS} {\bfseries ICRC2023} (2023) 1173}.

\bibitem{IceCube:2023ies}
{\scshape IceCube} collaboration, \emph{{Search for neutrino lines from dark matter annihilation and decay with IceCube}}, \href{https://doi.org/10.1103/PhysRevD.108.102004}{\emph{Phys. Rev. D} {\bfseries 108} (2023) 102004} [\href{https://arxiv.org/abs/2303.13663}{{\ttfamily 2303.13663}}].

\bibitem{ElAisati:2015ugc}
C.~El~Aisati, M.~Gustafsson and T.~Hambye, \emph{{New Search for Monochromatic Neutrinos from Dark Matter Decay}}, \href{https://doi.org/10.1103/PhysRevD.92.123515}{\emph{Phys. Rev. D} {\bfseries 92} (2015) 123515} [\href{https://arxiv.org/abs/1506.02657}{{\ttfamily 1506.02657}}].

\bibitem{Kawasaki:2017bqm}
M.~Kawasaki, K.~Kohri, T.~Moroi and Y.~Takaesu, \emph{{Revisiting Big-Bang Nucleosynthesis Constraints on Long-Lived Decaying Particles}}, \href{https://doi.org/10.1103/PhysRevD.97.023502}{\emph{Phys. Rev. D} {\bfseries 97} (2018) 023502} [\href{https://arxiv.org/abs/1709.01211}{{\ttfamily 1709.01211}}].

\bibitem{MERTIG1991345}
R.~Mertig, M.~Böhm and A.~Denner, \emph{Feyn calc - computer-algebraic calculation of feynman amplitudes}, \href{https://doi.org/https://doi.org/10.1016/0010-4655(91)90130-D}{\emph{Computer Physics Communications} {\bfseries 64} (1991) 345}.

\bibitem{Shtabovenko:2023idz}
V.~Shtabovenko, R.~Mertig and F.~Orellana, \emph{{FeynCalc 10: Do multiloop integrals dream of computer codes?}}, \href{https://doi.org/10.1016/j.cpc.2024.109357}{\emph{Comput. Phys. Commun.} {\bfseries 306} (2025) 109357} [\href{https://arxiv.org/abs/2312.14089}{{\ttfamily 2312.14089}}].

\bibitem{Alwall:2014hca}
J.~Alwall, R.~Frederix, S.~Frixione, V.~Hirschi, F.~Maltoni, O.~Mattelaer et~al., \emph{{The automated computation of tree-level and next-to-leading order differential cross sections, and their matching to parton shower simulations}}, \href{https://doi.org/10.1007/JHEP07(2014)079}{\emph{JHEP} {\bfseries 07} (2014) 079} [\href{https://arxiv.org/abs/1405.0301}{{\ttfamily 1405.0301}}].

\end{thebibliography}\endgroup

\end{document}